\documentclass[useAMS,usenatbib]{mn2e}

\usepackage{graphicx}

\def\sqdeg{\,deg$^2$}
\def\etal{{et~al.~}}

\def\msol{\,{\rm M_{\odot}}}
\def\kpch{\,{\it h}^{-1}\, {\rm kpc}}
\def\kpc{\,{\rm kpc}}

\def\mpc{\,{\rm Mpc}}

\def\kms{{\rm \,km \, s^{-1}}}

\def\mpc3h3{\,{\it h}^{-3}\, {\rm Mpc}^{3}}
\def\G3C{G$^3$C}

\title[GAMA: close-pairs and mergers]{Galaxy And Mass Assembly (GAMA): Galaxy close-pairs, mergers,  and the future fate of stellar mass}
\author[A.S.G. Robotham~\etal]{
A.S.G.~Robotham$^1$, 
S.P.~Driver$^{1,2}$,
L.J.M.~Davies$^1$,
A.M.~Hopkins$^3$,
I.K.~Baldry$^4$,
\newauthor
N.K.~Agius$^5$,
A.E.~Bauer$^3$,
J.~Bland-Hawthorn$^6$,
S.~Brough$^3$,
M.J.I.~Brown$^7$,
\newauthor
M.~Cluver$^8$,
R.~De Propris$^9$,
M.J.~Drinkwater$^{10}$,
B.W.~Holwerda$^{11}$,
L.S.~Kelvin$^{12}$,
\newauthor
M.A.~Lara-Lopez$^3$,
J.~Liske$^{13}$,
\'A.R.~L\'opez-S\'anchez$^{3,14}$,
J.~Loveday$^{15}$,
S.~Mahajan$^{10,16}$,
\newauthor
T.~McNaught-Roberts$^{17}$,
A.~Moffett$^1$,
P.~Norberg$^{17}$,
D.~Obreschkow$^1$,
M.S.~Owers$^3$,
\newauthor
S.J.~Penny$^7$,
K.~Pimbblet$^{7,18}$,
M.~Prescott$^{19}$,
E.N.~Taylor$^{20}$,
E.~van~Kampen$^{13}$,
\newauthor
\smash{S.M.~Wilkins$^{15}$}\\
$^1$ICRAR, The University of Western Australia, 35 Stirling Highway, Crawley, WA 6009, Australia\\
$^2$School of Physics \& Astronomy, University of St Andrews, North Haugh, St Andrews, KY16 9SS, UK\\
$^3$Australian Astronomical Observatory, PO Box 915, North Ryde, NSW 1670, Australia\\
$^4$Astrophysics Research Institute, Liverpool John Moores University, Egerton Wharf, Birkenhead, CH41 1LD, UK\\
$^5$Jeremiah Horrocks Institute, University of Central Lancashire, Preston PR1 2HE, UK\\
$^6$Sydney Institute for Astronomy, School of Physics, University of Sydney, NSW 2006, Australia\\
$^7$School of Physics, Monash University, Clayton, Victoria 3800, Australia\\
$^8$ACGC, Department of Astronomy, University of Cape Town, Rondebosch, 7701, South Africa\\
$^9$Finnish Centre for Astronomy with ESO, University of Turku, V\"ais\"al\"antie 20, Piikki\"o, 21500 Finland\\
$^{10}$Department of Physics, The University of Queensland, Brisbane, QLD 4072, Australia\\
$^{11}$University of Leiden, Sterrenwacht Leiden, Niels Bohrweg 2, NL-2333 CA Leiden, The Netherlands\\
$^{12}$Institut f\"ur Astro- und Teilchenphysik, Universit\"at Innsbruck, Technikerstra{\ss}e 25, 6020 Innsbruck, Austria\\
$^{13}$European Southern Observatory, Karl-Schwarzschild-Str.~2, 85748 Garching, Germany\\
$^{14}$Department of Physics and Astronomy, Macquarie University, NSW 2109, Australia\\
$^{15}$Astronomy Centre, University of Sussex, Falmer, Brighton BN1 9QH\\
$^{16}$Indian Institute of Science Education and Research Mohali, Knowledge City, S.A.S Nagar, Manauli, PO 140306, India\\
$^{17}$Institute for Computational Cosmology, Department of Physics, Durham University, South Road, Durham DH1 3LE, UK\\
$^{18}$Department of Physics and Mathematics, University of Hull, Cottingham Road, Kingston-upon-Hull, HU6 7RX, UK\\
$^{19}$Astrophysics Group, Department of Physics, University of the Western Cape, 7535 Bellville, Cape Town, South Africa\\
$^{20}$School of Physics, University of Melbourne, 3010 VIC, Australia
}

\begin{document}

\date{\noindent18/04/2014}

\pagerange{\pageref{firstpage}--\pageref{lastpage}} \pubyear{2014}

\maketitle

\label{firstpage}

\begin{abstract}

We use a highly complete subset of the GAMA-II redshift sample to fully describe the stellar mass dependence of close-pairs and mergers between $10^8\msol$ and $10^{12}\msol$. Using the analytic form of this fit we investigate the total stellar mass accreting onto more massive galaxies across all mass ratios. Depending on how conservatively we select our robust merging systems, the fraction of mass merging onto more massive companions is $2.0$\%--$5.6$\%.

Using the GAMA-II data we see no significant evidence for a change in the close-pair fraction between redshift $z=0.05$--$0.2$. However, we find a systematically higher fraction of galaxies in similar mass close-pairs compared to published results over a similar redshift baseline. Using a compendium of data and the function $\gamma_{M}=A(1+z)^m$ to predict the major close-pair fraction, we find fitting parameters of $A=0.021 \pm 0.001$ and $m=1.53 \pm 0.08$, which represents a higher low-redshift normalisation and shallower power-law slope than recent literature values.

We find that the relative importance of in-situ star-formation versus galaxy merging is inversely correlated, with star-formation dominating the addition of stellar material below $\mathcal{M}^*$ and merger accretion events dominating beyond $\mathcal{M}^*$.

We find mergers have a measurable impact on the whole extent of the GSMF, manifest as a deepening of the `dip' in the GSMF over the next $\sim$Gyr and an increase in $\mathcal{M}^*$ by as much as 0.01--0.05~dex.

\end{abstract}


\pagebreak

\section{Introduction}

The material that resides in modern day (redshift $z\sim0$) galaxies is believed to have been built up through a number of distinct physical mechanisms. Some of it assembled at high redshift  during localised instances of monolithic collapse of gas forming ancient bound structures including central bulges and globular clusters \citep{serl78}. As time progressed more stellar material was assembled through the cooling and accretion of gas directly onto galaxy disks \citep{fall80}, providing a slow growing and long term source of renewed stars \citep[see][for a discussion regarding how these two major processes dominate the intrinsic radiation and stellar mass of galaxies today]{driv12,driv13}. In unison with these two dominant mechanisms for building stars in galaxies, other processes have played important roles in redistributing this mass between galaxies. Whilst some stellar material might become unbound by strong tidal forces between galaxies, the more typical scenario is the concentration of mass into fewer galaxies via galaxy-galaxy merging \citep[for an extensive review on hierarchical structure formation see][]{baug06}.

Galaxy mergers are expected to be a common occurrence over the lifetime of the Universe within the cold dark matter cosmological paradigm \citep{whit78}. The significant role of mergers in building up mass is seen both in dark matter simulations \citep[see][for the role of mergers in pure dark matter N-body simulations]{gene09,ste09} and hydrodynamical simulations \citep[see][for extensive discussion of the role of mergers in dark matter + gas hydrodynamical simulations]{mura02,mal06}. Also, they are believed to have a role in the production of AGN \citep{barn91,hopk08a,darg10b,coti13}, the transformation of galaxy morphology \citep{toom72,barn92b,hopk08b} and an associated impact on the apparent size of galaxies \citep{per14}, and are likely to have a significant role in modifying the star formation history \citep{barn91,barn92a,darg10b,xu12b,per14}. The modification of star formation is a complex process, and is likely to be a mixture of enhanced star formation in the late stages of major mergers \citep{ower07,hopk13,robo13} and the net suppression of star formation in minor close-pair galaxies during the earlier stages of galaxy-galaxy interactions before the final merger takes place \citep{robo13}.

Putting aside the galaxy scale effects of mergers, the process of combining stellar material in ever fewer halos and galaxies has important cosmological implications. Most obviously, mergers change the number of galaxies as a function of stellar mass, thus modifying the galaxy stellar mass function (GSMF) and luminosity function \citep{whit78}. The GSMF is considered a gold-standard product of cosmological galaxy formation simulations and models \citep[e.g. ][]{crot06,bowe06}, where the standard approach is to optimise the numeric or semi-analytic recipes to achieve the observed GSMF. Empirically and numerically measuring the fraction of galaxies that are currently merging as a function of stellar mass is a strong test of our understanding of the Universe since this offers an extra evolutionary characteristic for models to reproduce. The physics that produces such a `merger function' is highly complex, wrapping in dark matter clustering, halo occupancy (itself a combination of fuelling and feedback) and baryon dominated dynamical friction.

There is an advantage to making this measurement at low redshift since the quality of the data used to make such measurements (photometry and spectroscopy) is at its most complete and the complexity of simulating all the physics that have brought us to the present day Universe is at its highest. Measuring the merger rate in the low redshift Universe also allows us to make a direct prediction regarding the near future of the GSMF. Whilst merger estimates are likely to be most accurate at low redshift, concerted efforts have been made to make similar measurements at redshifts beyond 1, e.g.\ \citet{cons03,lotz08a,ryan08,bund09,brid10} {\it et seq}.

However, there are complexities to making this observational merger rate estimate over a large range of epochs \citep[e.g.][]{wil11}, even in simulations where we have an arbitrarily large quantity of information regarding the state of the system \citep[e.g.][]{gene09}. The two obvious routes are by analysing pre-merger states or post-merger products. A pre-merger state is a configuration of dynamically close galaxies, i.e.\ galaxies that are close in both projected position and velocity space. Much work has been carried out using the commonality of close-pairs to predict the near future merger of galaxies, where much of the groundwork for this sort of analysis is described in detail in \citet{patt00,patt02,depr05,masj06,ber06,depr07,masj08,ryan08,depr10,lope11} {\it et seq}. In this approach a small dynamical window of projection and velocity separation between galaxies is used to extract `close-pairs'. Once the commonality of such systems is known, various dynamical friction recipes can be applied to map these populations onto typical merger timescales \citep[e.g.][]{binn87,patt02,kitz08}. The uncertainty in this mapping is often similar to the implied timescales involved. Anecdotally this is clear once we recognise that the merging galaxies will have similar dynamical configurations (projected spatial separations and velocity separations) multiple times over the course of a single merger event, and clearly some of these are significantly closer to the moment of coalescence than others.

The second approach (post-merger products) considers galaxies with temporary signs of disturbance due to the kinematically violent nature of galaxies merging \citep[see e.g.][Casteels et~al. in prep. {\it et seq.}, and discussions therein]{abr03,cons03,lotz04,dokk05,lotz08b,cons09,lotz11,holw11}. A vast range of techniques have been considered for measuring asymmetry including the Gini coefficient \citep{abr03}, $M_{20}$ \citep{lotz04} and the Concentration Asymmetry and Smoothness \citep[CAS,][]{cons03} of galaxies, but all share a common theme of trying to identify post-merger distortion signatures. Mapping the commonality of these galaxy flux asymmetries onto merger timescales is a complex process. The mass ratios of galaxies involved in the merger has a direct impact on the longevity of tidally disrupted signatures in merger product galaxies \citep{lotz10a}, and even the gas content (i.e.\ wet versus dry merging) can affect the timescales of such light asymmetries by factors of a few \citep{lotz10b,holw11}. On top of these physically driven uncertainties there are significant observational effects that limit the confidence we can put on the estimate. Chief among these is the depth of imaging used in the analysis, where \citet{ji14} have recently demonstrated that the observed timescales for asymmetry is a strong function of the surface brightness limit of the data. Forward propagation of simulations that incorporate observational constraints is the best guide on how to make this timescale mapping, which limits the conclusions that can be drawn from a purely empirical analysis.

In between the pre and post-merger phases there is a short lived period of rapid merging of stellar material, causing signatures such as tidal tails, bridges and shells \citep[e.g.][]{hern05,patt05,depr07,darg10a}. This period, typically a few 100 Myr \citep{hern05}, is when merging galaxies are on a dissipative transfer orbit, i.e.\ orbital angular momentum of the galaxy-galaxy pair is rapidly transferred to stellar angular momentum within the product galaxy.

There is a caveat to the visual disturbance seen pre-merger, in that it is not always an extremely short-lived phase. Various dynamically loose configuration orbits can create long lived signs of disturbed stellar material \citep{lotz11,patt13}. however the expectation is that highly disturbed close-pairs will, on average, be more likely to merge on shorter timescales than close-pairs with no signs of visual disturbance \citep{hern05,depr07}. This is a reasonable proposition given that {\it all} galaxies will be tidally disturbed at some point immediately prior to merger. Visually dramatic pre-merger events were correctly interpreted as merger signatures in the early simulation work of \citet{toom72}, however the timescale of such phases is, on average, shorter compared to the longer lived progenitor galaxy orbits and product galaxy profile asymmetries. For this reason merger rate estimates have usually been based on measurements of the typicality of pre- and post-merger states.

The two routes of using pre or post-merger states have their advantages and disadvantages. Constructing a complete sample of dynamically close-pairs is observationally expensive due to the required spectroscopy (which ideally should be complete on small angular scales), however it has an advantage in that the stellar masses of the merging galaxies are directly observed. By looking at post-merger disturbance evidence the input stellar masses involved are strongly obfuscated, however it is observationally quite efficient since spectra are not necessarily a requirement--- for many purposes photo-z redshifts estimated from multi-band imaging suffice. In both cases there are strong caveats on how to convert the raw quantity measured (e.g.\ the fraction of galaxies in close-pairs or the degree of disturbance in the galaxy light distribution) into a merger {\it rate}, i.e.\ the number of events per unit volume per unit {\it time}. It is the time part that is especially hard to quantify, since our view of the Universe is effectively a static snapshot of a complex evolving baryonic process. Computer modelling and galaxy dynamical friction estimates give a guide to the likely timescale for a close-pair to become transformed into a `merged' but disturbed galaxy, however this mapping is highly uncertain/variable \citep[e.g.][{\it et seq.}]{cons06,kitz08,gene09,lotz10a,lotz10b,lotz11,holw11}.

This work makes use of the GAMA survey, a highly complete spectroscopic survey (discussed in detail below). The aim of this paper is to measure the analytic form of the stellar mass pre-merger close-pair distribution function, and use this to make predictions regarding the likely result of mergers soon to occur in the local Universe. In particular we are interested in using the stellar mass dependence of galaxy mergers to make an estimate of the net evolution that galaxy mergers will cause on the GSMF.

Section \ref{sec:data} discusses the data products used for this work. Section \ref{sec:correct} details the various biases and corrections that have to be considered in this analysis. Section \ref{sec:galmerge} presents the main empirical observations for galaxy close-pairs. Section \ref{sec:closepairfits} presents the analytic fits to the empirical observations, and the implications these have for the mass contained in mergers and the future evolution of the galaxy stellar mass function. Section \ref{sec:conclusions} summarises the major conclusions of this work.

For distances and densities we assume the same cosmology as used to generate our mock catalogues (which in turn was based on the Millennium simulation parameters), i.e.\ $\Omega_\Lambda=0.75$, $\Omega_M=0.25$ and $H_{0}=100\,{\rm km}\,{\rm s}^{-1}\,{\rm Mpc}^{-1}$. The exception to this are the stellar mass calculations, which use $\Omega_\Lambda=0.7$, $\Omega_M=0.3$ and $H_{0}=70\,{\rm km}\,{\rm s}^{-1}\,{\rm Mpc}^{-1}$. It is more common to see stellar masses quoted with close to native values (i.e.\ using $H_{0}=70\,{\rm km}\,{\rm s}^{-1}\,{\rm Mpc}^{-1}$ rather than $H_{0}=100\,{\rm km}\,{\rm s}^{-1}\,{\rm Mpc}^{-1}$), so we do not scale the stellar masses, and hence comoving mass densities have an $h^3$ rather than $h$ dependency, where $h=H_{0}/100\,{\rm km}\,{\rm s}^{-1}\,{\rm Mpc}^{-1}$. Regarding the disjoint in cosmology used, even at our high redshift extreme of $z=0.3$ the distances agree to $\sim$1\%, so this will not be the dominant error contribution to any of the parameter fits discussed later, and makes a negligible difference to quoted values of stellar mass and distance.

\section{Data}
\label{sec:data}

\subsection{GAMA}

The Galaxy And Mass Assembly project (GAMA) is a major new
multi-wavelength photometric and spectroscopic galaxy survey \citep{driv11}. The final
redshift survey will contain $\sim$300,000 redshifts to $r_{\rm AB}=19.8$~mag
over $\sim280$~\sqdeg, with a survey design aimed at providing an
exceptionally uniform spatial completeness
\citep{robo10,bald10,driv11}. 

Extensive details of the GAMA survey characteristics are given in
\citet{driv11}, with the survey input catalogue described in
\citet{bald10}, the spectroscopic processing outlined in \citet{hopk13b}, and the spectroscopic tiling algorithm explained in
\citet{robo10}. The first 3 years of data obtained are referred to as GAMA-I. The survey was extended
into GAMA-II, which has recently completed 3 of its 5 fields--- the 3 Northern
equatorial strips. This complete Northern equatorial data (called GAMA-II-N) are used in this work.
The GAMA-II-N redshifts used have been measured using the {\sc AUTOZ} code presented
in \citet{bald14}.

Briefly, the GAMA-II-N survey covers three regions  
each $12 \times 5$ degrees centred at 09h, 12h and 14h30m
(respectively G09, G12 and G15 from here). The GAMA-II
equatorial regions used for this work covers $\sim180$~\sqdeg\ to $r_{\rm AB}=19.8$
All regions are more than 98\% complete within this magnitude limit 
\citep[see][for details]{driv11}, with special
emphasis on a high close-pair completeness, which is greater than 97\%
for all galaxies at the physical scales investigated in this work. The GAMA-II-N data
is presented in full in Liske et al. 2014 (in prep).

\subsection{Pair Catalogue}
\label{sec:paircat}

Close galaxy interactions play a significant role in the evolution of galaxies \citep{robo13}. The process of creating the GAMA Galaxy Group Catalogue (\G3C) involved the construction of all galaxy pairs \citep[see][for details]{robo11}. These pairs include galaxies with cluster-scale radial (velocity: $\sim$1,000 ${\rm km}\,{\rm s}^{-1}$) and tangential (spatial: $\sim$Mpc) separations. Using the full pair catalogue would include galaxy pairs with very large dynamical times. Instead we select a narrow window of interaction phase space in order to preferentially extract galaxies that will be most affected by close galaxy-galaxy interactions, and that are most likely to merge in the {\it near} future (next few Gyrs). The pair sample selected here is based on that presented in \citet{robo12,robo13}, where we aimed to recover galaxy pairs that are similar to the MW Magellanic Clouds system.

We made three selections using different thresholds of projected spatial separation $r_{\rm sep}$ and radial velocity separation $v_{\rm sep}$: 

\begin{equation}
\begin{array}{lll}
\label{eqn:Pselect}
P_{r20v500}&=&\{r_{\rm sep}<20\kpch  \land v_{\rm sep}<500 \kms \}\\
P_{r50v500}&=&\{r_{\rm sep}<50\kpch  \land v_{\rm sep}<500 \kms \}\\
P_{r100v1000}&=&\{r_{\rm sep}<100\kpch \land  v_{\rm sep}<1000 \kms \}
\end{array}
\end{equation}

All three selections are commonly used in the literature. The input data was the full GAMA-II-N data taken from Tiling Catalogue 40, with a requirement that redshifts had to be greater than 0.01, and the galaxy SURVEY\_CLASS was greater than or equal to 3 (i.e.\ GAMA main survey, see Baldry et al. 2010 and Driver et al. 2011 for details).
$P_{r20v500}$ contains 3,057 galaxy-galaxy pairs, $P_{r50v500}$ contains 10,470 galaxy-galaxy pairs and $P_{r100v1000}$ contains 29,428 galaxy-galaxy pairs. These selections represent supersets of possible pairs. Extra cuts (discussed later in this paper) are applied to ensure a volume complete and unbiased pair catalogue.

\subsection{Mock Catalogue}
\label{sec:mockcats}
To test how well our assumptions about the physics of the Universe match reality, GAMA has access to a suite of mock catalogues \citep{mers13}. The mock catalogues were constructed by first populating the dark matter halos of the Millennium Simulation \citep{spri05} with galaxies, the positions and properties of which were predicted by the \citet{bowe06} description of the Durham semi-analytical model, GALFORM, and adjusted to match the GAMA survey luminosity function of \citet{love12}.

Nine mock catalogues were produced that have the same geometry and survey selection as GAMA-I. These mock catalogues were used extensively in the original construction and testing of the \G3C \citep{robo11}, and in this work they again play a vital role--- allowing us to determine which aspects of the data are expected given our best efforts at modelling the Universe. For detailed discussion of the mock catalogues the interested reader should refer to \citet{robo11}.

\subsection{Stellar Mass Selection}
\label{sec:selection}

The stellar masses used for this work are the latest versions of the type described in \citet{tayl11}.
For the 2.2\% of objects which are missing stellar masses because of the fitting code missing data,
the stellar masses are approximated using the $g-i$ relation calculated in \citet{tayl11}, this is given by:

\begin{equation}
\begin{array}{lll}
\mathcal{M}(z,g,i)&=&-0.4i\\
&&+2\log_{10}DM(z)\\
&&+\log_{10}(1+z)\\
&&+(1.2117-0.5893z)\\
&&+(0.7106-0.1467z)(g-i)
\end{array}
\end{equation}

\noindent where $\mathcal{M}$ is our notation for total stellar mass, $z$ is the galaxy redshift, $g$ is the observed GAMA $g$-band apparent Kron magnitude, $i$ is the observed GAMA $i$-band apparent Kron magnitude \citep[see][for details regarding the GAMA photometric processing]{hill11} and $DM(z)$ is the luminosity distance modulus for our chosen cosmology and a redshift $z$. This relation naturally corrects for redshift k-corrections and the self attenuation of galaxy light by dust, so returns close to an intrinsic stellar mass estimate with $\sim$0.1~dex error \citep[see][for extensive details of the GAMA stellar masses and fidelity tests]{tayl11,cluv14}.

\begin{figure}
\centerline{\mbox{\includegraphics[width=3.7in]{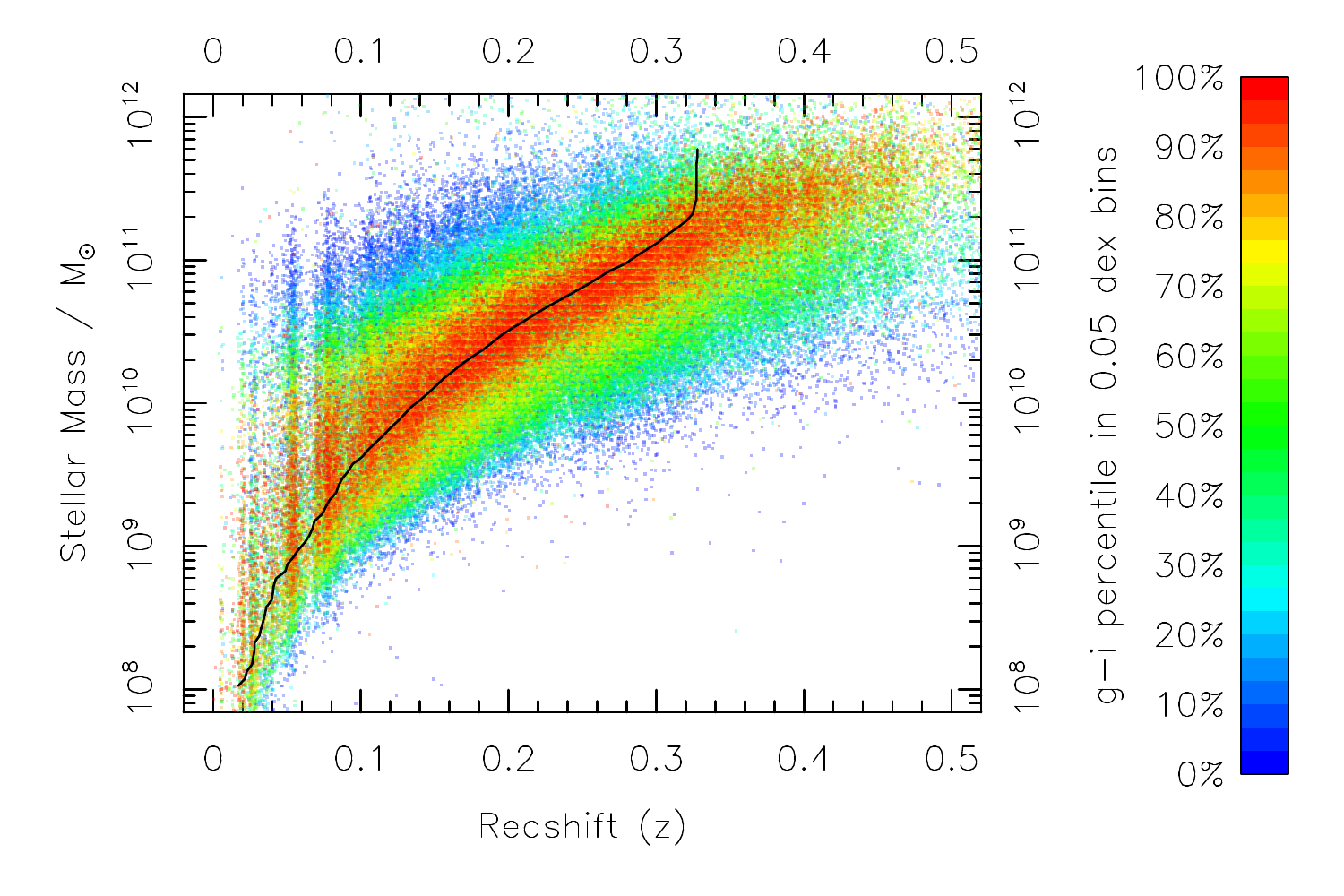}}}
\caption{\small The variable stellar mass redshift selection limit. In each vertical $M\pm0.025$~dex bin the cumulative density function (CDF) of apparent $g-i$ colour is calculated. Depending on where a galaxy appears in the CDF given its stellar mass the colour of the data point is different: relatively bluer (redder) $g-i$ colour galaxies are bluer (redder) data points. The black line shows the selection limit and is imposed at the $95^{th}$ percentile for each bin.}
\label{fig:SMselect}
\end{figure}

The next step is to estimate a reasonable redshift limit for a given stellar mass. At low redshift we observe galaxies with close to their intrinsic rest-frame colours, and as we move out in redshift on average optical colours become redder due to the typical shape of galaxy spectra. Moving to even higher redshift we can deduce from the above equation that stellar mass has a strong dependence on $g-i$ colour. The effect is such that at higher redshifts we are able to see only the {\it bluer} galaxies of a given stellar mass because of the $+(0.7106-0.1467z)(g-i)$ term in the stellar mass equation above. This strongly biases the sample towards blue galaxies, rather than being representative of the ensemble of galaxies. This colour bias can be seen in Figure \ref{fig:SMselect}, where in horizontal slices we show the $g-i$ quantile of each galaxy in that slice, from 0\% (bluest in the stellar mass bin selected) to 100\% (reddest in the stellar mass bin selected). To conservatively select galaxies by stellar mass we find the low redshift 95\% extreme of the $g-i$ distribution at all stellar masses. This produces the black line in Figure \ref{fig:SMselect}.

To ensure that this selection is robust against the possible effects of evolution out to the redshifts probed, we also investigated a constant $z=0.1$ limit above stellar masses of $4\times 10^{9}\msol$. The main close-pair parameter fits (discussed later) were unchanged within-errors, and the constraints on the parameterisation had errors a factor $\sim3$ larger. This gives us confidence in using the larger sample provided by using the sliding redshift limit shown in Figure \ref{fig:SMselect} for the following work.

\begin{figure}
\centerline{\mbox{\includegraphics[width=3.7in]{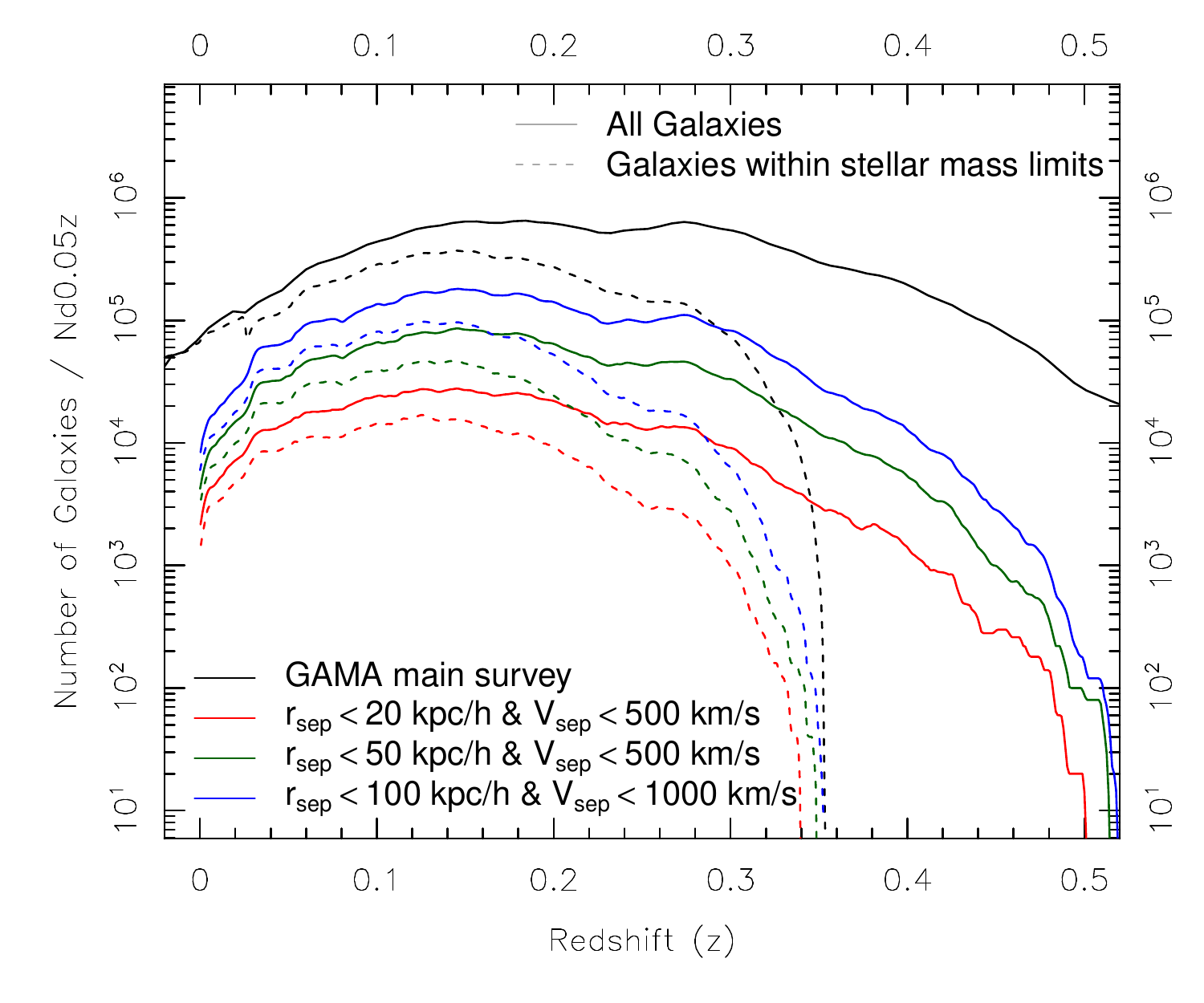}}}
\caption{\small N(z) distribution for galaxies in the GAMA-II catalogue used in this work, with redshift quality $Q>=3$ in $\Delta z=0.05$ bins. Solid lines show all GAMA galaxies prior to stellar mass filtering. The dashed lines shows galaxies that are also within the stellar mass complete redshift limits shown as the solid black line in Figure \ref{fig:SMselect}. The black lines show all GAMA-II-N galaxies available to the survey, the various coloured lines show the different pairwise dynamical selections as specified in the bottom-left legend.}
\label{fig:Nz}
\end{figure}

Figure \ref{fig:Nz} shows how much of our sample is removed by virtue of this conservative stellar mass selection limit. To be highly complete in terms of the stellar mass selection, a large fraction of all objects beyond $z=0.3$ are removed from the sample ($\sim$87\%). The effect for close-pairs broadly mimics that for all available GAMA galaxies, with the stellar mass selection removing comparatively more galaxies at higher redshifts and $N(z)$ peaking close to $z=0.15$.

\begin{figure}
\centerline{\mbox{\includegraphics[width=3.7in]{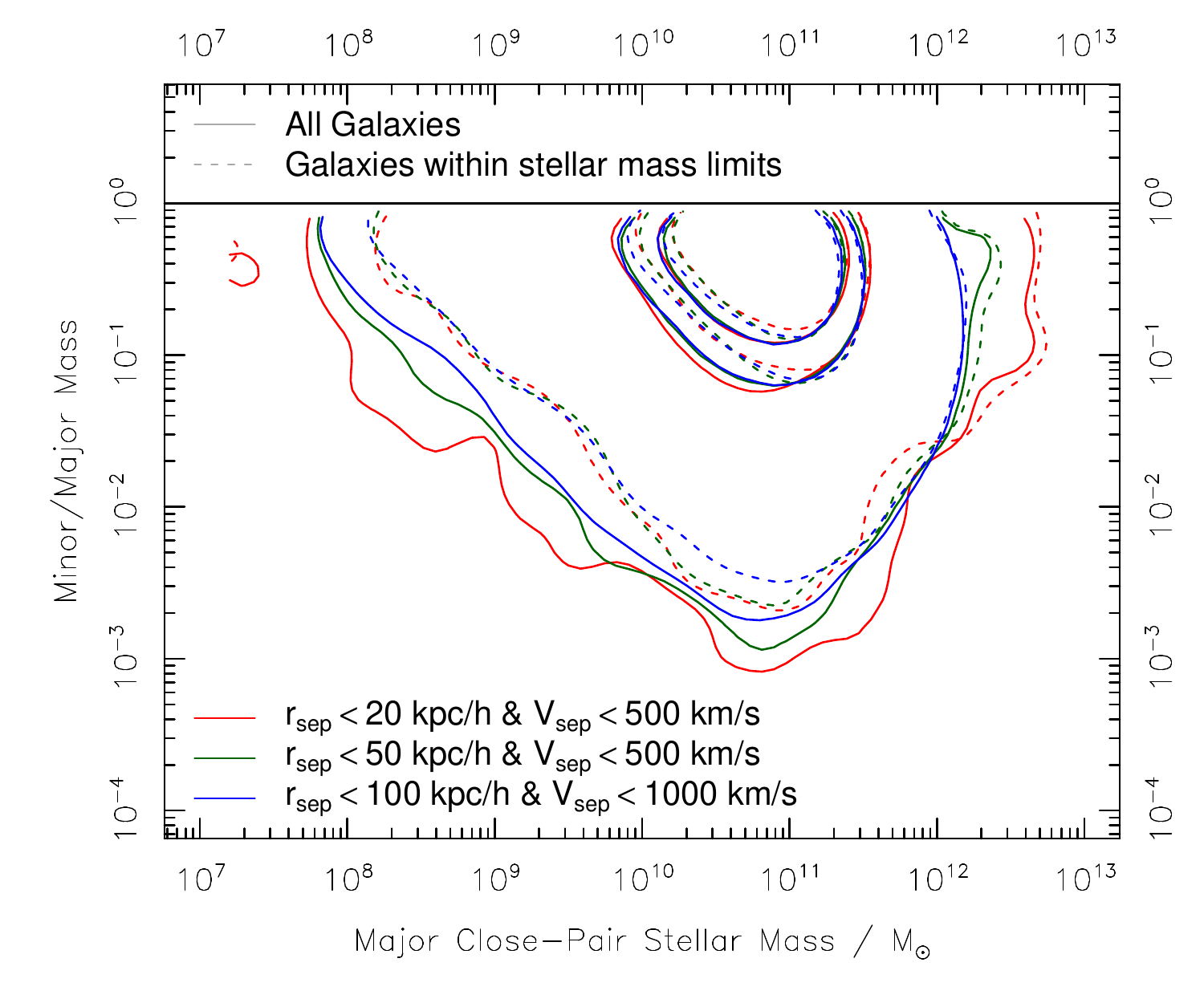}}}
\caption{\small 2D density contours showing the highest density regions containing 50\% (inner contours), 68\% (middle) and 99\% (outer) of the data when comparing the major close-pair stellar mass (x-axis) to the minor/major stellar mass ratio (y-axis). The line definitions are as per Figure \ref{fig:Nz}.}
\label{fig:SMvRatio}
\end{figure}

Figure \ref{fig:SMvRatio} shows the raw number density of major close-pair stellar mass (i.e.\ the stellar mass of the more massive close-pair galaxy) versus the minor/major mass ratio. The dashed lines show the effect of applying the stringent stellar mass criterion, where in all cases we are systematically shifted towards observing more massive galaxies with a lower stellar mass ratio close-pair companion. The number density of close-pairs peaks near $\mathcal{M}^*$ \citep[$10^{10.66}\msol$,][]{bald12} for all selections, and favours mass ratios nearer to 1 (i.e.\ major close-pair systems with similar stellar mass for the two galaxies). However, we do retain data all the way out to 1,000+:1  mass ratios even after selecting for stellar mass completeness, which should allow us to explore the extreme minor merger population of galaxies with high fidelity.

The subset $PS_{r20v500} \subset P_{r20v500}$, where both galaxies of each pair are above our stellar mass-redshift limit (shown as a solid line in Figure \ref{fig:SMselect}) contains 1,434 pairs. Equivalently $PS_{r50v500} \subset P_{r50v500}$ contains 4,741 pairs and $PS_{r100v1000} \subset P_{r100v1000}$ contains 13,496 pairs. In all cases this is slightly less than half of the number of pairs in the respective supersets.

\subsection{Visual Classifications}

\label{sec:visclass}

To assess the types of interactions recovered by different selections of paired galaxies we investigated the galaxy morphologies in $PS_{r20v500}$, $PS_{r50v500}$ and $PS_{r100v1000}$, focussing on signs of visual disturbance. 22,728 galaxies were selected for visual inspection by virtue of being in a pair adhering to the $PS_{r100v1000}$ stellar mass and dynamical state selection criteria (this is an inclusive selection since $PS_{r20v500}$ and $PS_{r50v500}$ are both fully contained by this larger selection window).

It is important we determine a background `disturbed' fraction for galaxies known not to be in a close-pair configuration. To achieve this we created four control samples of non-paired galaxies where $10^8<\mathcal{M}/\msol<10^9$ (all available: 50 galaxies), $10^9<\mathcal{M}/\msol<10^{10}$ (random: 200 galaxies), $10^{10}<\mathcal{M}/\msol<10^{11}$ (random: 200 galaxies) and $10^{11}<\mathcal{M}/\msol<10^{12}$ (random: 200 galaxies). This added a further 650 galaxies, bringing the total for visual inspection to 23,378. The control samples were selected in regions where the spectroscopic completeness was 100\% out to 100$\kpch$ in projection. The control sample galaxy images were added to the parent close-pair sample, and were analysed by classifiers at the same time. Classifiers were not made aware of the presence of a control sample.

A sophisticated scheme was developed that ensured maximal internal consistency between different classifiers, and that removed the most serious subjective classification biases. This is described in detail in Appendix \ref{sec:visclassdetails}, and concludes with the generation of optimal objective classification weights for each classifier, and ultimately an objective mean classification for each galaxy in the sample analysed.

\begin{figure}
\centerline{\mbox{\includegraphics[width=3.7in]{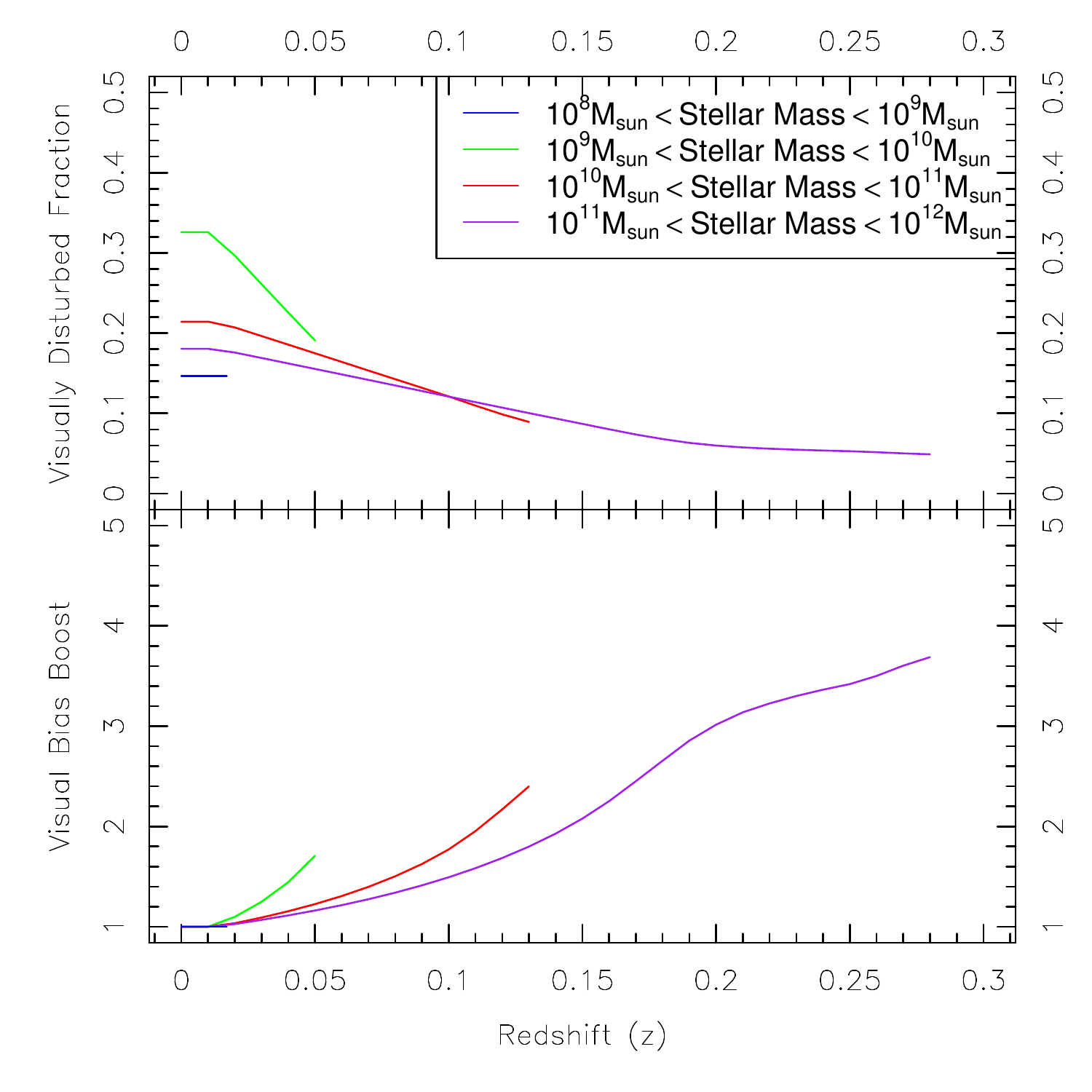}}}
\caption{\small Top panel shows the mean visual disturbance rates as a function of redshift for different stellar mass samples {for all galaxies in our close-pair and control sample}. The assumption used for correction purposes is that galaxy mergers should be no less common at higher redshift \citep[this is a conservative assumption, since they are generally observed to be more common, e.g.][]{brid10}. We treat the disturbance ratio between the lowest redshift bin and any other as the required mean correction factor for binned samples. The effect of this correction factor is shown in the bottom panel, with the general trend that it increases quite linearly with redshift (i.e.\ inverse physical resolution).}
\label{fig:disturbcorrect}
\end{figure}

Having applied our optimal objective classification weights we find the main artefact that negatively affects the visual classification process is the loss of resolution as galaxies are observed at higher redshifts, where the same physical scale is represented by fewer pixels.

The galaxies analysed are extracted from the full visual classification sample (close-pair galaxies and the control samples). The effect of this selection is that we tend to observe the same stellar mass galaxy with a less massive companion at lower redshift due to our stellar mass selection. However, even restricting the sample to close-pairs with a greater than 3:1 mass ratio we find the dominant bias is the redshift of observation, not the mass ratio of the pair.

To account for this we analysed the mean `disturbed' rate as a function of redshift for different subsets of stellar mass.

Figure \ref{fig:disturbcorrect} shows the result of this analysis, where all `disturbed' fractions drop systematically with redshift. The assumption we make to correct for this bias is that the merger rate is not evolving strongly over the 3 Gyr baseline shown \citep{kart07}. The bottom panel of Figure \ref{fig:disturbcorrect} displays the boost required for a galaxy with a given stellar mass at a given redshift. Within a redshift of 0.1 the bias boost is less than a factor 2, but by redshift 0.3 it is $\sim4$. The size of these corrections can therefore be substantial and error prone, so for clarity of presentation later results are presented with and without the visual classifications applied.

\section{Galaxy Pair Corrections}

\label{sec:correct}

There are a number of corrections that need to be applied to any pair catalogue to account for observational artefacts and contamination. This Section discusses each bias and correction in detail.

\subsection{Photometric Confusion}

\label{ref:Pcomp}

\begin{figure}
\centerline{\mbox{\includegraphics[width=3.7in]{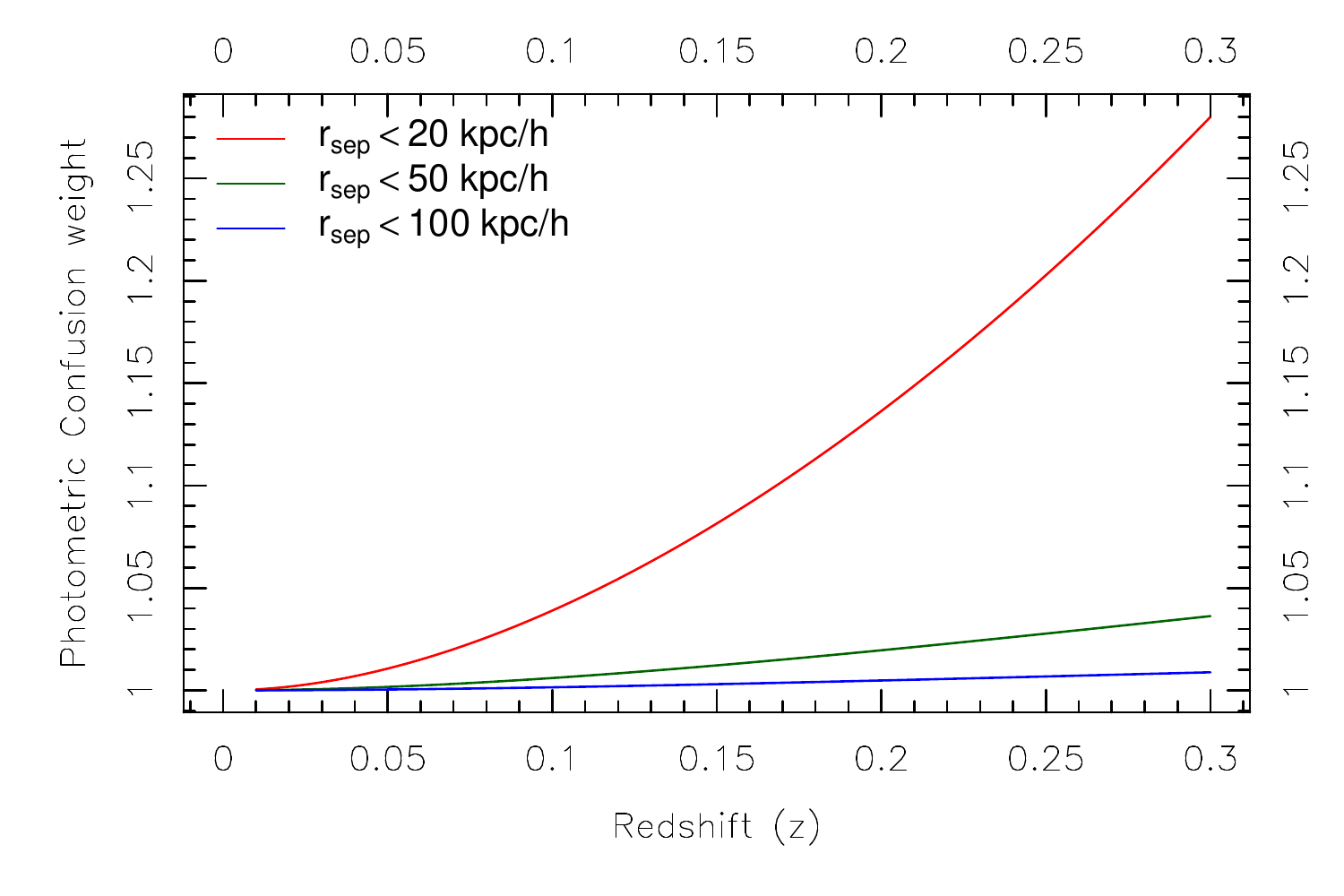}}}
\caption{\small The weighting applied to an observed pair at a given redshift within specified projected separation limits. This accounts for the sharp fall in close-pairs within 3'' on the sky, which at a redshift of 0.3 can be a substantial fraction of the pair search radius.}
\label{fig:photocomp}
\end{figure}

The first substantial completeness correction that should be considered when analysing pairs is the effect of photometric confusion. The effect we observe for pair galaxies is that as they become closer in angular separation they become harder to distinguish into distinct components by automated deblending algorithms. Since the GAMA main survey is defined from the SDSS $r$-band, we are really witnessing the limits of the SDSS deblender. The effect is relatively simple to quantify since we expect little evolution in the distribution of physical galaxy-galaxy separations between $0<z<0.1$. What we observe with our sample is a fairly linear drop in very compact pairs as a function of redshift, with a sharp deficit of close-pairs within $3''$ regardless of redshift. To numerically correct for this effect we weight all pair counts at a given redshift by the fraction of the projected close-pair area lost to deblending, i.e.\

\begin{equation}
W_{photo}(D_{proj},z)=\frac{2 \pi D_{proj}^2}{2 \pi \cos \left(\frac{D_{ang}(3'',z)}{D_{proj}}\right) D_{proj}^2}
\end{equation}

\noindent for $D_{proj}>D_{ang}(3'',z)$, where $D_{proj}$ is the projected pair limit of interest (e.g.\ 20 $h^{-1}$kpc) and $D_{ang}(3'',z)$ returns the projected physical size of $3''$ at a redshift of $z$.  Figure \ref{fig:photocomp} shows the effect of this correction for the 3 separation limits $r_{sep}$ used in this paper. At low redshifts, where $D_{ang}(3'',z)$ is much less than $D_{proj}$, the weight is close to 1 for all separations. However, by a redshift of 0.3 a substantial fraction ($\sim20\%$) of the pair separation area for galaxies within 20 $h^{-1}$kpc of each other on the sky is lost to deblending artefacts. For pairs within 50 $h^{-1}$kpc and 100 $h^{-1}$kpc the photometric correction is small throughout the range investigated in this work. This correction implicitly assumes the fraction of the projected pair radius covered by the 3'' deblending window has a 1--1 relationship with the corresponding drop in close-pairs, which is true if all orientations with respect to our line-of-sight are equally likely for any given radius.

\subsection{Spectroscopic Fibre Collision Incompleteness}

\label{sec:Scomp}

A priori, galaxies in close-pairs have a higher probability of being spectroscopically incomplete than isolated galaxies. The main source of this potential bias is fibre collisions on the 2dF instrument used to collect GAMA redshifts on the AAT. Fibres cannot be allocated within $30''$ of another fibre in a given configuration, so a single pass survey (such as 2dFGRS and SDSS, both of which had small amounts of overlap) can suffer from quite pronounced anti close-pair bias. GAMA was designed with close-pair work in mind, so the target tiling was optimised to minimise this bias \citep[see][]{robo10}. Every area of sky was observed on average $\sim10$ times, so even the most complex clusters of targets were completely unpicked by the conclusion of the survey \citep[see][for analysis of the close-pair bias in the GAMA-I survey]{driv11}.

Because of this approach to conducting the survey, the equatorial GAMA-II regions are more than 97\% complete on all angular scales. The caveat is when two galaxies are within a single fibre (2''). Work on occulting line-of-sight pairs (Holwerda et~al. in prep) finds that only 0.2\% of GAMA galaxies share a fibre, so this effect is considered small enough to be ignored for future analysis in this work. To remove any (very small) remaining bias we compute a close-pair correction. In the simplest form we compute for every galaxy the redshift success fraction for potential GAMA-II main survey targets within the 3 angular separations investigated in this work. The reciprocal of this number becomes the weight $W_{spec}$, i.e.\ if 8/10 galaxies were observed within 20 $h^{-1}$kpc of a galaxy we can say it is 80\% complete on this scale, and it inherits a weight of 1.25. All observed close-pairs in our

$PS_{r20v500}$, $PS_{r50v500}$ and $PS_{r100v1000}$ samples have the corresponding mean completeness correction applied. This correction is very small in practice since we are more than 97\% complete at all angular scales, offering a $\sim 3$\% boost to the observed close-pair numbers. Since the parent population is also slightly incomplete (although better than 98\%) the final close-pair fraction boost is $\sim 98/97=1.01$, i.e.\ $\sim1$\%.

\subsection{Pair Complex Correction}

\label{sec:PCcor}

A numerical correction that must be considered is the occurrence of galaxies in multiple galaxy-galaxy pairs. Table \ref{tab:pairstat} shows the frequency of different complex multiplicity for the three different pair samples investigated in this work. Unsurprisingly the larger dynamical windows often find a substantial number of close-pairs for the same galaxy, with two cases where a single galaxy is in a close pair with 15 others. All of these pairs should be counted, but to conserve mass between different pair subsets a galaxy that is observed in multiple pairs should be down-weighted by the number of pair systems it is found in, i.e.\ the stellar mass of the two galaxies in 15 close-pairs should not be counted 15 times.

This weight ($W_{complex}$) is simply the reciprocal of the number of pair systems (given the sample limits imposed) that the galaxy is found in, so if it is found in 3 pairs $W_{complex}=1/3$. The analysis was carried out with and without the complex correction made, where the dominant effect of the complex correction is to reduce the observed normalisation for the close-pair space density. Results below are presented {\it without} the complex correction made. Later in the paper a more flexible analytic estimate of the required complex correction is presented, which allows the end user to apply the scaling for close-pair scenarios not explicitly presented in this work. This is of practical importance since {\it exact} complex corrections will vary depending upon the stellar mass range investigated, and can only be calculated using all of the close-pair data.

\begin{table}
\begin{center}
\begin{tabular}{rrrr}
Close-pair companions  	& $PS_{r20v500}$ & $PS_{r50v500}$ & $PS_{r100v1000}$\\
\hline
1	&	2,505	&	6,200	&	9,843	\\
2	&	178		&	1,032	&	3,106	\\
3	&	7 		&	187		&	1,253	\\
4	&	0		&	35		&	529		\\
5	&	0		&	19		&	240		\\
6	&	0		&	0		&	132		\\
7	&	0		&	0		&	70		\\
8	&	0		&	0		&	44		\\
9	&	0		&	0		&	24		\\
10	&	0		&	0		&	16		\\
11	&	0		&	0		&	5		\\
12	&	0		&	0		&	7		\\
13	&	0		&	0		&	5		\\
14	&	0		&	0		&	3		\\
15	&	0		&	0		&	2		\\
\end{tabular}
\end{center}
\caption{Number of galaxies with $N$ close-pair companions for the different subsets investigated.}
\label{tab:pairstat}
\end{table}

\subsection{True Pair Corrections}

\label{sec:TPcor}

Having corrected for the empirical bias inherent in the data, we can also correct for biases in the types of systems our pair criteria actually selects. Since we are ultimately interested in which galaxy close-pairs will merge, and the future fate of the stellar mass function due to mergers, the requirement is that our pairs should represent interacting galaxies and not spurious projected systems.

We have chosen to be conservative with our correction schemes and include the results of three different analyses that should bracket the extreme limits: direct analysis of all close-pairs (this will produce the highest close-pair fractions and merger rates); close-pairs corrected for statistical biases implied by analysis of the mock catalogues; close-pairs where data is weighted by how visually disturbed component galaxies appear to be (this will produce the lowest close-pair fractions and merger rates). Including the uncorrected data lends the possibility that the results can be remapped at a later date using more modern simulations.

\subsubsection{Mock Estimate}

\label{sec:mockcorrect}

A standard contaminant in close-pair work are small velocity separations between galaxies generated by cosmological effects (i.e.\ the galaxies might not be physically close along the line-of-sight, but appear close in velocity separation because of their respective motions on the Hubble flow) versus true pair-wise velocity separations. The first way of accounting for the likely magnitude of such chance projections is by analysing the GAMA mock catalogues (see Section \ref{sec:mockcats} above for a more detailed discussion of the mock catalogues).

For each of our close-pair subsets ($PS_{r20v500}$, $PS_{r50v500}$ and $PS_{r100v1000}$) we calculate the fraction of real space pairs within the radial limit that are recovered (the positive-positive fraction, $PP$), and the fraction of redshift pairs that are spurious (the false-positive fraction, $FP$). In terms of the pair fraction recovered the implied weight is $W_{TP-mock}=FP/PP$. If the selection criteria misses 50\% of the real pairs, but also contains 50\% false pairs, the combination cancels out (i.e.\ our pair fraction reflects the true pair fraction).

For $PS_{r20v500}$ $W_{TP-mock}=0.961$, for $PS_{r50v500}$ $W_{TP-mock}=0.891$ and for $PS_{r100v1000}$ $W_{TP-mock}=0.646$. In all cases the bias is towards recovering too many pairs by default. For clarity we show all results with and without the mock catalogue based correction applied. This is pragmatic since the mock catalogues have the most uncertainty at the smallest scales due the complexities of modelling baryonic physics and dynamical friction, neither of which are fully described by semi-analytic models. For more information regarding the discrepancies between the GAMA mock catalogues and observations at small spatial separations see the discussion in \citet{robo11}.

\subsubsection{Visual Disturbance Estimate}

\label{sec:visualcorrect}

The final method used for determining true close-pairs is to consider the requirement that the galaxies in these pairs should look physically disturbed if they are currently interacting, and therefore likely to merge on a shorter timescale on average.

This is modulo the caveats discussed in the introduction, in particular that not {\it all} visually disturbed close-pairs are guaranteed to merge on a rapid timescale. We consider the visually disturbed population in addition to the uncorrected and the mock catalogue corrected analysis, where this correction offers a very conservative lower limit on the true close-pair fraction due to the shallow nature of the SDSS imaging data used in the analysis \citep{ji14}.

\begin{figure}
\centerline{\mbox{\includegraphics[width=3.7in]{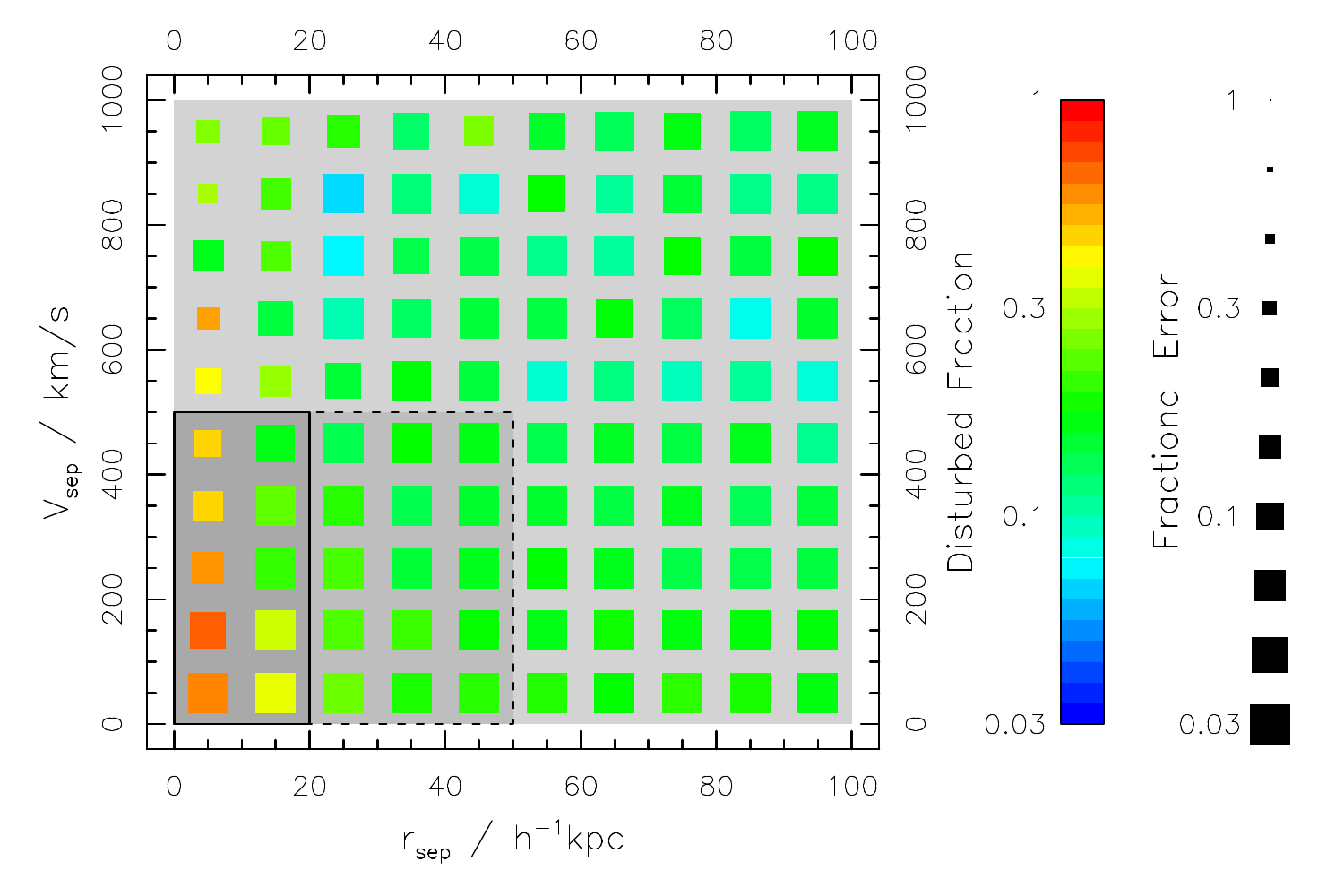}}}
\caption{\small Fraction of galaxies in each dynamical bin showing signs of disturbance. This is corrected for the simplest stellar mass dependent redshift bias (see Figure \ref{fig:disturbcorrect}). The darkest grey box with solid lines shows the $PS_{r20v500}$ sample, the next darkest box with dashed lines shows the $PS_{r50v500}$ sample, and the entire plotting region shows the $PS_{r100v1000}$ sample. The size of the box indicates the error, where smaller means less significant, i.e.\ more error in the measurement.}
\label{fig:disturbgrid}
\end{figure}

We use the classifications discussed in Appendix \ref{sec:vc}, and calculate for any subset of pair galaxies the mean of the debiased disturbance rates (discussed in Section \ref{sec:visclass}) in that selection, giving $W_{TP-vc}$. This approach will reduce the pair fraction to those that either recently interacted with the other galaxy in the pair, or that have recently undergone a merger. For $PS_{r20v500}$ the disturbance fraction is generally very high (see Figure \ref{fig:disturbgrid}), however it drops to $\sim10\%$ for $PS_{r100v1000}$. This approach conservatively recovers the subset of galaxies that are likely to merge soonest. For clarity when it is used, we present results with and without the visual disturbance rate correction applied.

Figure \ref{fig:disturbgrid} summarises the mean debiased disturbance fraction for different dynamical windows of close-pairs. The immediate result is clear, galaxies that are dynamically close (very compact pairs) are much more likely to appear visually disturbed. We find the disturbance fraction of galaxies in close-pairs is comparable to the 40\%$\pm$11\% level observed in \citet{patt05}. If all close-pair galaxies in the $PS_{r20v500}$ sample are considered we find $\sim20$\% are disturbed, but if we apply the same factor 2 correction for `false-pairs' as \citet{patt05} we find the same $\sim40$\% disturbed fraction for our dynamically closest pairs. This Figure also demonstrates that significant close-pair disturbance is visible even for fairly large radial velocity separations. In fact the radial separation needs to be above $\sim$700km/s for tangentially close pairs ($r_{proj} \le 10h^{-1}$kpc) before the disturbance fraction returns to the background level seen for larger projected separations and our isolated control sample.

The fact there is a measurable background disturbed fraction of $\sim$10\%, i.e.\ it does not just fall to zero at large separations, is important because this reflects the fraction of galaxies that look disturbed because they have just undergone a merger, have intrinsic asymmetries \citep[e.g.][]{holw12}, or because they have been misclassified (we do not have the data to disentangle the dominant causes). This same background fraction of $\sim$10\% is also measured in the isolated control sample and it is very similar to the 9\%$\pm$3\% level observed for isolated galaxies in \citet{patt05}. This figure is also in good agreement with the spectroscopically corrected `strongly disturbed' fraction of 6--9\% measured in \citet{darg10a} for their volume limited bright sample. \citet{hern05} find a similar consistency between the disturbed fraction (and properties) of isolated galaxies versus those in dynamically loose pair configurations, i.e.\ the background we see at wider separations. The exact interpretation of these numbers between different surveys due to the effect of surface brightness limits in identifying asymmetric structure \citep{ji14}. In later analysis we make use of this background level, considering the excess above this level as the approximate signifier of the fraction of galaxies that are disturbed because they are dynamically interacting in a close-pair.

The disturbance levels observed here are notably lower than the $\sim$70\% disturbed levels for field ellipticals described in \citet{dokk05}. That work used substantially deeper photometric data, revealing much lower surface brightness tidally disrupted features. Since surface brightness plays such a key role in the detection of disturbance, the safest interpretation is that the disturbance fractions should only be compared in a relative sense once the background has been subtracted. In any case later analysis is always discussed with and without the visual disturbance corrections applied.

The mock catalogue and visual disturbance `True Pair' corrections ($W_{TP-mock}$ and $W_{TP-vc}$ respectively) are attempting to account for the same effect: pairs that are close in dynamical phase space are not actually spatially close and interacting. For this reason the corrections are never applied in combination, with either no correction, or $W_{TP-mock}$ or $W_{TP-vc}$ being applied in turn. Of the three approaches, it will be the case that no correction will lead to over-estimates in the pair fraction (and associated merger rates etc), whilst $W_{TP-vc}$ will likely lead to under-estimates since true pairs that are pre first-passage will likely not display easily observable asymmetries \citep[e.g.][]{toom72}.

\subsection{Summary of Corrections}

Above we have listed a large number of corrections that need to be considered to properly account for biases and artefacts in close-pair data. Whilst this might dissuade the casual reader from the veracity of the following results, they should be reassured that the typical amount of correction is small. Indeed the 1$\sigma$ range of $W_{photo}.W_{spec}$ (the other corrections are applied separately and explicitly for clarity) only spans the range 1.01 to 1.27 in the sample $PS_{r20v500}$ pairs which requires the biggest corrections on average because of the compact angular separation. The broad results are highly robust to the application of these corrections. The corrections only have a small (but measurable) impact on the normalisation of the close-pair  space density, but not on the shape of the distribution with respect to stellar mass.

\section{Observed Galaxy Close-Pairs}
\label{sec:galmerge}

\begin{figure}
\centerline{\mbox{\includegraphics[width=3.7in]{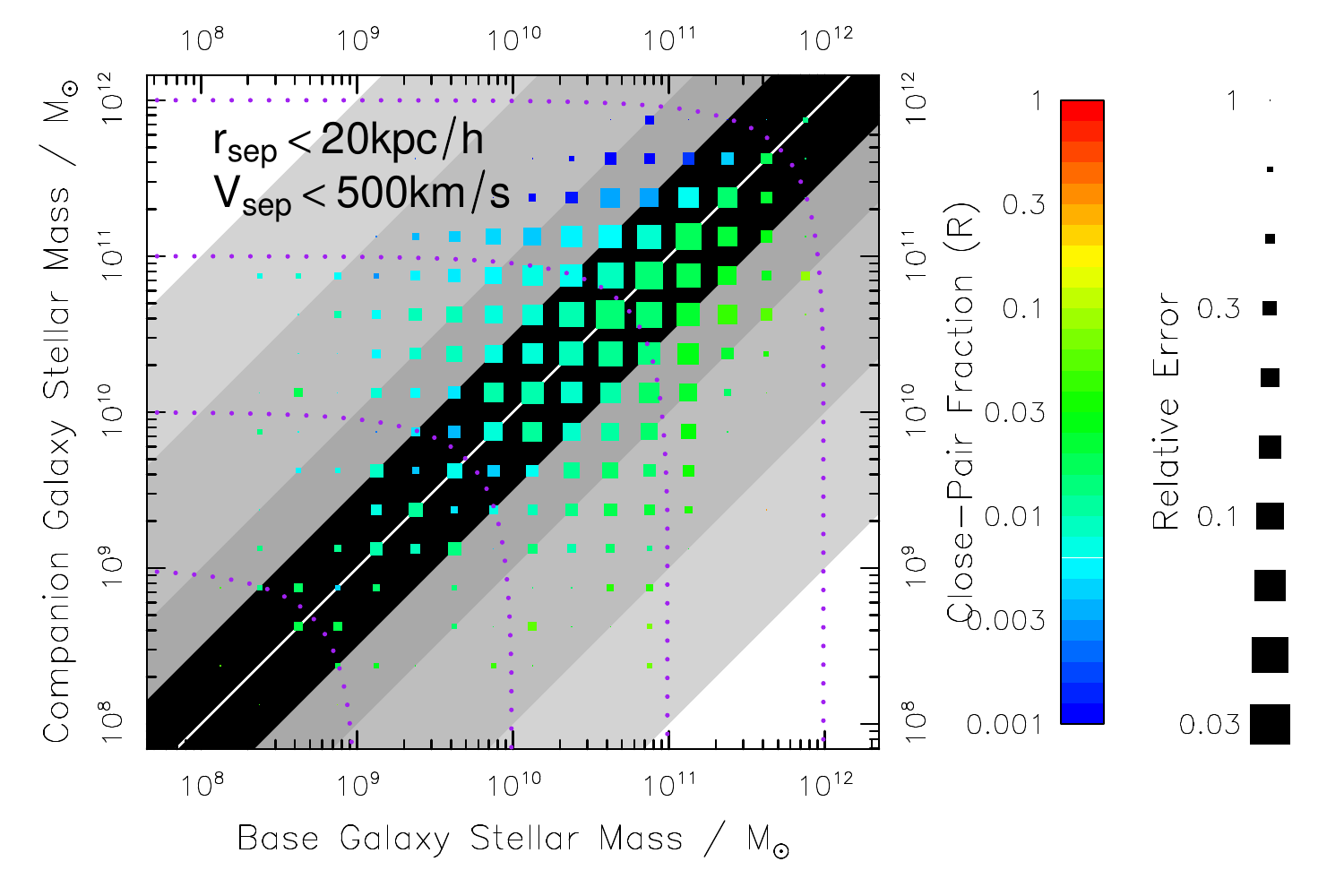}}}
\centerline{\mbox{\includegraphics[width=3.7in]{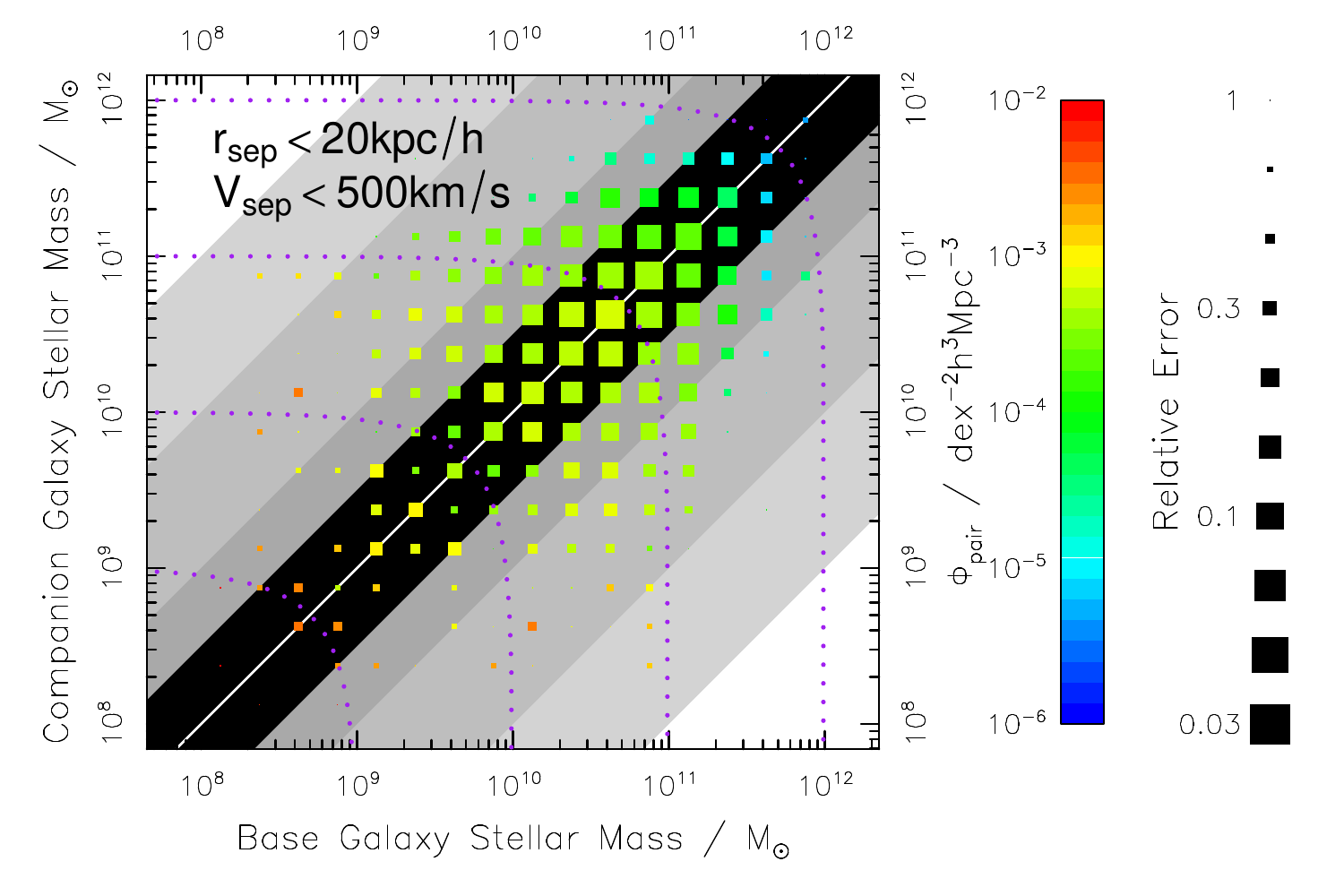}}}
\centerline{\mbox{\includegraphics[width=3.7in]{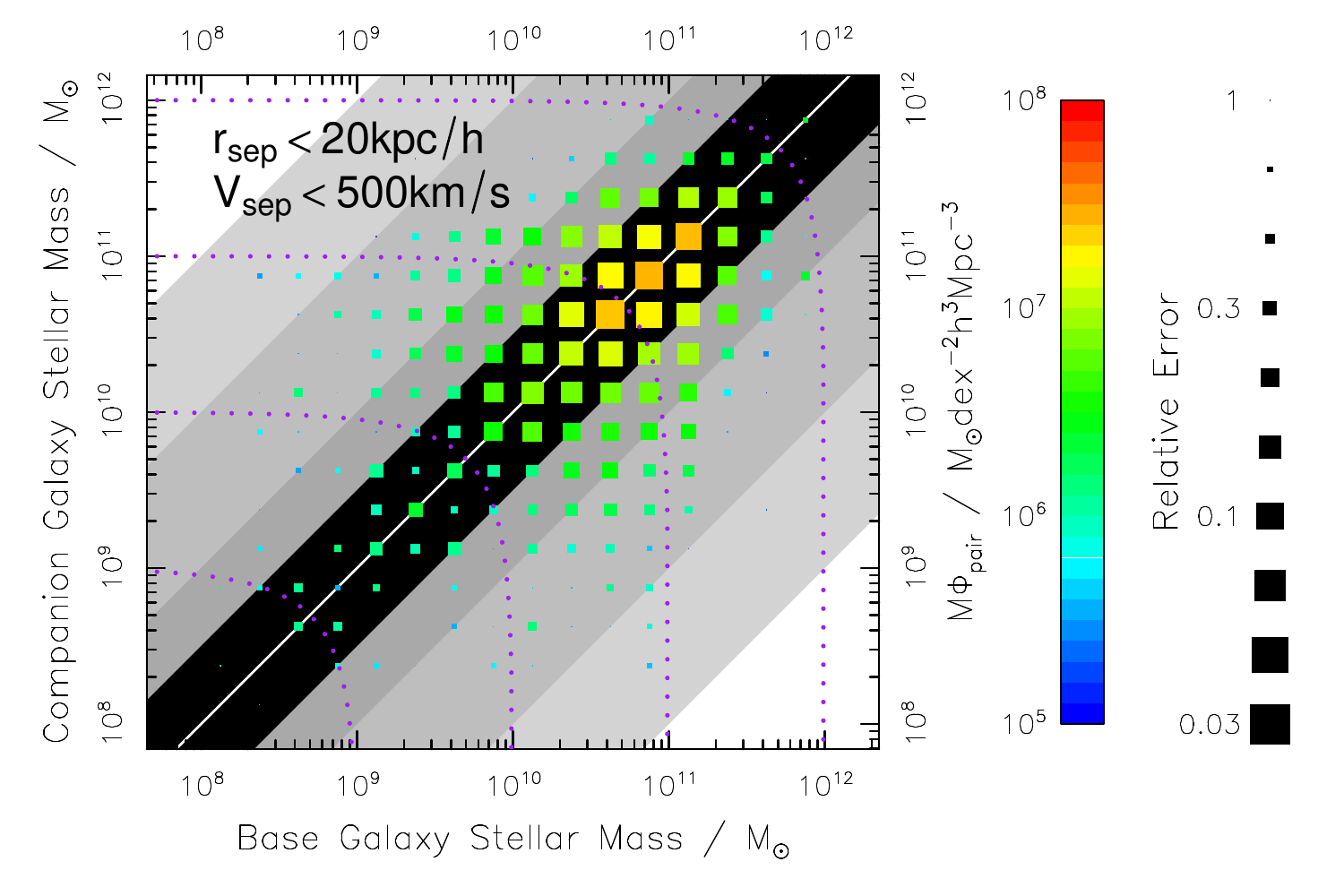}}}
\caption{\small Pair properties for $r_{\rm sep}<20\kpc$ and $v_{\rm sep}<500 \kms$ sample ($PS_{r20v500}$) true pair corrected using the mock catalogues ($W_{TP-mock}$, see Section \ref{sec:TPcor}). Top panel shows the observed pair fraction for the y-axis companion galaxy per stellar mass interval of the x-axis base galaxy. Middle panel shows the observed pair number density for the y-axis companion galaxy per stellar mass interval of the x-axis base galaxy. Bottom panel shows the observed minor accreting mass stellar mass density for the y-axis pair galaxy per stellar mass interval of the x-axis base galaxy (the number density of the pairs multiplied by the mass of the less massive galaxy in any close-pair. In all panels the black background region shows the regime of major mergers (mass ratio within a factor 0.5~dex), subsequent lightening grey regions show increasing decades in merger mass ratio. The purple lines show different merger mass products ranging from $10^8 M_{\sun}$--$10^{12} M_{\sun}$.
}
\label{fig:MergeGridA20V500}
\end{figure}

\begin{figure}
\centerline{\mbox{\includegraphics[width=3.7in]{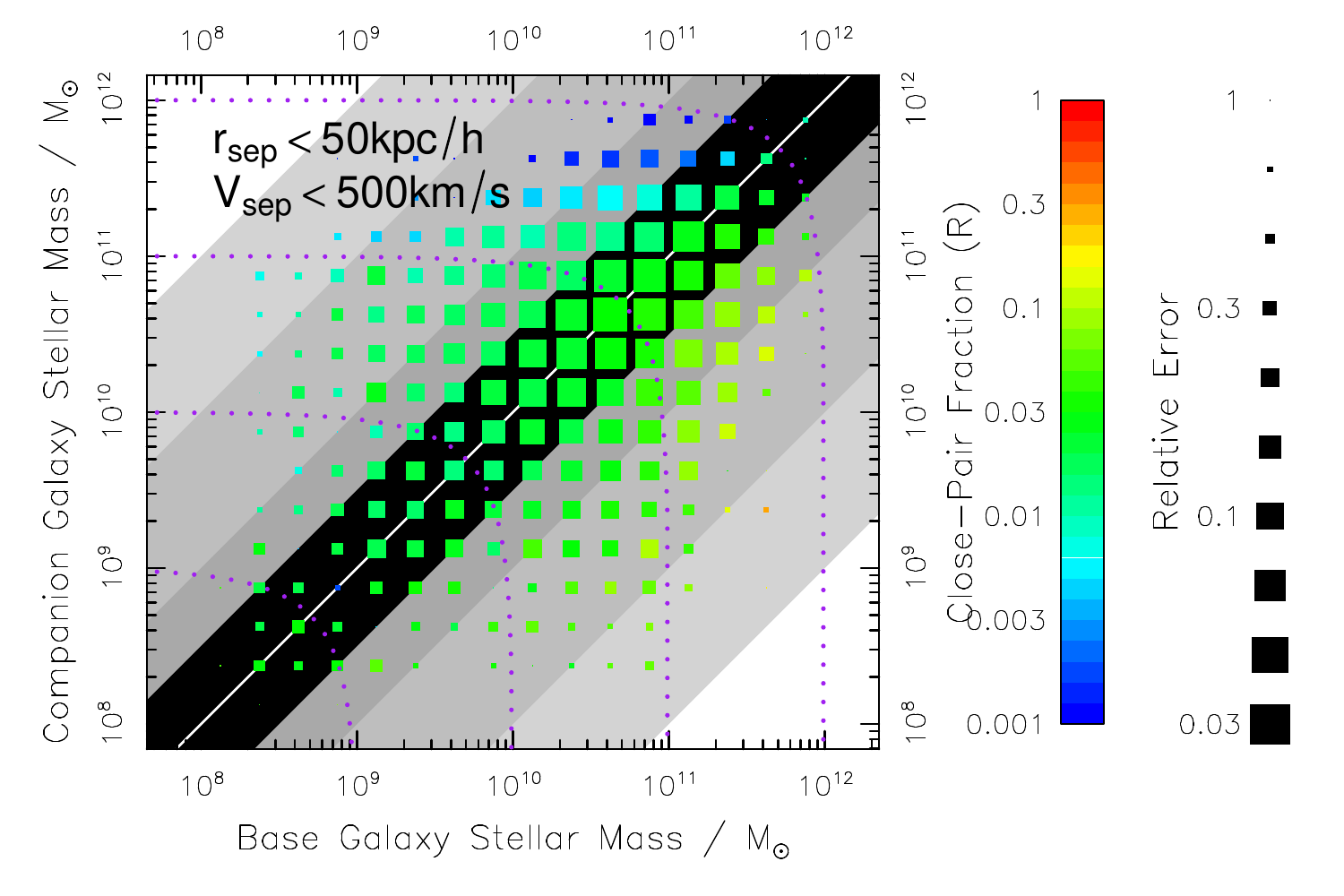}}}
\centerline{\mbox{\includegraphics[width=3.7in]{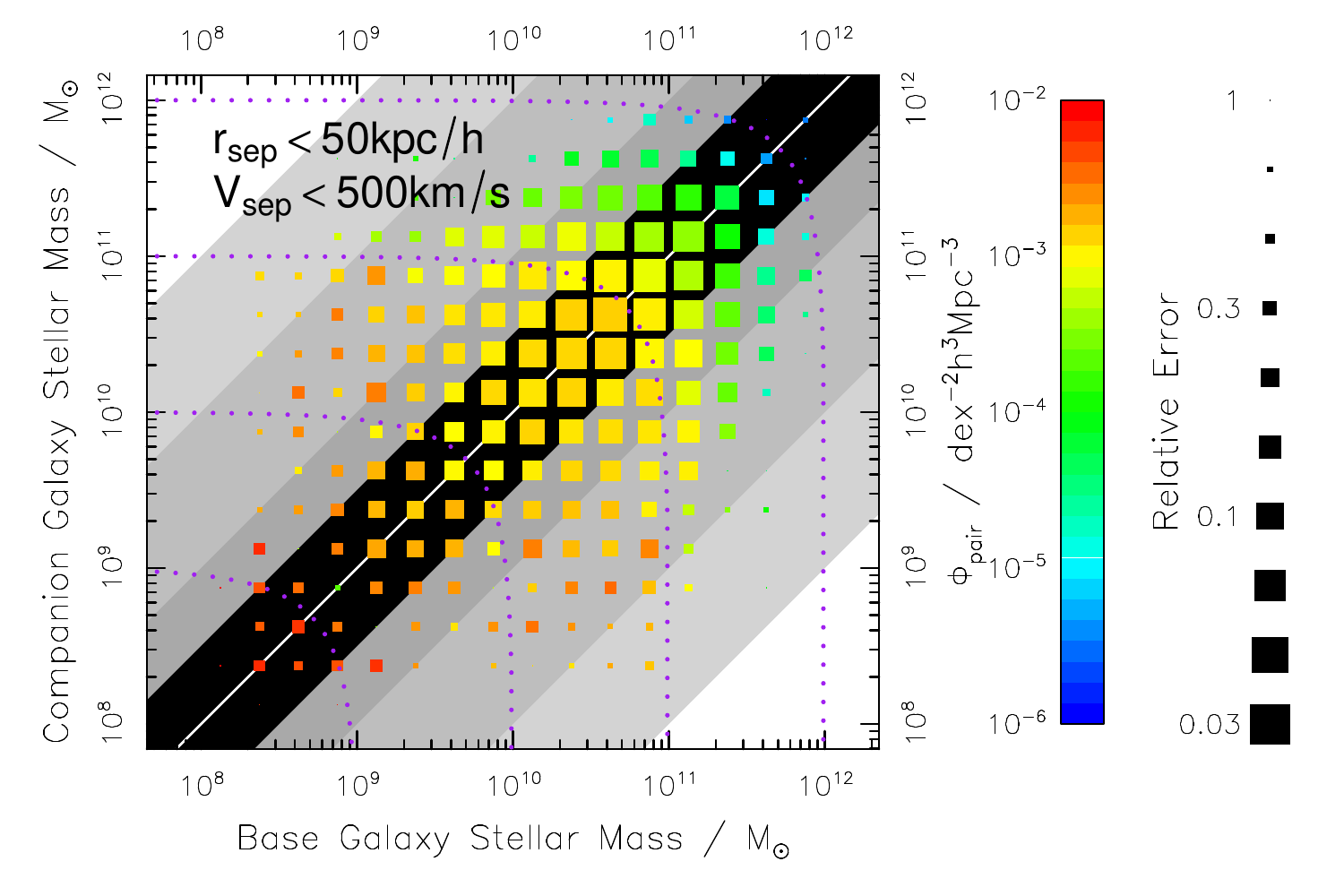}}}
\centerline{\mbox{\includegraphics[width=3.7in]{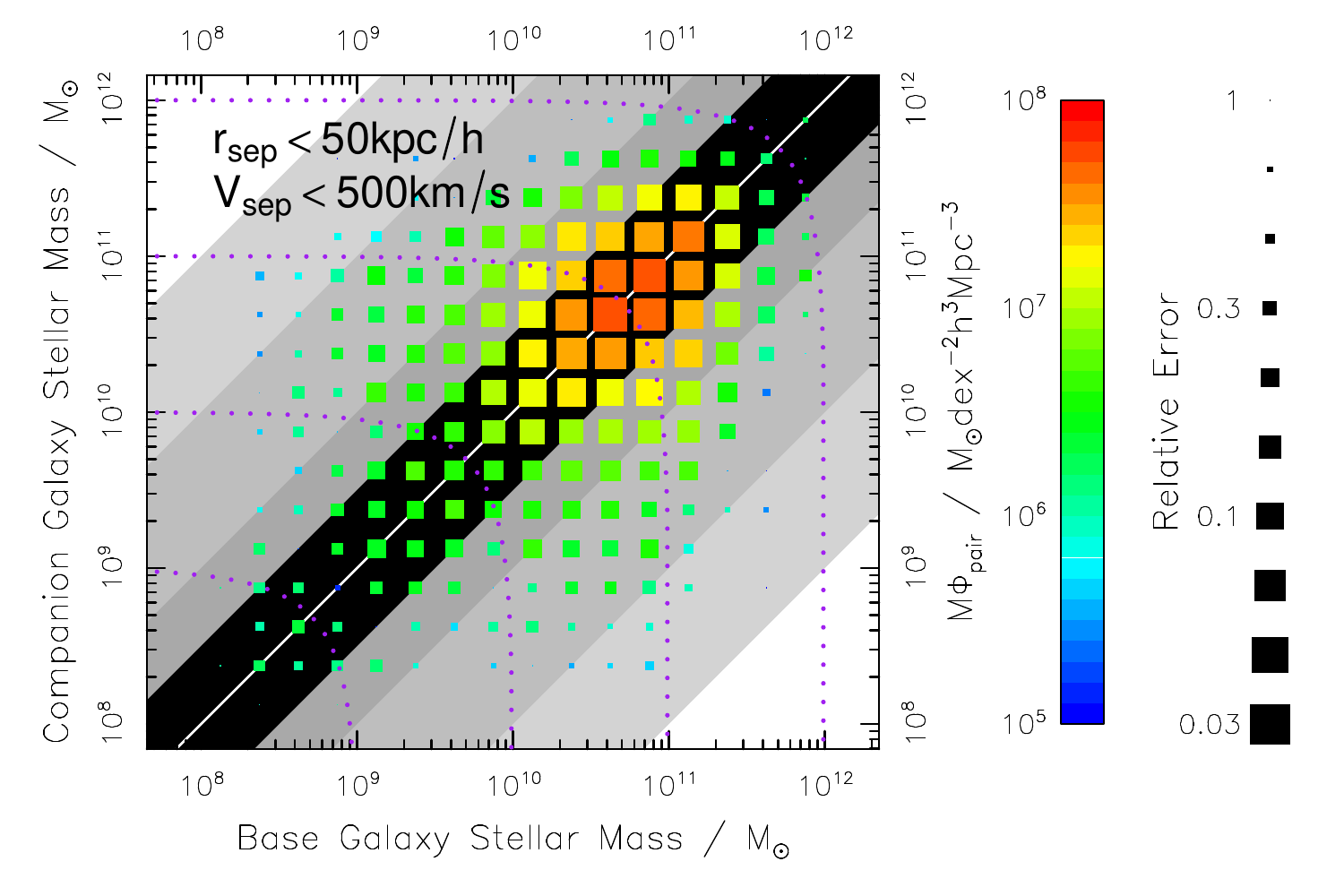}}}
\caption{\small Pair properties for $r_{\rm sep}<50\kpc$ and $v_{\rm sep}<500 \kms$ sample ($PS_{r50v500}$) true pair corrected using the mock catalogues ($W_{TP-mock}$, see Section \ref{sec:TPcor}). For more information see the caption for Figure \ref{fig:MergeGridA20V500}.}
\label{fig:MergeGridA50V500}
\end{figure}

\begin{figure}
\centerline{\mbox{\includegraphics[width=3.7in]{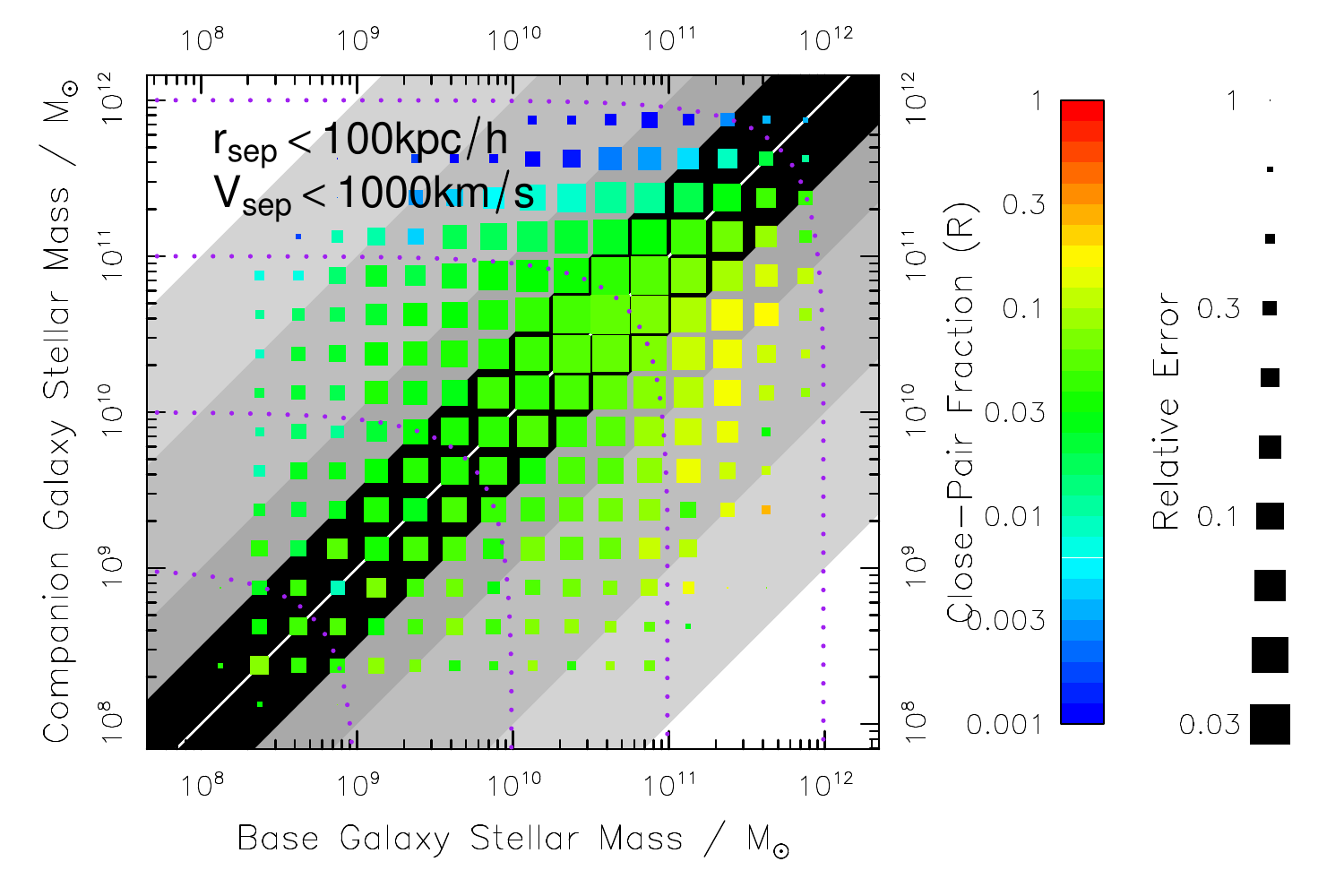}}}
\centerline{\mbox{\includegraphics[width=3.7in]{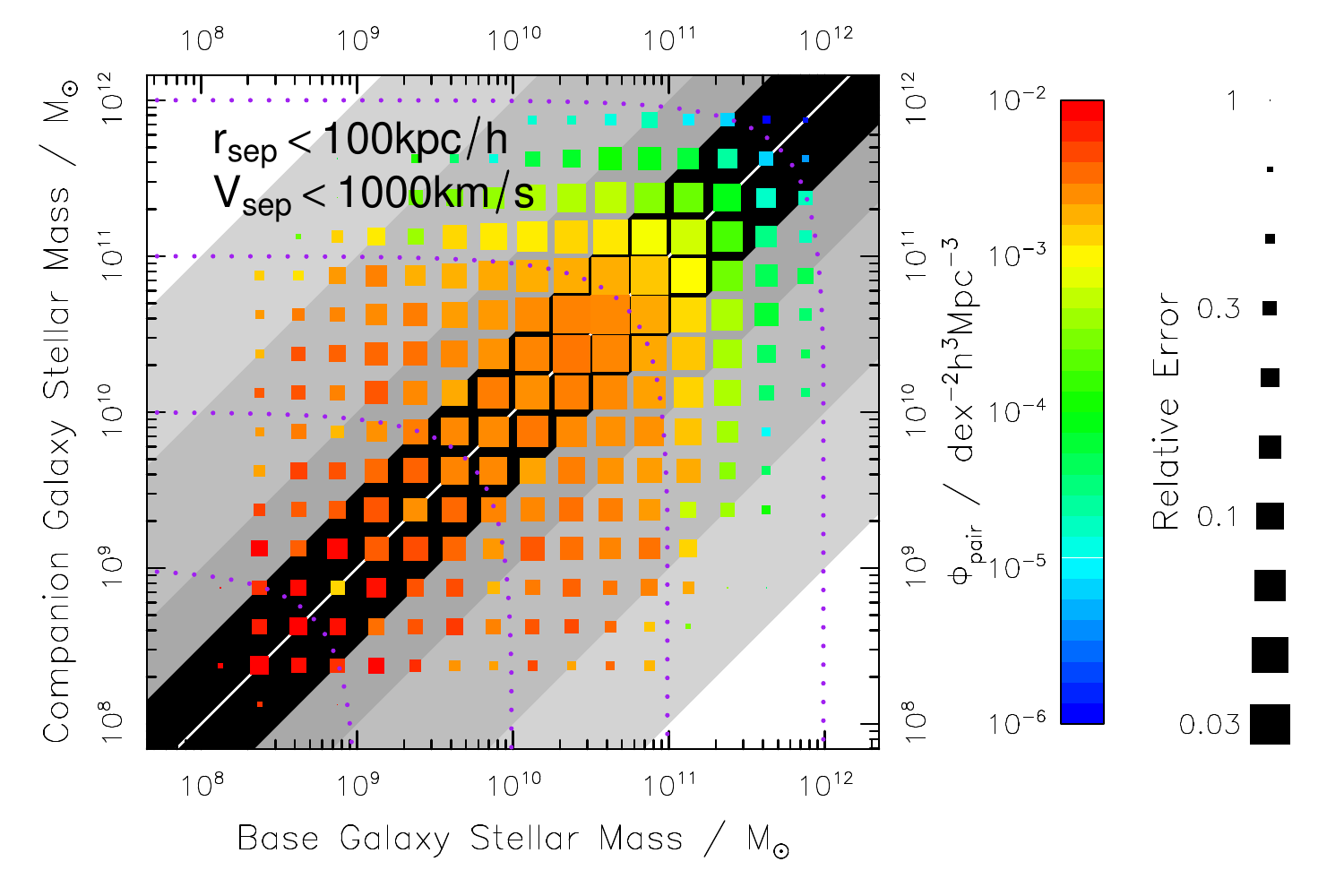}}}
\centerline{\mbox{\includegraphics[width=3.7in]{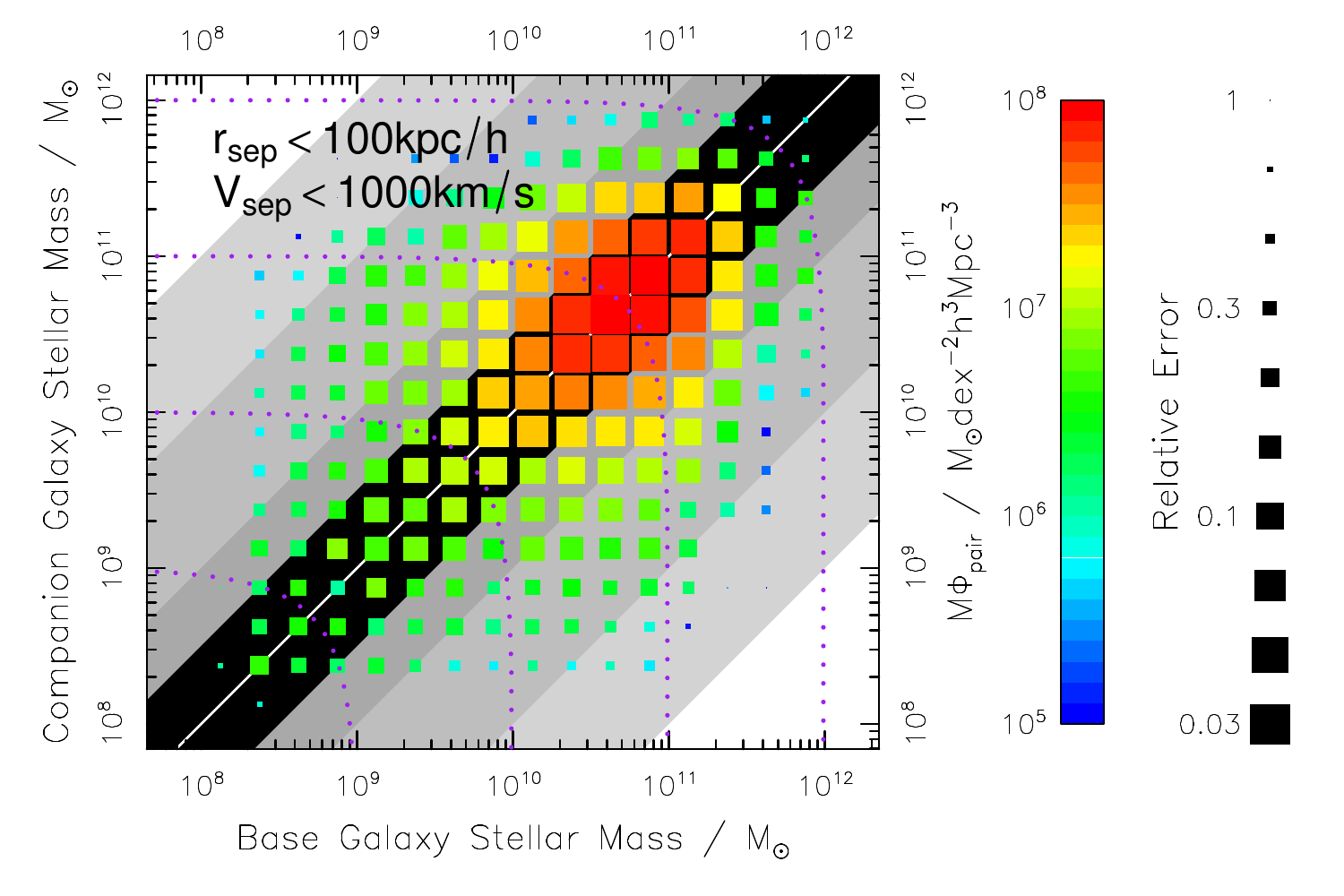}}}
\caption{\small Pair properties for $r_{\rm sep}<100\kpc$ and $v_{\rm sep}<1000 \kms$ sample ($PS_{r100v1000}$) true pair corrected using the mock catalogues ($W_{TP-mock}$, see Section \ref{sec:TPcor}). For more information see the caption for Figure \ref{fig:MergeGridA20V500}.}
\label{fig:MergeGridA100V1000}
\end{figure}

This Section contains the main close-pair observations for our three close-pair selections: $PS_{r20v500}$, $PS_{r50v500}$ and $PS_{r100v1000}$.
Figures \ref{fig:MergeGridA20V500}, \ref{fig:MergeGridA50V500} and \ref{fig:MergeGridA100V1000} display the full range of stellar mass pairs from $10^8\msol$--$10^{12}\msol$ for the $PS_{r20v500}$, $PS_{r50v500}$ and $PS_{r100v1000}$ samples respectively.
Each Figure shows the close pair fraction (top panel) comoving close-pair density (middle panel) and the mass density in close-pairs (bottom panel).

The Figures show results with true pair corrections using the mock catalogues only ($W_{TP-mock}$), i.e.\ they do not show the visual disturbance corrections ($W_{TP-vc}$, see Section \ref{sec:TPcor}). Each bin shown is volume limited by the redshift limits calculated in Section \ref{sec:selection}, where we take the lowest stellar mass possible in the abscissa and ordinate bins of interest to determine the redshift limit to apply to the sample. Further to this, all Figures include the various corrections discussed in Section \ref{sec:correct}, explicitly:

\begin{equation}
R_{i,j}=\frac{NP(i,j)}{N(i)}\overline{W}_{photo}(i)\overline{W}_{spec}(i)
\end{equation}

\noindent where $R_{i,j}$ is the close-pair fraction with stellar masses in the bins i (x-axis, `Base' galaxies) and j (y-axis, `Companion' galaxies) cell in the Figure, $NP(i,j)$ is the number of pairs with galaxy stellar masses in the bins i and j (in either order), $N(i)$ is the number of galaxies with with stellar masses in the bin i, $\overline{W}_{photo}(i)$ is the mean photometric confusion weight for all galaxies with stellar masses in the bin i and $\overline{W}_{spec}(i)$ is the mean spectroscopic fibre collision weight for all galaxies with with stellar masses in the bin i. The $\bar{W_{complex}}$ factor (the mean complex correction for all galaxies that contribute to $NP(i,j)$) is not explicitly applied to the results, instead we later make use of an analytic approximation for this correction discussed in detail later in the paper).

To give an idea of the impact, the mean scaling applied by the $\overline{W}_{photo}.\overline{W}_{spec}$ factor for the $PS_{r20v500}$ sample across all cells $i,j$ is 1.17, with 25\%, 50\% and 75\% quartile ranges of 1.09, 1.12 and 1.18 respectively. The number count densities vary smoothly in a well-behaved manner over the full grid of observations. Importantly, we do not see evidence of unusual discontinuities at 1--1 stellar masses (the diagonal values). This is where we might expect photometric errors to cause artefacts if an appreciable number of spatially close galaxies have apertures that erroneously overlap, creating false 1--1 stellar mass close-pairs. We also show later that the observed corrected 2D distribution can be very well fit by a simple 3 parameter model. All this information suggests that whilst the calculation of these correction terms might be relatively onerous, they generally only have a small impact on our results and behave in the correct manner. This is in a large part thanks to the extremely high spectroscopic completeness for close-pairs in GAMA-II-N, consistent photometric apertures applied across multiple bands \citep{hill11} and careful stellar mass measurements \citep{tayl11}.

In the Figure panels we use the terms `Base Galaxy Stellar Mass' and `Companion Galaxy Stellar Mass', where a pair is formed by the combination of a `Base' and `Companion' galaxy. For the bottom two panels the data is symmetric about the diagonal since it is a requirement that we conserve number counts and mass, no matter which way round we treat a pair. For the top panel the distinction is important: the colouring shows the number of pairs per decade of stellar mass for the `Base' galaxy. For example, let us assume that there are 1,000 galaxies with stellar masses in the bin around $10^8\msol$ and 100 galaxies with stellar masses in the bin around $10^{10}\msol$. Let us further assume that there are 10 close-pairs with one galaxy in the $10^8\msol$ and one galaxy in the $10^{10}\msol$ bin.
This means depending which of these masses is treated as the `Base' and `Companion' the pair fraction is either 1/100 (when $10^8\msol$ is the `Base') or 1/10 (when $10^{10}\msol$ is the `Base'). This is why the top panels in Figures \ref{fig:MergeGridA20V500}--\ref{fig:MergeGridA100V1000} are asymmetric about the diagonal: the pair fraction cares about which stellar mass it is being compared to since there are more less-massive galaxies per cosmic volume \citep{bald12}.

In Figures \ref{fig:MergeGridA20V500}--\ref{fig:MergeGridA100V1000} the black diagonal band shows the region containing potential future `major mergers', which uses a popular literature definition of a 3:1 close-pair mass ratio \citep[e.g.][]{hopk08a}. The lighter grey shaded regions highlight increasing decades in stellar mass ratio for galaxies in close-pairs.

We will now discuss the details of Figures \ref{fig:MergeGridA20V500}--\ref{fig:MergeGridA100V1000} at the example of Figure \ref{fig:MergeGridA20V500}, which presents the pairwise results for the $PS_{r20v500}$ selection. For this sample the lower mass ratio limits are $\sim10:1$ for $\mathcal{M}\sim10^9\msol$, $\sim30:1$ for $\mathcal{M}\sim10^{10}\msol$ and$\sim100:1$ for $\mathcal{M}\sim10^{11}\msol$. The fall off beyond this is expected due to the sharp drop in the GSMF at high masses, and the small volume in which we could possibly observe low stellar mass galaxies. The general effect we see is that for a given `Base' stellar mass, galaxies are more likely to be in a close-pair with less massive `Companion' galaxies. This argument seems entirely reasonable given the monotonic decline of the GSMF in \citet{bald12}, where less massive galaxies are increasingly common (the low mass end exhibiting a steep power law slope of $-1.47$). However, these Figures alone do not reveal whether the increasing probability of being in a pair closely tracks the exact shape of the GAMA GSMF from \citep{bald12}, this will be investigated in detail later in this paper.

The middle panel in Figure \ref{fig:MergeGridA20V500} represents the close-pair number per unit volume in GAMA. This is constructed by multiplying the close-pair fractions in each cell in the top panel by the GSMF value for the x-axis stellar mass. We do this because the top panel shows y pair galaxies per unit x. By construction this Figure is a mirror image about the diagonal. There is clear evidence that the close-pair number per unit volume is consistently highest for lower stellar masses, i.e.\ if we consider `Base' stellar masses at $\mathcal{M}\sim10^{11}\msol$ they have a higher space density of close-pair when the `Companion' has a stellar mass $\mathcal{M}\sim10^{9}\msol$ rather than $\mathcal{M}\sim10^{12}\msol$. This is despite the fact that the close-pair fraction per galaxy peaks at around $\mathcal{M}^*$ \citep[$10^{10.66}\msol$,][]{bald12}, and is due to the huge number of lower mass galaxies at the low end of the GSMF.

For predicting the likely future of the GSMF, the bottom panel is of key importance. This is the product of the close-pair density per unit volume and the stellar mass of the minor accreting mass in any pair (i.e.\ the lower of the two stellar masses in any close-pair). This panel identifies the stellar mass of galaxies that contain stars whose orbits will be most strongly affected by being in a closely interacting pair, and for the $PS_{r20v500}$ selection the accreting mass due to a likely future merger. Throughout, galaxies with stellar masses between $\mathcal{M}^*<\mathcal{M}<10^{11}\msol$ comfortably dominate the mass undergoing close interactions and mergers. Since only mass ratios close to 1:1 (3:1 major mergers and closer in mass) will cause large changes to the component stellar mass of close-pair galaxies, major mergers should comfortably dominate the movement of {\it mass} due to mergers in the $z=0$ Universe. However, minor mergers could still have a significant role in the redistribution of {\it number counts}. We shall look at the role of major and minor mergers in more detail in Sections \ref{sec:mergevarmass}, \ref{sec:mergevarz} and \ref{sec:inoutmerge}, using these results to predict the redistribution of mass and number counts around the GSMF.

\section{Parameterising Galaxy Close-Pairs}
\label{sec:closepairfits}

Having measured the observed close-pair distributions for our three different $PS$ selections we now investigate whether there is a meaningful manner of parameterising the distributions. Such a process is important since if the fitting function is a good match to the data then a lot of information can be conveyed with relatively few numbers. Further, if the appropriate fitting function continues to behave sensibly beyond the range of our data then useful extrapolated properties can be derived. In this work we wish to know how mass will redistribute itself in the GSMF, which required knowledge of how major and minor mergers behave outside the stellar mass range of our observations. We also wish to know how much mass is contained in accreting material, i.e.\ the integral of all accreting stellar mass between 0 and $\infty$. For reference, the double-Schechter form of the GSMF \citep{bald12} can be specified by:

\begin{equation}
\begin{array}{lll}
\label{eqn:GSMF}
\phi_{G}(\mathcal{M}) \equiv \frac{dn}{d\mathcal{M}}=\\\\
e^{(-\mathcal{M}/\mathcal{M}^*_G)}\left[\phi^*_{G1}\left(\frac{\mathcal{M}}{\mathcal{M}^*_G}\right)^{\alpha_{G1}}+ \phi^*_{G2}\left(\frac{\mathcal{M}}{\mathcal{M}^*_G}\right)^{\alpha_{G2}}\right]\\
\end{array}
\end{equation}

\noindent where $\mathcal{M}^*_G$, $\phi^*_{G1}$, $\phi^*_{G2}$, $\alpha_{G1}$ and $\alpha_{G2}$ take the standard definitions of $\mathcal{M}^*$, $\phi^*_{1}$, $\phi^*_{2}$, $\alpha_{1}$ and $\alpha_{2}$ in \citet[see][]{bald12}.

To manipulate the empirical results of Section \ref{sec:galmerge} we now introduce an analytic fit of the close-pair stellar mass function (CPSMF), defined as the volume density of close pairs as a function of the stellar masses $\mathcal{M}_1$ and $\mathcal{M}_2$ of the two galaxies in a pair. The function is necessarily symmetric with respect to the exchange of $\mathcal{M}_1$ and $\mathcal{M}_2$. By inspection of the above Figures and the parameterisation of the GSMF given in Equation \ref{eqn:GSMF}, an appropriate functional form to investigate appeared to be a multiplicative Schechter function,

\begin{equation}
\begin{array}{lll}
\label{eqn:CPSMF}
\phi_{CP}(\mathcal{M}_1,\mathcal{M}_2) \equiv \frac{\partial^2n}{\partial\mathcal{M}_1\partial\mathcal{M}_2}=\\\\
e^{-(\mathcal{M}_1+\mathcal{M}_2)/\mathcal{M}^*_{CP}}\left[\phi^*_{CP}\left(\frac{\mathcal{M}_1}{\mathcal{M}^*_{CP}}\right)^{\alpha_{CP}}\times \phi^*_{CP}\left(\frac{\mathcal{M}_2}{\mathcal{M}^*_{CP}}\right)^{\alpha_{CP}}\right]\\
\end{array}
\end{equation}

\noindent where $\mathcal{M}^*_{CP}$ is the knee for the 2D close-pair distribution, $\phi^*_{CP}$ is the normalisation and $\alpha_{CP}$ is the low-mass slope. It is notable that this function only is the multiplication of a single power-law slope version of the Schechter function, rather than the double component form preferred for the GSMF in \citet{bald12}. During detailed investigations of the most appropriate form for the 2D close-pair distribution, a single power-law form was overwhelmingly preferred when comparisons of the Marginal-Log-Likelihood of the posterior distributions were made. For this reason we will only present the results of the single power-law fits.

With the 2D number density of close-pairs specified as above, and using the double power-law analytic form of the GSMF, we can specify the close-pair {\it fraction} (close-pairs per unit galaxy, so close-pair number density per unit galaxy number density), as:

\begin{equation}
\gamma_{CP}(\mathcal{M}_B,\mathcal{M}_C) \equiv \frac{\phi_{CP}(\mathcal{M}_B,\mathcal{M}_C)}{\phi_{G}(\mathcal{M}_B)}
\end{equation}

\noindent where $\phi_{G}(\mathcal{M}_B)$ is the GSMF for the `Base' galaxy in a  pair, and $\phi_{CP}(\mathcal{M}_B,\mathcal{M}_C)$ is as specified in Equation \ref{eqn:CPSMF} for the `Companion' and `Base' galaxies in a pair respectively.
This leaves us to calculate the free parameters for $\phi_{CP}$, which we will do for the 3 different dynamic windows $PS$ specified by Equation \ref{eqn:Pselect} and the stellar mass selections detailed in Section \ref{sec:selection}. We directly use the measured empirical GSMF rather than its double-Schechter approximation specified in Equation \ref{eqn:GSMF} \citep[i.e.\ we use the published values in][]{bald12}, but we note that both give compatible results.

A caveat to this calculation is that we will end up counting galaxies more than once in some situations, because they potentially appear in more than one close-pair. This effect is particularly likely when calculating close-pairs in the largest dynamical window. By treating the likelihood of being in a close-pair as an independent event, we can use the sum of a geometric series formula to rescale the close-pair fraction for {\it unique} close-pairs (so this means there should not be more close-pairs than galaxies). This rescaling assumes independent close-pair occurrences (hence the use of the sum of the geometric series) and is therefore only approximately true (in reality close-pairs are more likely to have another close-pair than a random galaxy), but it produces accurate results up to the largest dynamic window investigated here, even when using close-pair fractions within a multi-decade versus multi-decade stellar mass window. Making these assumptions, the appropriate rescaling factor to use is $1/(1+\gamma_{CP}(\mathcal{M}_B,\mathcal{M}_C))$ for the specified fitting parameters. This factor guarantees that, at most, 100\% of all galaxies (and all galaxy mass) are in close-pairs. For calculating the total mass contained in close-pairs, this factor must be used. This factor will vary depending on the 2D stellar mass window of interest, so it is left to the user to construct appropriately in general cases. For the $PS_{r20v500}$ selection the rescaling tends to be only a few percent, so it can often be ignored without introducing significant bias. For the $PS_{r50v500}$ and $PS_{r100v1000}$ selections it should generally be applied.

\subsection{Fitting the Data}

\label{sec:datafit}

The fitting posterior space was investigated, and the parameter probability distributions are well behaved covariate Gaussians, so the maximum likelihood and expectation for the fit parameters are in excellent agreement. Therefore, to optimally fit the data, we used a maximum likelihood analysis of the close-pair number density distributions, where the inverse of the Hessian about the mode in likelihood space becomes our covariance matrix. Since we fit to the close-pair number density distributions some results require scaling by the $1/(1+\gamma_{CP}(\mathcal{M}_B,\mathcal{M}_C))$ factor to ensure mass conservation. Since the scaling required necessarily varies depending on the stellar mass ranges of interest, this must be applied by the user. Specific results of interest are presented here with the required scaling applied explicitly.

\begin{figure*}
\centerline{\mbox{\includegraphics[width=6.0in]{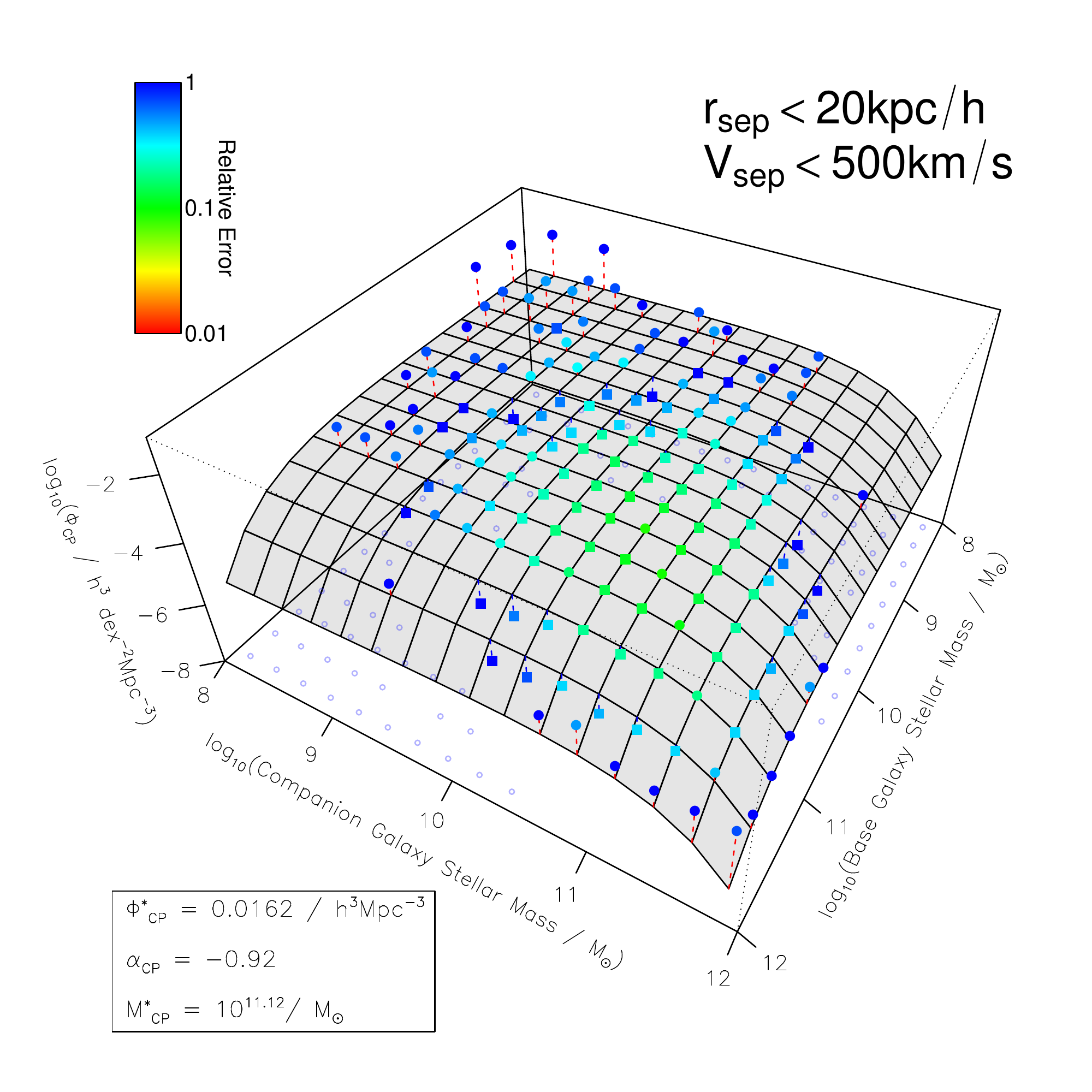}}}
\caption{\small Fit to the 2D close-pair density distribution for $r_{\rm sep}<20\kpc$ and $v_{\rm sep}<500 \kms$ sample ($PS_{r20v500}$) as observed, i.e.\ {\it not} true pair corrected using the mock catalogues (see Section \ref{sec:TPcor}). The fit is represented by the grey shaded 2D manifold. The colour of the binned data represents the estimated relative error (so redder points dominate the fit more). Circles on the base of the 3D plot represent missing data.}
\label{fig:MergeFitA20V500}
\end{figure*}

\begin{figure*}
\centerline{\mbox{\includegraphics[width=6.0in]{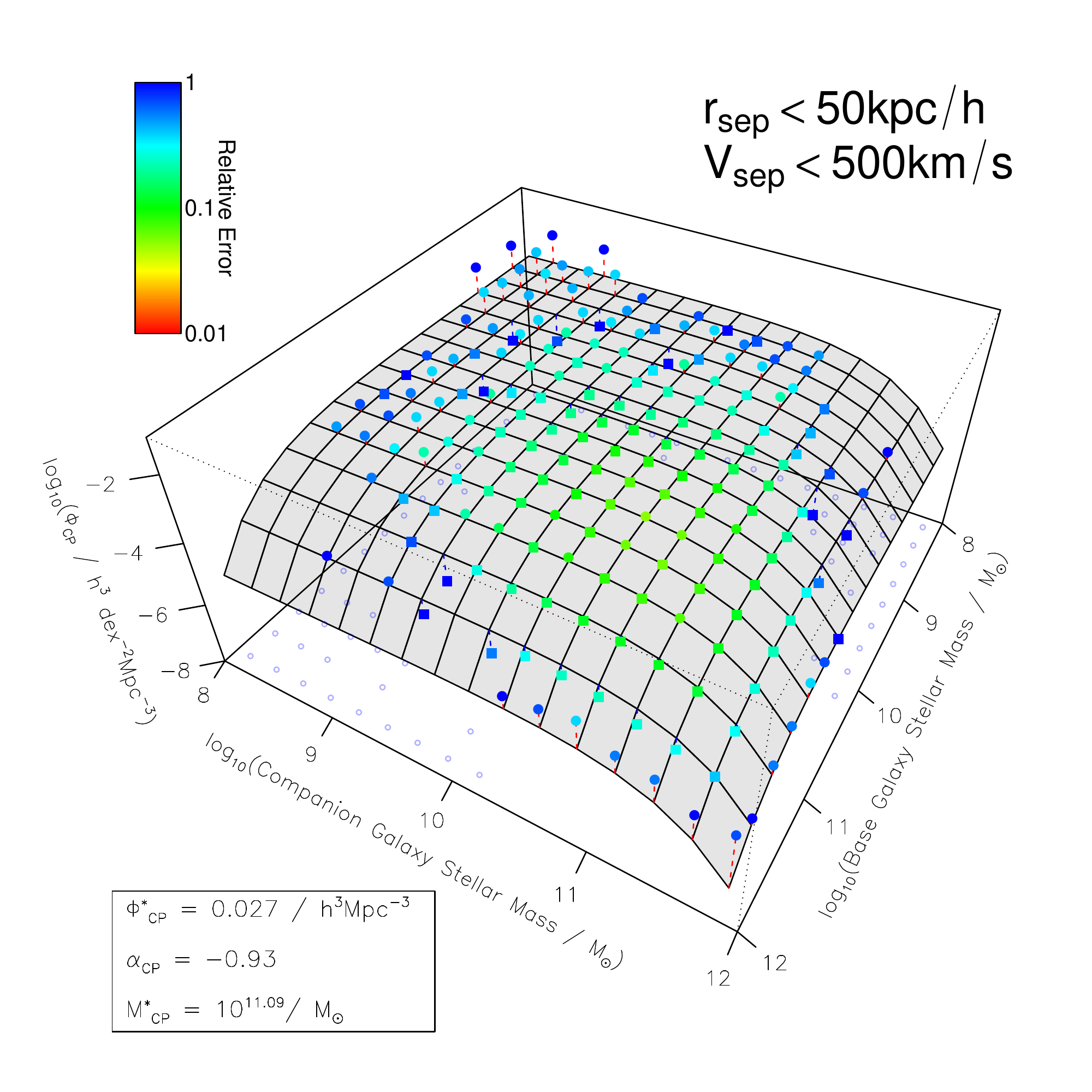}}}
\caption{\small Fit to the 2D close-pair density distribution for $r_{\rm sep}<50\kpc$ and $v_{\rm sep}<500 \kms$ sample ($PS_{r50v500}$) as observed, i.e.\ {\it not} true pair corrected using the mock catalogues (see Section \ref{sec:TPcor}). See Figure \ref{fig:MergeFitA20V500} for details.}
\label{fig:MergeFitA50V500}
\end{figure*}

\begin{figure*}
\centerline{\mbox{\includegraphics[width=6.0in]{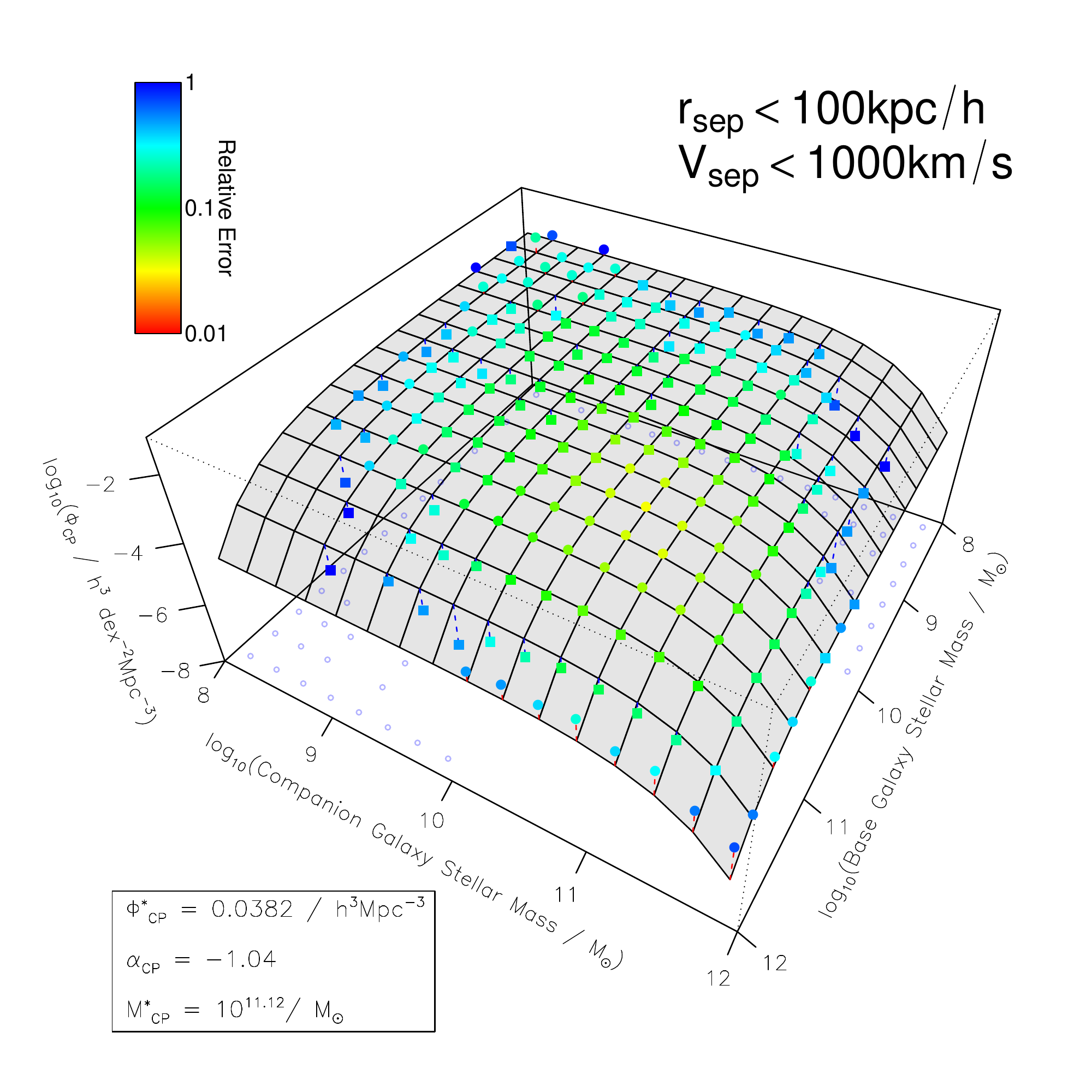}}}
\caption{\small Fit to the 2D close-pair density distribution for $r_{\rm sep}<100\kpc$ and $v_{\rm sep}<1000 \kms$ sample ($PS_{r100v1000}$) as observed, i.e.\ {\it not} true pair corrected using the mock catalogues (see Section \ref{sec:TPcor}). See Figure \ref{fig:MergeFitA20V500} for details.}
\label{fig:MergeFitA100V1000}
\end{figure*}

In all cases the fit was made to the un-binned number densities (the coloured data-points shown in Figures \ref{fig:MergeFitA20V500}, \ref{fig:MergeFitA50V500} and \ref{fig:MergeFitA100V1000}) via a Nelder-Mead uphill gradient search of the likelihood space (since we measure maximum likelihood). The local parameter covariance was calculated as part of the fitting process. In all cases the single parameter variance dominates, so only this is presented here.

\begin{table}
\begin{center}
\begin{tabular}{| l | l | l | l |}
				& $\mathcal{M}^*_{CP}$ 	& $\phi^*_{CP}$ 	& $\alpha_{CP}$ 	\\
				& $/\msol$			& $/h^3{\rm Mpc}^{-3}$		&					\\
\hline
$PS_{r20v500}$  	& $10^{11.12 \pm 0.03}$			& $0.0162 \pm 0.0008$	& $-0.92 \pm 0.05$ 		\\
$PS_{r50v500}$ 	& $10^{11.09 \pm 0.02}$			& $0.0270 \pm 0.0008$	& $-0.93 \pm 0.04$		\\
$PS_{r100v1000}$ 	& $10^{11.12 \pm 0.01}$			& $0.0382 \pm 0.0008$	& $-1.04 \pm 0.02$		\\

\end{tabular}
\end{center}
\caption{Table of best-fit CPSMF fitting parameters for the three different dynamical selections used in this work.}
\label{tab:bestfit}
\end{table}%

Table \ref{tab:bestfit} shows the best fit parameters for the three dynamical selections used in this work, as shown in Figures \ref{fig:MergeFitA20V500}, \ref{fig:MergeFitA50V500} and \ref{fig:MergeFitA100V1000}. The values for $\mathcal{M}^*_{CP}$ are extremely consistent for all selections, agreeing within the error ranges determined. This suggests that the close-pairs stellar mass function is very well described by a fixed value of $\mathcal{M}^*_{CP}\sim11.1$ for all dynamical windows. $\phi^*_{CP}$ varies strongly with the dynamical window used, as should be expected since larger comoving volumes should contain more pairs by chance alone, regardless of other physical processes further enhancing this number. $\alpha_{CP}$ is similar for the two smallest dynamical selections, where values of $\alpha_{CP} \ge -1$ indicate most stars in close-pairs are found within galaxies of stellar masses around $\mathcal{M}^*_{CP}$. $\alpha_{CP}$ is larger for the largest dynamical window, but this is barely significant given the calculated errors, and perhaps not physically notable.

Using the analytic parameterisations presented in Table \ref{tab:bestfit} we can extrapolate to stellar masses beyond the range $10^{8} \le \mathcal{M} / \msol \le 10^{12}$ used to constrain the fits. Of clear interest is the implied accreting mass in mergers, which can be thought of as the total mass of sub-dominant components in close-pairs (i.e.\ the mass of the smaller galaxy in close-pairs). This is straightforward to calculate analytically given the fits to the data, and in all three dynamical windows investigated the mass is well bound within the mass range explored in this work. Using the appropriate scaling factors to account for galaxies being in potentially more than one close-pair (so to avoid double counting the accreting mass), we find the $PS_{r20v500}$ sample has a comoving accreting mass density of $0.038\times10^6 / (\msol/\mpc3h3$), $PS_{r50v500}$, $PS_{r20v500}$ has $0.100\times10^6 / (\msol/\mpc3h3$) and $PS_{r100v1000}$ has $0.300\times10^6 / (\msol/\mpc3h3$). This compares to a total comoving stellar mass density for all galaxies of $0.651\times10^6 / (\msol/\mpc3h3$) using the GSMF measured in \citet{bald12}. This means that, e.g.\, $\sim 6$\% of all stellar mass is available for minor merger accretion on the shortest dynamical timescale investigated here (the $PS_{r20v500}$ selection).

These numbers simply reflect the sub-dominant mass in close-pairs, which is not to say that all of this mass will merge on a rapid timescale. The simplest correction we can make is to account for the fraction of galaxies  separated radially, but seen as a close-pair in projection by a coincidence between cosmological redshift and peculiar pair-wise velocity. This correction has been estimated from analysis of mock catalogue pairs (discussed in Section \ref{sec:mockcorrect}) and implies a scaling factor $W_{TP-mock}$ of 0.961, 0.891 and 0.646 for the $PS_{r20v500}$, $PS_{r50v500}$ and $PS_{r100v1000}$ selections respectively. We can also make an adjustment for the de-biased fraction of visually disturbed galaxies in pairs $W_{TP-vc}$ (see Section \ref{sec:visualcorrect}), which we interpret to be a sign that they are in a real interaction and might shortly merge. This implies scaling factors of 0.44, 0.27 and 0.22 for the $PS_{r20v500}$, $PS_{r50v500}$ and $PS_{r100v1000}$ selections respectively. If we assume a background disturbed fraction of $\sim 0.1$ due to post recent-merger disturbances \citep[this is suggested by the background seen at large dynamical scales in Figure \ref{fig:disturbgrid} and for the isolated control sample of galaxies, but is also in good agreement with the fractions found in][]{patt05,darg10a} then these factors become 0.34, 0.17 and 0.12.

\begin{table*}
\begin{center}
\begin{tabular}{| l | l | l | l |}
Sub-dominant mass in close-pairs 	& All close-pairs					& Mock corrected 			& Visual disturbance corrected 		\\
						& $\msol/\mpc3h3$ (\% of all mass)		& $\msol/\mpc3h3$	(\% of all mass)		&  $\msol/\mpc3h3$	(\% of all mass)		\\
\hline
$PS_{r20v500}$  			& 0.038	(5.8\%)					& {\bf 0.036 (5.6\%)	}				& 0.013 (2.0\%)					\\
$PS_{r50v500}$ 			& 0.100	(15.4\%)					& 0.089 (13.7\%)					& 0.017 (2.6\%)					\\
$PS_{r100v1000}$ 			& 0.300	(46.1\%)					& 0.195 (30.0\%)					& 0.036 (5.5\%)					\\

\end{tabular}
\end{center}
\caption{Table of comoving density of sub-dominant mass in close-pairs. Columns show uncorrected results, mock catalogue corrected results (see Section \ref{sec:mockcorrect}) and visual disturbance corrected results (see Section \ref{sec:visualcorrect}).}
\label{tab:massmerge}
\end{table*}%

Table \ref{tab:massmerge} shows various estimates of the sub-dominant mass in close-pairs corrected for the various observational effects discussed above. To estimate the likely mass in future mergers we should, at a minimum, apply the $W_{TP-mock}$ mock catalogue corrections for spurious cosmological redshift coincidence (giving the results in the middle column). Being conservative, we can go further and estimate the minor mass in near-future mergers by applying the de-biased visual disturbance excess $W_{TP-vc}$ above the normal background fraction of $\sim 0.1$ (giving the results in the far right column). Taking the $PS_{r20v500}$ selection and $W_{TP-mock}$ selection, this suggests that $\sim$5.6\% of galaxy stellar mass is likely to accrete onto larger galaxies in the near future (within this dynamical window `near future' implies $\sim$Gyrs). This figure is in broadly good agreement with the $\Delta M/M=0.09 \pm 0.04$ Gyr $^{-1}$ presented in \citet{dokk05}, which was considering the accretion of galaxies onto the red sequence through the analysis of post-merger tidal disturbance. The work presented here considers all galaxies, not merely those on the red sequence. Therefore this lower number reflects the fact that some fraction of galaxies will not be on the red sequence, and therefore are less likely to have recently undergone a recent merger event.

Given the fits to the data, we can calculate the fraction of sub-dominant mass in close-pairs that could accrete in major mergers i.e.\ mergers with a stellar mass ratio below 3:1). The $PS_{r20v500}$ selection has 63\% of all sub-dominant mass in major close-pairs, the $PS_{r50v500}$ selection also has 63\% in major close-pairs and the $PS_{r100v1000}$ selection has 61\%. This means in all cases the majority of mass accreting onto more massive galaxies does so in a major merger event (in this work we term the `accreting mass' as the mass of the minor close-pair galaxy, even when the masses are similar). This is significant since it is these events that will most dramatically reorganise the distribution of {\it mass} in the GSMF, since the product of such an event can have a hugely different mass (up to a factor 2 increase, by definition). However, the redistribution of {\it number counts} in the GSMF might still be hugely affected by minor mergers, despite the minority effect they have on the movement of mass. This fraction of mass likely to merge in a major-merger event is similar to the 75\% found in \citet{lope11} using VVDS data ($z<1$), however they use a 4: mass ratio threshold to define `major-mergers' and only consider minor merger events down to 10:1 mass ratios. Given the flexibility of the analytic fits we can recalculate quantities using these thresholds and approximate the dynamical window used in that work. Doing this we find 79\% of merger mass in major close-pairs, i.e. slightly more mass is concentrated into major-mergers at lower redshift, but the difference is not statistically significant.

Using the raw observational data and our fits to it, we can take this analysis further and investigate the regions of the GSMF that are undergoing the most merger activity.

\subsection{Major Close-Pair Fraction Variation with Stellar Mass}

\label{sec:mergevarmass}

The literature on mergers usually concentrates on major mergers. These events happen on the most rapid timescales due to efficient dynamical friction when masses are equal \citep[e.g.][]{boyl08,kitz08}, they also tend to be the most spectacular, producing enhanced star formation rates and are the progenitors for the most luminous sub-mm galaxies \citep{ricc10}. The term `major merger' is potentially ambiguous, but in this work we use the term to mean stellar mass ratios below 3:1. Since we cannot be certain that close-pairs with mass ratios below 3 will certainly merge we wish to avoid labelling them as `major merger' close-pairs. Instead from here we will use the term `major close-pair' to refer to such systems, with the corollary `minor close-pair' for systems where the close-pair mass ratio is above 3:1.

\begin{figure}
\centerline{\mbox{\includegraphics[width=3.7in]{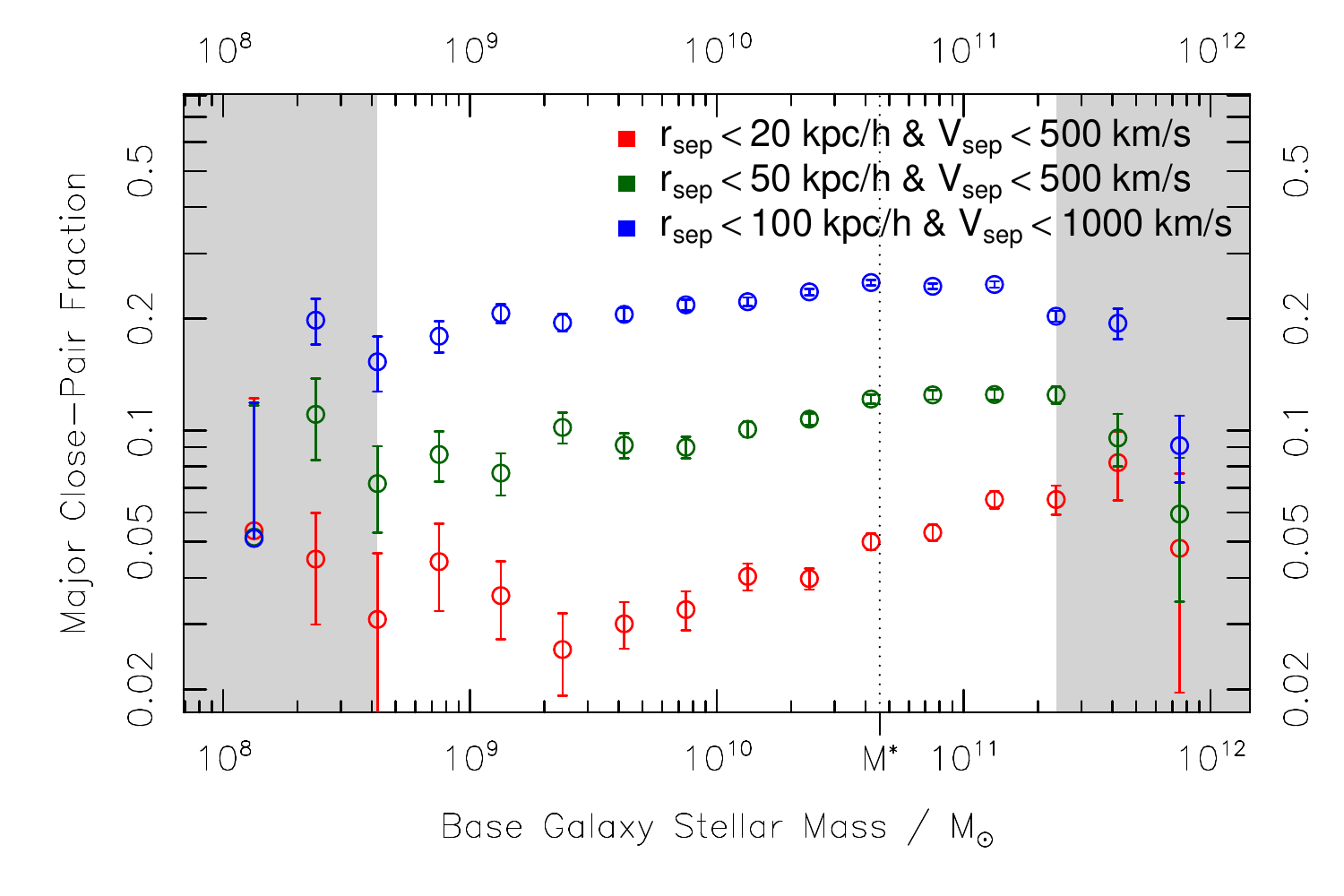}}}
\caption{\small Major merger fractions as a function of stellar mass for different dynamical pair selections. This is a simplified representation of the information presented in Figures \ref{fig:MergeGridA20V500}, \ref{fig:MergeGridA50V500} and \ref{fig:MergeGridA100V1000}.}
\label{fig:majmerge}
\end{figure}

Figure \ref{fig:majmerge} shows how the major close-pair fraction varies as a function of stellar mass. In this Figure no true pair corrections have been applied ($W_{TP-mock}$ or $W_{TP-vc}$, see above), so these results can be considered as hard upper limits on the possible merger fraction. In all cases we can see a significant enhancement in the fraction of galaxies experiencing major mergers as a function of stellar mass. The strength of this variation changes with the dynamical window being applied, where the smallest dynamical selection ($PS_{r20v500}$) shows a very strong gradient, changing by a factor $~\sim$3 between $2\times10^9\msol$ and $2\times10^{11}\msol$. The largest dynamical window ($PS_{r100v1000}$) is considerably flatter over this same range, but also suggests a hint of a turnover at the highest stellar masses. Such a turnover should be expected from dynamical friction arguments where merger time scales are most rapid for more massive galaxies in pairs when the mass ratios are closer to unity \citep{binn87}.

Figure \ref{fig:majmerge} has important implications for how major close-pair fractions are compared at different redshifts and across different surveys, since potentially even a small shift in the stellar mass at which the major merger fraction is being measured could result in a large increase or decrease in the resultant fraction. The highest $S/N$ measurement of galaxy parameters in apparent magnitude selected surveys is usually close to $\mathcal{M}^*$, so most surveys are, in effect, made at this point in Figure \ref{fig:majmerge} (the vertical dashed line indicates this point at $z\sim0$). This Figure only demonstrates the potential bias at $z\sim0$, but \citet{bund09} find very similar trends at higher redshifts with roughly a doubling in the close-pair fraction between $10^{10}\msol$ and $10^{11}\msol$ for a sample selection approximately similar to our $PS_{r20v500}$ sample.

\subsection{Major Merger Rates}

\label{sec:mergerate}

There is a lot of complexity in correctly mapping galaxy close-pairs into a galaxy merger rate. Even once we have applied mock catalogue true pair corrections, or corrected for signs of visual disturbance, we still have to estimate how rapidly the remaining close-pair will merge to know how often such events occur per unit volume per unit time. Earlier on in the field of galaxy close-pair analysis this mapping was approximated via simple dynamical friction arguments \citep{patt00,patt02}, mostly of the form presented in \citet{binn87}. More recently, effort has been invested into better estimating the complex physical processes by mapping close-pair properties onto large N-body simulations \citep{boyl08,kitz08}. These timescales tend to be significantly longer than those implied by the analytic arguments of \citet{binn87}, in general bringing historically published values of galaxy merger rates down by a factor of a couple.

\begin{figure}
\centerline{\mbox{\includegraphics[width=3.7in]{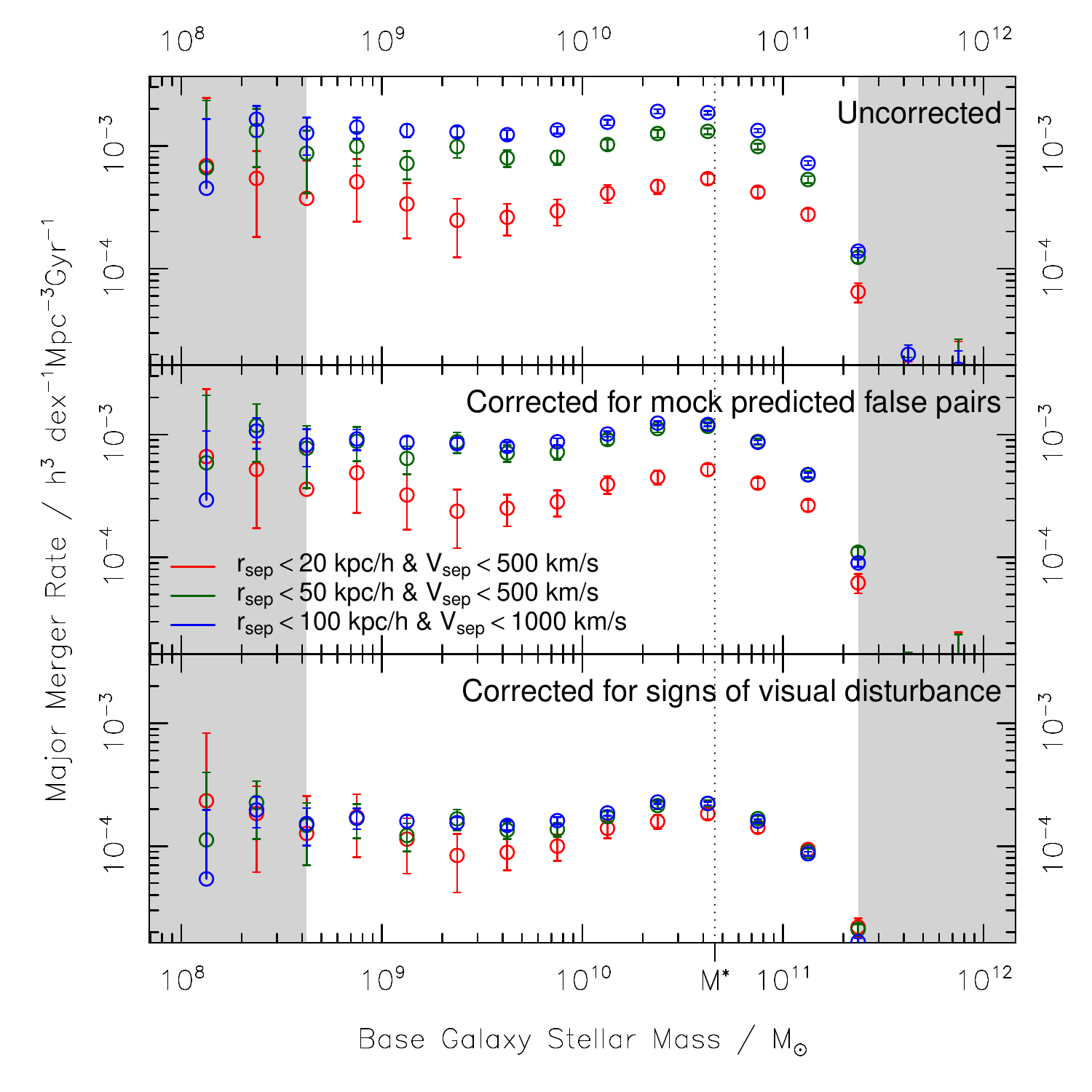}}}
\caption{\small Major merger rates as a function of stellar mass for different dynamical pair selections. This is a simplified representation of the information presented in Figures \ref{fig:MergeGridA20V500}, \ref{fig:MergeGridA50V500} and \ref{fig:MergeGridA100V1000} combined with estimates of merger timescale for different stellar mass close-pairs taken from \citet{kitz08}. The \citet{kitz08} mappings naturally account for false close-pairs due to their calibration to simulated data, so we expect the top-panel (applying it to our `uncorrected close-pair fractions) to be the most representative of the true merger rates.}
\label{fig:majmergerate}
\end{figure}

We use our predicted close-pair number densities for major-mergers presented above and apply the merger timescales suggested by equations 10 and 11 of \citet{kitz08}. Being agnostic to the reliability of such mappings, we are also careful to apply this merger timescale mapping to all three samples: the uncorrected close-pairs; the close-pairs corrected for mock catalogue estimated false pairs; and the close-pairs corrected for signs of visual disturbance. Because of the origin of the mapping presented in \citet{kitz08} the mock catalogue corrections we have presented {\it should} already be folded in. As such, applying the \citet{kitz08} mappings should be most appropriate for the {\it uncorrected} close-pair data. Since we wish to be conservative in this analysis, in Figure \ref{fig:majmergerate} we present the main results of applying these mappings to all of our different dynamical selections and merger corrections.

Applying the \citet{kitz08} timescales to the different close-pair selections the results converge together relative to the differences we see when we show the rawer major close-pair fraction, i.e.\ compare the top panel of Figure \ref{fig:majmergerate} to Figure \ref{fig:majmerge}. Compared to Figure \ref{fig:majmerge} it is immediately noticeable that once dynamical merging timescales are folded in, $\mathcal{M}^*$ galaxies are experiencing the highest rate of merger events per unit volume per unit time, i.e.\ they inhabit the stellar mass domain of maximal merger activity. Below this stellar mass the merger rate drops appreciably and then plateaus or possibly even rises slightly again, the distinction being difficult to confirm with the data available and the uncertainty in the merger timescale prescription applied. All dynamical selections and merger corrections see a very strong decline in the merger rate above $\mathcal{M}^*$, revealing that such massive major events are extremely unlikely in the local Universe.

It is interesting to observe that applying the different corrections to the pair data brings the different merger rate measurements into much closer alignment. Indeed, scaling the data by the observed prevalence of visual disturbance brings all dynamical selections in to broad agreement (given the errors). From inspection of Figure \ref{fig:disturbgrid} this should not be entirely surprising--- once we subtract the `background' disturbed rate (itself a combination of post-merger disturbance, intrinsically disturbed galaxies and false identification) we generally add few extra additional close-pairs as we move from $PS_{r20v500}$ to $PS_{r100v1000}$. This suggests we are converging on a selection of galaxies that are in a visually dramatic stage of the merger process.

The errors on the predicted merger rate distributions are likely to be even larger than depicted in Figure \ref{fig:majmergerate} since the dominant form of error is almost certainly that due to the forward mapping of galaxy close-pair properties onto a merger timescale. In reality, we can probably be confident of these mappings to within a factor of a couple, and ongoing work is being invested in better estimating these mappings using high resolution N-body simulations that systematically map out a useful subset of close-pair parameter space (discussed in Section \ref{sec:futurework}).

\subsection{Major Close-Pair Fraction Variation with Redshift}

\label{sec:mergevarz}

Having explored the effect of measuring the major close-pair fraction at different stellar masses, we will now investigate the apparent evolution with redshift. To be consistent with comparative literature we will make this calculation at $\mathcal{M}^*$ \citep[from][we take $\mathcal{M}^*=10^{10.66}$]{bald12}, where the GAMA selection limits allow us to calculate the major merger fraction out to $z\sim0.2$. To compare to previous work covering a large range of redshift \citep[using the modified compilation or major merger fractions published in][]{xu12a} we scale the merger fraction by the pair projection bias discussed above. For this comparison the data has been standardised to consider a 3:1 threshold for major mergers, and a dynamical selection window of $20\kpch$ projected separation and 500km/s velocity separation. Due to the nature of the surveys used, the data is dominated by galaxies near $\mathcal{M}^*$. There will be residual variation due to the exact mass ranges considered (as seen in Figure \ref{fig:majmerge}), but we choose not to attempt post-hoc corrections to the presented values.

Some of our earlier corrections made either strong or weak assumptions regarding likely close-pair evolution over the GAMA baseline, but for the reasons outlined these should not undermine our measurement. To determine the angular separation that most affected the SDSS deblender in Section \ref{ref:Pcomp} we investigated the origin of close-pair incompleteness out to $z=0.1$, however the actual correction we subsequently applied is independent of redshift information, only correcting for the fraction of the projected close pair that a $3''$ aperture covers. Also, since we make this comparison of the major merger fractions without applying the redshift dependent visual classification de-biasing, these results are not dependent on any earlier assumptions of non-evolution over the redshift baseline used in GAMA.

\begin{figure}
\centerline{\mbox{\includegraphics[width=3.7in]{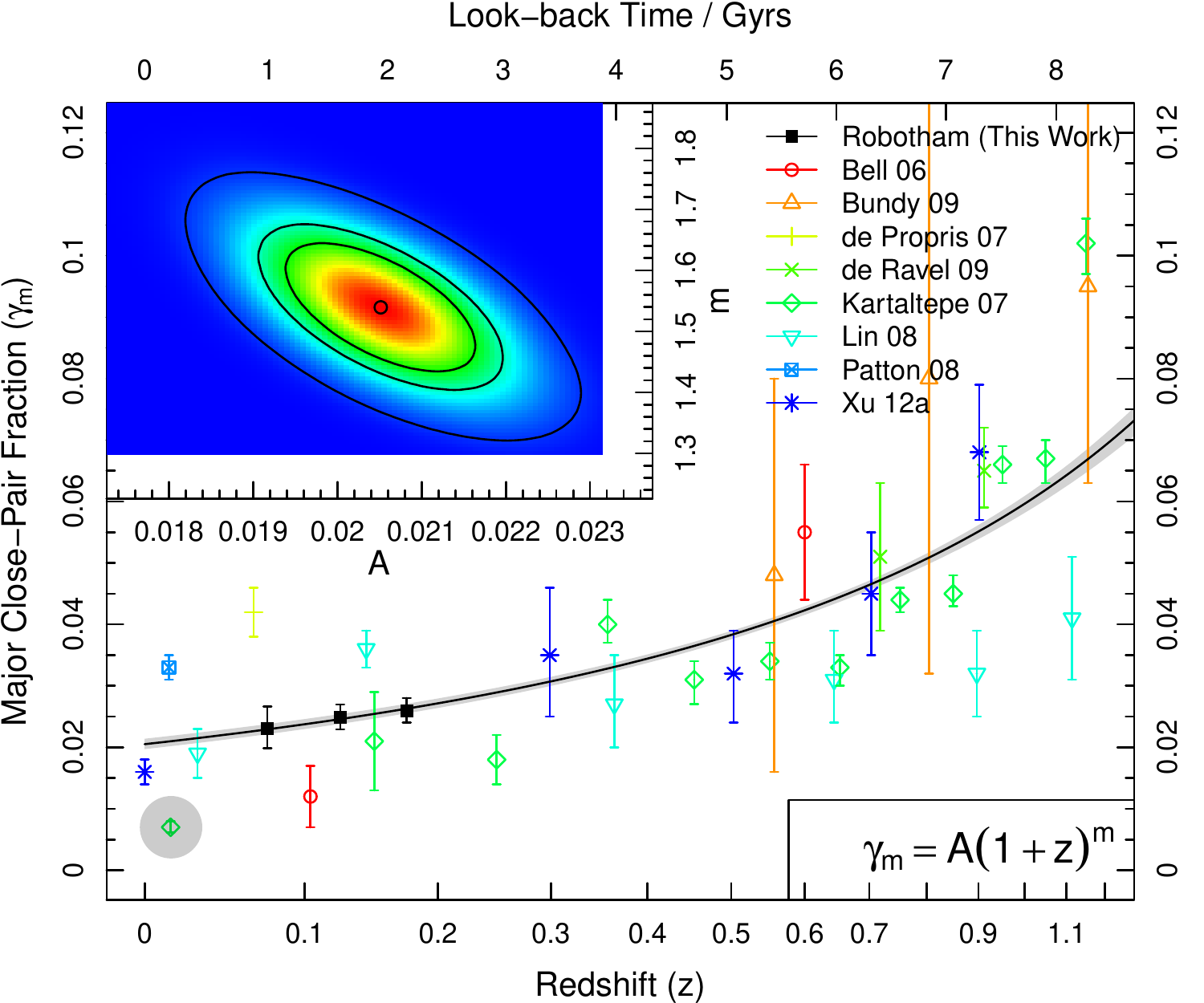}}}
\caption{\small Major merger close-pair fraction as a function of redshift (bottom) or look-back time (top). The stellar mass range explored is limited to $z<0.2$ in GAMA since we are still conservatively complete to $\mathcal{M}^*$ galaxies out to this redshift. This Figure uses data taken from \citet{bell06,bund09,depr07,dera09,kart07,lin08,patt08,xu12a} as presented in \citet{xu12a}, where results are scaled so as to use a common $20\kpch$ projected separation, 500km/s velocity separation and major merger definition of a 3:1 mass ratio close to $\mathcal{M}^*$. The solid black line shows the Bayesian expectation for our simple 2 parameter model, and the grey shaded region shows the 1-$\sigma$ marginalised range of allowed fits. The top-left inset panel shows the 50\%, 1$\sigma$ and 95\% percentile range contours for our preferred model using the posterior MCMC chains.
}
\label{fig:majmergeevo}
\end{figure}

Figure \ref{fig:majmergeevo} shows a compendium of major close-pair fractions published in \citet{xu12a} using data taken from \citet{bell06,bund09,depr07,dera09,kart07,lin08,patt08,xu12a}. The black data points show our GAMA major close-pair fractions in three redshift bins spanning $z=0.05$--$0.2$. The GAMA results have by far the strongest constraints of published values in the redshift range explored due to the huge number of close-pairs available in the survey in this regime. These values are largely consistent with the mixture of published values covering the same range of redshift. We see very mild evidence for major close-pair fraction evolution (increase with redshift) over the range investigated, but the results are consistent with the fraction remaining constant between $z=0.05$--$0.2$.

Combining these data together, it is useful to attempt to find the optimal parameterisation of the evolution of major close-pair fractions. Comparing the marginalised log-likelihoods of a number of simple models (including $Az+C$, $A(1+z)+C$ $A(1+z)^m$, $A(1+z)^m+C$ and $A(1+z)^m e^{c(1+z)}$), we find we prefer a simple two parameter model of the type:

\begin{equation}
\gamma_{m} = A(1+z)^m,
\end{equation}

\noindent where $\gamma_{M}$ is the major close-pair fraction, $z$ is the observed redshift, and $A$ and $m$ are parameters to be fitted. Using standard Metropolis Markov Chain Monte Carlo (MCMC) sampling we can estimate the posterior of the likelihood space, with the likelihood model based on the Gaussian density of the data given the model. This returns the expectation of the three parameters we wish to fit, along with the standard deviations and covariance. The `most likely' parameterisation of the evolution of close-pair major mergers is found to be: $A=0.021 \pm 0.001$ and $m=1.53 \pm 0.08$ where we censor out the highest tension low-$z$ \citet{kart07} data point (this is indicated by a transparent grey circle on Figure \ref{fig:majmergeevo}, and is discussed in detail below). The grey shaded region in Figure \ref{fig:majmergeevo} shows the full 1$\sigma$ range of allowed fits using our model parameterisation.

The data analysed here does not extend to high enough redshifts to witness, and therefore fit for, any high redshift downturn in the major close-pair fraction \citep[i.e. the preferred $A(1+z)^m e^{c(1+z)}$ model presented in][]{cons06}. Our value for the power-law ($m$) is greater than the $m=0.41\pm0.20$ figure published in \citet{lin08}, but less than the $m=2.2\pm0.2$ in \citet{xu12a}. Most of the data used to constrain our fit overlaps with that presented in \citet{xu12a}, with the exception that we have much better constraints at low redshift through the addition of the three GAMA data points. The flatter slope is partly driven by the lack of significant evolution seen for the GAMA data.

We also note that \citet{xu12a} find a normalisation $A=0.013\pm0.001$, which is significantly less than the value we find. Inspecting their Figure 6, it is clear there fit is being heavily dragged down at low redshift by the lowest-$z$ \citet{kart07} data point. This data point also has by far the most tension with the best global fit, being more than 10$\sigma$ away from the preferred functional form. This is likely to be the main origin of the much lower normalisation and steeper power-law fit seen in \citet{xu12a}. To test this we attempted a fit without the GAMA data, and then either included or discarded the lowest redshift \citet{kart07} data point. Including the data point returns fit parameters $A=0.013\pm0.001$ and $m=2.26\pm0.08$, but discarding it returns $A=0.020\pm0.005$ and $m=1.55\pm0.21$.

The fit parameters we find when including the lowest redshift \citet{kart07} data point is entirely consistent with that found in \citet{xu12a}, whilst the fit when discarding it is in excellent agreement with our new parameterisation including the new GAMA data points, albeit with larger errors. This is expected given the poorer statistical constraints offered by the available data without the GAMA results. Collectively this suggests that the new fitting parameters ($A=0.021 \pm 0.001$ and $m=1.53 \pm 0.08$) are good estimates of the true power-law model, and previous estimates have been systematically biased by the low redshift SDSS derived \citet{kart07} data point. All but the lowest data point in \citet{kart07} is assembled from COSMOS HST data, however the lowest data point itself was derived from the close-pairs catalogue of \citet{all04}. This work utilised a very restrictive definition of close-pair, requiring spatial proximity relative to the physical size of the galaxies rather than simply a constant angular separation criterion. For this reason it seems likely that the derived estimate is much lower than the intrinsic value.

Both the \citet{lin08} and \citet{xu12a} parameterisations are strongly rejected once we include our new GAMA data (see the posterior contours in the top-left inset panel of Figure \ref{fig:majmergeevo}). \citet{brid10} also consider a compendium of data including the \citet{kart07} data points, and they find $m=2.83\pm0.29$. However, when using only their CFHTLS-Deep data they find $m=2.33\pm0.72$, which brings their result for the power law slope into statistical agreement with the figure presented here.

\subsection{Merger Inputs and Outputs}

\label{sec:inoutmerge}

Because of the large stellar mass dynamic range explored in GAMA, we can make a detailed analysis of the stellar masses of galaxies both entering mergers (as implied by the close-pairs fractions) and of galaxies produced by merger events. By looking at these merger inputs and outputs we can make an estimate of the likely evolution of the GSMF due to the effect of mergers alone (i.e.\ separate to any evolution due to secular stellar evolution taking place in these galaxies, or smooth accretion of gas).

\begin{figure}
\centerline{\mbox{\includegraphics[width=3.7in]{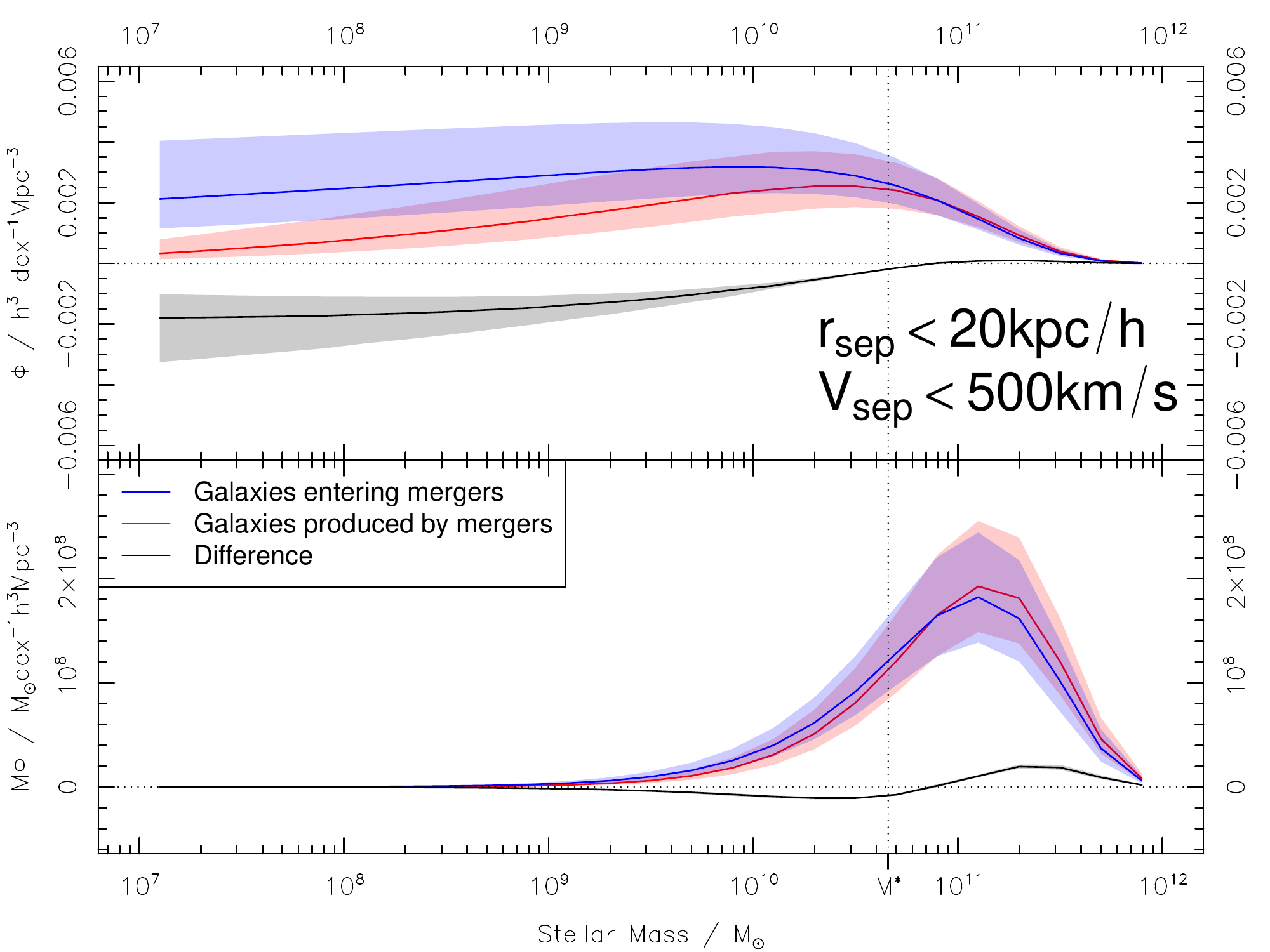}}}
\centerline{\mbox{\includegraphics[width=3.7in]{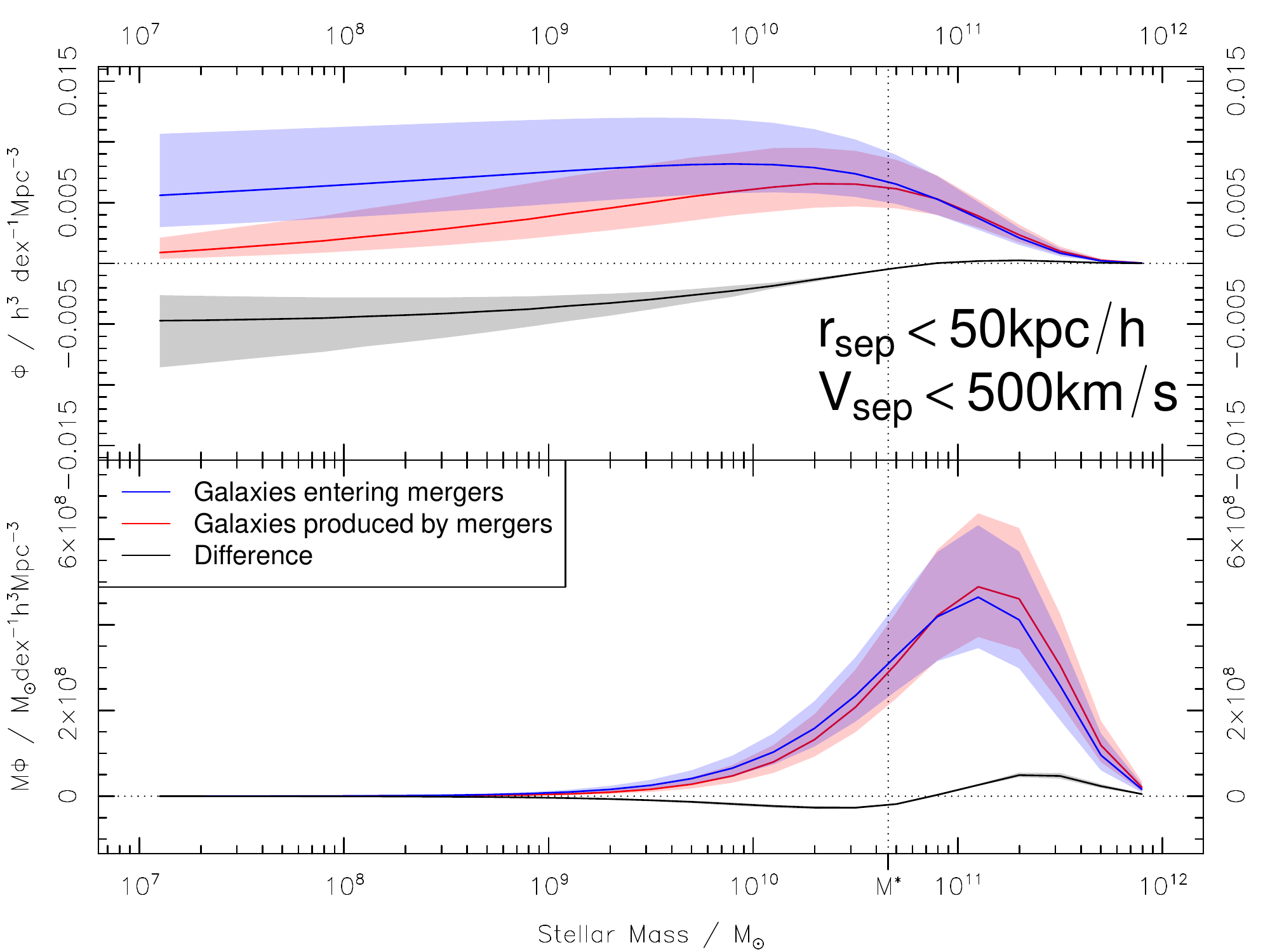}}}
\centerline{\mbox{\includegraphics[width=3.7in]{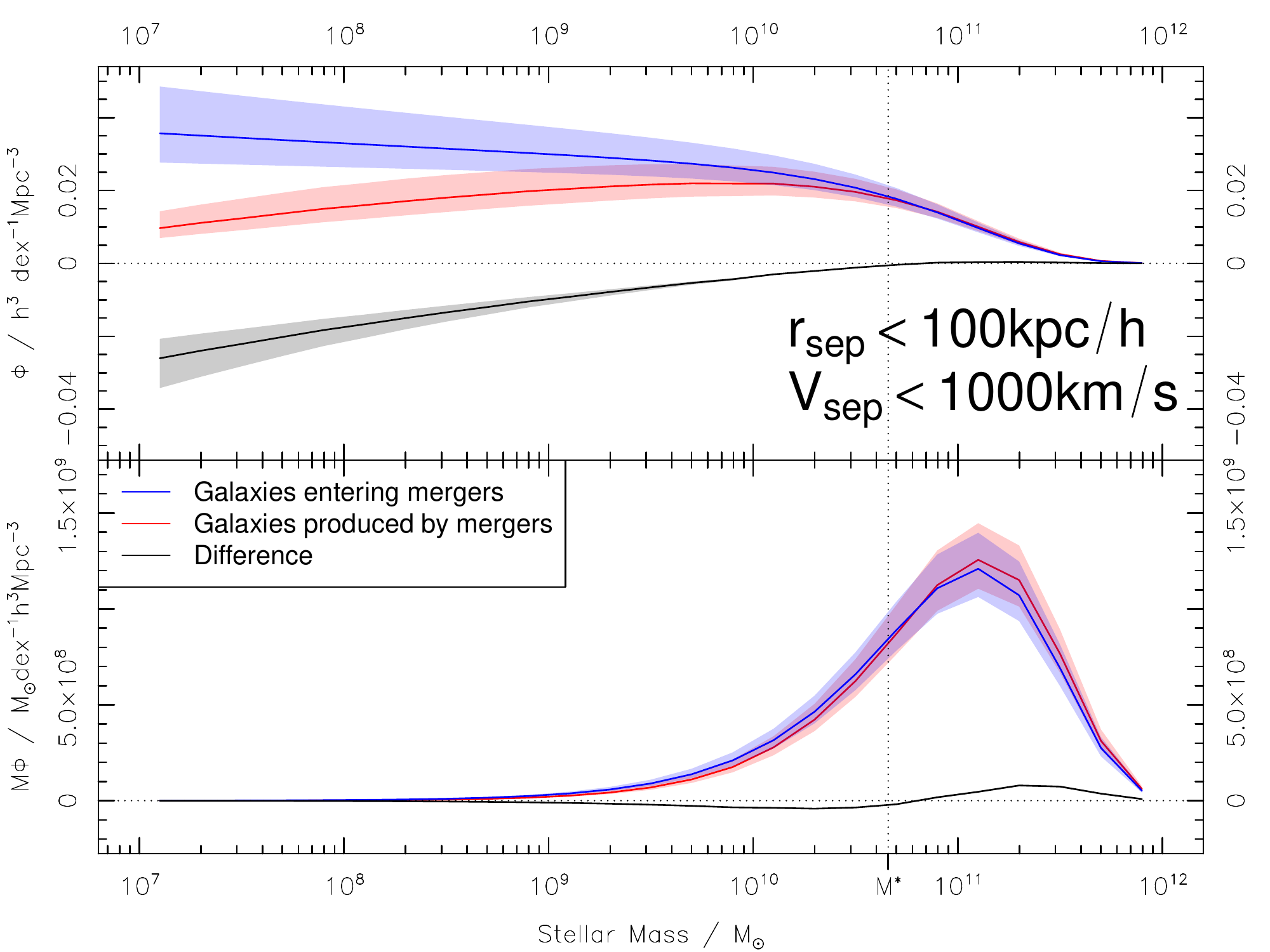}}}
\caption{\small Depiction of the number density of mergers in different stellar mass bins and the resultant product of all of these mergers for different dynamical pair selections. 
In all cases the lines shown are for the simplest observational incompleteness corrections and corrections for galaxies appearing in multiple close-pairs. Further corrections can be made for mock catalogue projection effects and galaxy visual disturbance fractions, as discussed in Section \ref{sec:datafit}. A general observation is that in all cases we see stellar mass being moved from sub-$\mathcal{M}^*$ to super-$\mathcal{M}^*$ regions of the GSMF.}
\label{fig:numdiffmerge}
\end{figure}

The inputs and outputs from mergers can be self-consistently (i.e.\ guaranteed to conserve mass) assessed by using the analytic fits to the GAMA close-pair data presented in Section \ref{sec:closepairfits}. The comoving density of inputs and outputs to close-pairs uncorrected for projection effects or visual disturbance is shown in Figure \ref{fig:numdiffmerge}.

The $PS_{r20v500}$ and $PS_{r50v500}$ selections are very similar modulo a difference in the normalisation (the $PS_{r50v500}$ selection has a larger number of inputs to and outputs from mergers). At low masses the net effect is that the GSMF is depleted by merger activity (see black lines in each Figure panel), and at higher masses the net effect is the GSMF is enhanced. The transition point (where the inputs and outputs are equal) is very close to $\mathcal{M}^*$ for the GSMF (indicated by the vertical dashed line). As we saw for Figure \ref{fig:majmergerate}, this suggests that $\mathcal{M}^*$ is the key region of interest in terms of merger activity.

Since the $\alpha_{CP}$ of the fits is greater than -1 for these two dynamical selections, the comoving number density of mass entering mergers actually has a maximum at moderate stellar mass ($\sim4\times10^{9}\msol$). This is close to the dip in the GSMF seen clearly in \citet{bald12}. Inevitably, this means the dip in the GSMF will become more prominent after these likely future close-pair mergers take place, and the double-Schechter characteristics of the GSMF will be enhanced. It is important to note that a single-Schechter form of the GSMF modified by a CPSMF that has a value of $\alpha_{CP}$ or $\mathcal{M}^*_{CP}$ that differs to the equivalent GSMF parameters will necessarily become a double-Schechter function. Depending on whether $\mathcal{M}^*_{CP}$ is larger (smaller) the resulting GSMF will have a dip (hump) below $\mathcal{M}^*$. The CPSMF $\alpha_{CP}$ determines the degree to which mass movement occurs primarily due to minor merges (more negative) or major mergers (less negative).

The $PS_{r100v1000}$ selection is not as well constrained, with the inputs to close-pair mergers still diverging at lower masses. It also has a broader region of enhanced galaxy creation, with net production at stellar masses less than $\mathcal{M}^*$. If we make the reasonable assumption that, on average, galaxies in the $PS_{r100v1000}$ selection will merge on longer timescales than the other dynamical windows investigated in this work, the implication is that the typical mass of the minor galaxy in a merger will become less with time.

The implication of likely merger inputs and outputs can be seen in light of galaxy transformations. In the regime where galaxies are more likely to be entering mergers than to be products of lower mass mergers, i.e.\ stellar masses below $10^{10}\msol$, any galaxy we observe is likely to {\it not} be the product of a recent merger. In contrast, if we consider galaxies with stellar masses more massive than $\mathcal{M}^*$ then it is increasingly likely that any given galaxy is the product of a recent merger. We can make these broad claims without explicitly specifying timescales due to the declining $\alpha_{CP}$ slope and because $\mathcal{M}^*_{CP}>\mathcal{M}^*$, i.e.\ this is merely a statistical argument. Recent work by \citet{kann13} suggests $\mathcal{M}^*$ is a transition point between HI-gas-rich bulge-less disk galaxies at lower masses and spheroid dominated HI-gas-poor massive galaxies. On the assumption that merger activity does have a role to play in morphologically transforming disks to spheroids, and also removes HI gas \citep[leading to the observed cessation of star formation in][]{robo13}, then the shape of our close-pair distribution function is in good qualitative agreement with these results.

\subsection{Star-Formation Versus Mergers}

As well as galaxy stellar mass increasing through the continual accretion of less massive galaxies, stellar mass can also be increased through the conversion of gas into stars. Since merger activity is occurring disproportionately at higher stellar masses \citep[seen in this work, but also in e.g.][]{bund09}, there should be a corresponding effect on the star formation activity of these galaxies. Particular subsets are likely to experience the effects of mergers differently. Work by \citet{darg10b} suggests that spiral galaxies see a doubling of their typical star-formation during major merger events, whilst the star-formation rates of quiescent galaxies is largely unchanged. Counter to this are simulations of wet mergers at $1<z<2$ presented in \citet{per14}. This work does not find any strong evidence for net star formation enhancement triggered by minor or major mergers.

A full investigation of the role merger activity has on specific star formation rates is deferred to a future collaboration paper, but the implication of the results in \citet{robo13} is that galaxies more massive than $\mathcal{M}^*$ might have their star formation more efficiently shut down due to close-pair interactions since $\mathcal{M}^*_{CP}>\mathcal{M}^*$. In this respect, close-pairs could naturally contribute towards the down-sizing signal at low redshift ($z<0.1$ in this work) since massive galaxies are more likely to be disturbed by galaxy-galaxy interactions, i.e.\ they are more likely to have had their star formation shut down and to be morphologically transformed \citep{robo13}.

The role of star-formation in building up stellar material has been comprehensively explored in \citet{baue13} which used data from the GAMA survey. We can directly compare the mass contribution arriving in galaxies via in-situ star formation \citep[we ignore the minor extra effect of star-formation triggering/suppression in interacting galaxies since the dominant effect is ambiguous][]{robo13,per14}.

\begin{figure}
\centerline{\mbox{\includegraphics[width=3.7in]{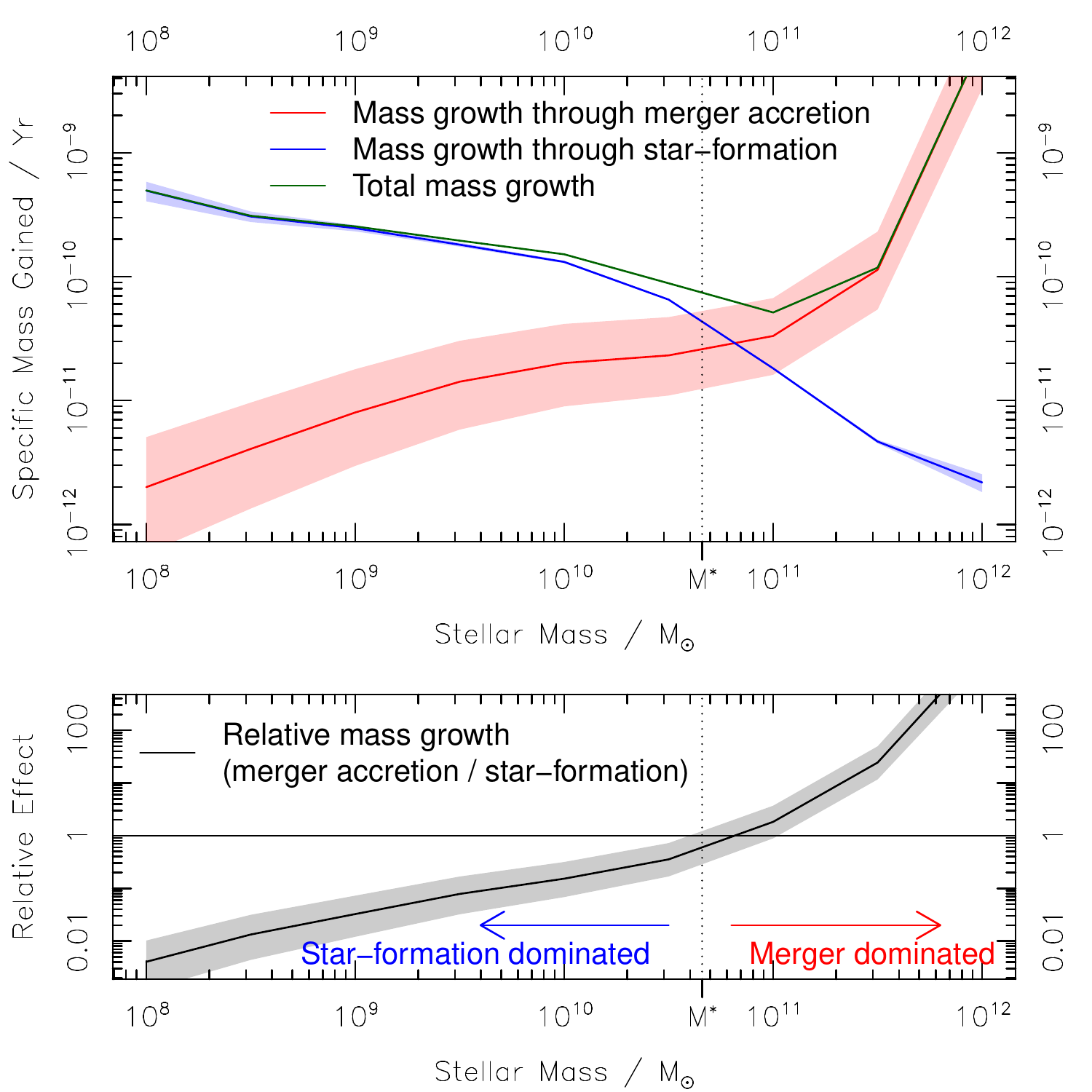}}}
\caption{\small Top panel shows the fraction of mass added to current galaxy stellar mass via various mechanisms. The red line shows the fraction of increase expected via minor galaxy accretion onto galaxies where we use close-pair fractions derived from our most dynamically compact $PS_{r20v500}$ close-pair sample, true pair corrected using the mock catalogues ($W_{TP-mock}$, see Section \ref{sec:TPcor}). The timescale is assumed to be 1 Gyr with a lower limit of 0.5 Gyr and upper limit 2 Gyr, where the red shaded region folds this uncertainty in with the errors in our parameterisation. The blue line shows the effect of taking the \citet{baue13} $sSFR$ at low redshift. The green line is the sum of the red and blue lines (merger accretion and star-formation). The bottom panel shows the mass increase through merger accretion relative to that from star-formation.}
\label{fig:mergevSF}
\end{figure}

Figure \ref{fig:mergevSF} shows the competing effects of minor mass accretion versus star formation for stellar masses ranging from $10^{8}\msol$ to $10^{12}\msol$ using specific star formation rates ($sSFR$) taken from \citet{baue13}. We can see that the star-forming rate dominates the mechanism for mass addition all the way up to $\mathcal{M}^*$, but beyond this point the majority of mass being added to galaxies arrives by virtue of galaxy accretion events, i.e.\ mergers. In this mass regime the majority of major merger events (which we now know dominate mergers in terms of mass involved) will be dry mergers, since quiescent galaxies are the most numerous type \citep{baue13,robo13}. In terms of properly understanding galaxy evolution at low redshift, mergers are significant in the redistribution of mass above $10^{10}\msol$, but barely relevant at all below $10^{9}\msol$. The observational results we present here are in good qualitative agreement with the simulation results presented in \citet{lhui12}. They find that smooth accretion, an event associated with star formation, is more prevalent for lower mass systems, whilst mergers dominate mass assembly at the largest stellar masses.

\subsection{Galaxy Stellar Mass Function Future due to Mergers}
\label{sec:GSMFfuture}

\begin{figure}
\centerline{\mbox{\includegraphics[width=3.7in]{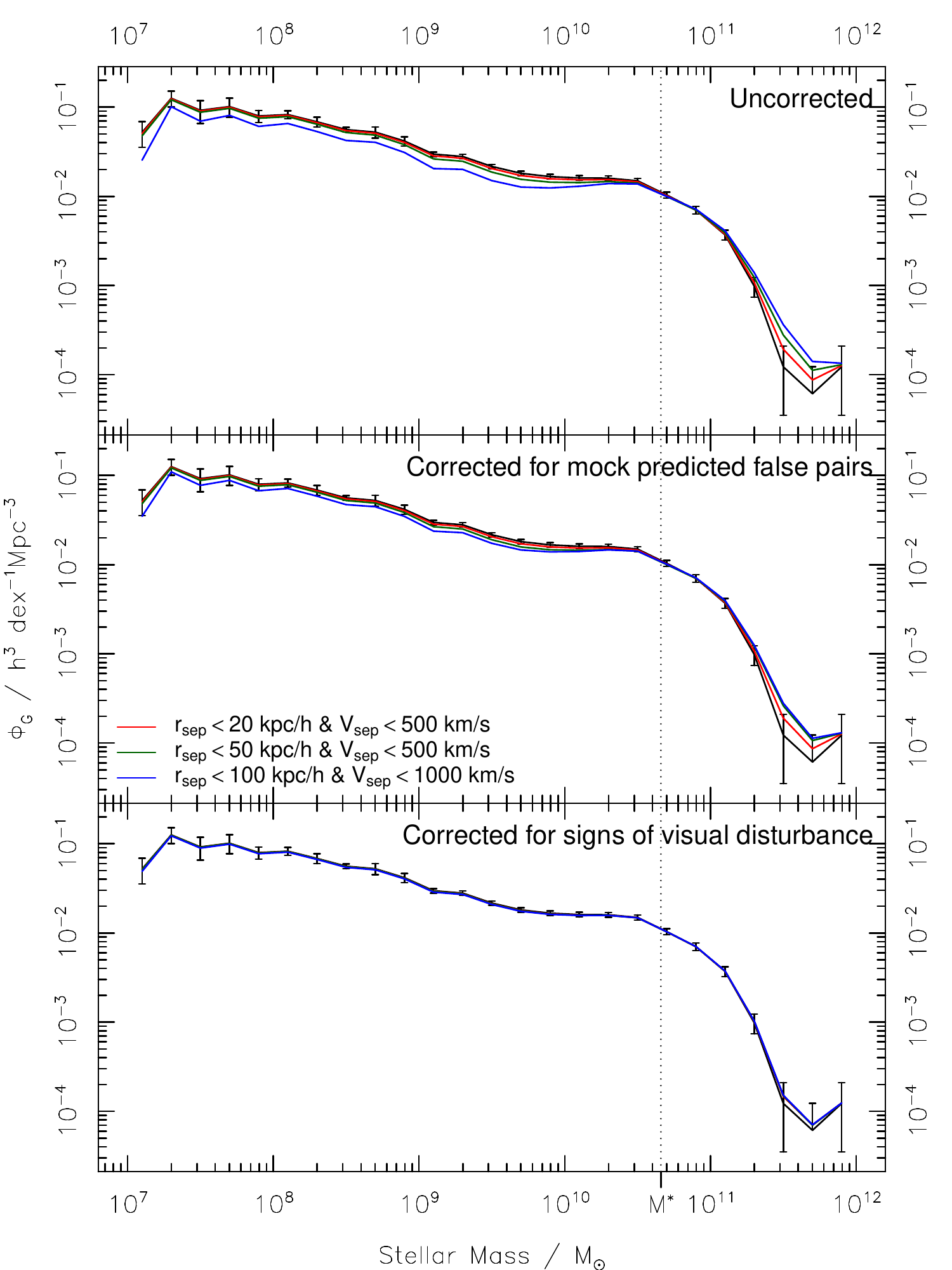}}}
\caption{\small Depiction of the effect on the GSMF presented in \citet{bald12} if the different close-pair selections shown in Figure \ref{fig:numdiffmerge} were assumed to merge on some unknown timescale. The top panel is only corrected for galaxies appearing in multiple close-pairs. The middle panel is true pair corrected from the mock catalogues using $W_{TP-mock}$ (see Section \ref{sec:TPcor}). The bottom panel is corrected for the fraction of visually disturbed galaxies in pairs using $W_{TP-vc}$ (see Section \ref{sec:TPcor}). In all cases mass is moved from moderate stellar mass galaxies ($10^8 M_{\sun}$--$10^{10} M_{\sun}$) to beyond $\mathcal{M}^*$ (although this is hard to identify in the bottom panel). This has the inevitable effect of enhancing the dip in this regime, increasing the significance of the double-Schechter shape in the GSMF and moving the future value of the GSMF $\mathcal{M}^*$ to larger masses.}
\label{fig:GSMFfuture}
\end{figure}

We can directly apply the inputs to and outputs from mergers to the GSMF. Figure \ref{fig:GSMFfuture} shows the effect of taking the net product curves presented in Section \ref{sec:inoutmerge} and Figure \ref{fig:numdiffmerge}. The top panel shows the direct application of the net product cures, with no scaling made for completeness biases in the close-pairs. The middle panel shows the effect of scaling each curve by the mock catalogue estimated false close-pair rate (see Section \ref{sec:TPcor}). The bottom panel shows the effect of scaling by the net visual disturbance of galaxies in each sample (see Section \ref{sec:TPcor}).

The top two panels show quite clearly how the dip in the GSMF just below $10^{10}\msol$ will become enhanced with time due to mergers alone, ignoring other processes (e.g.\ secular, AGN, gas infall etc) that may be taking place within galaxies. Considering the $PS_{r20v500}$ and $PS_{r50v500}$ selections, we can see that the number densities in the GSMF pivot around $\mathcal{M}^*$, net shifting mass from the sub-$\mathcal{M}^*$ to super-$\mathcal{M}^*$.

\begin{figure}
\centerline{\mbox{\includegraphics[width=3.7in]{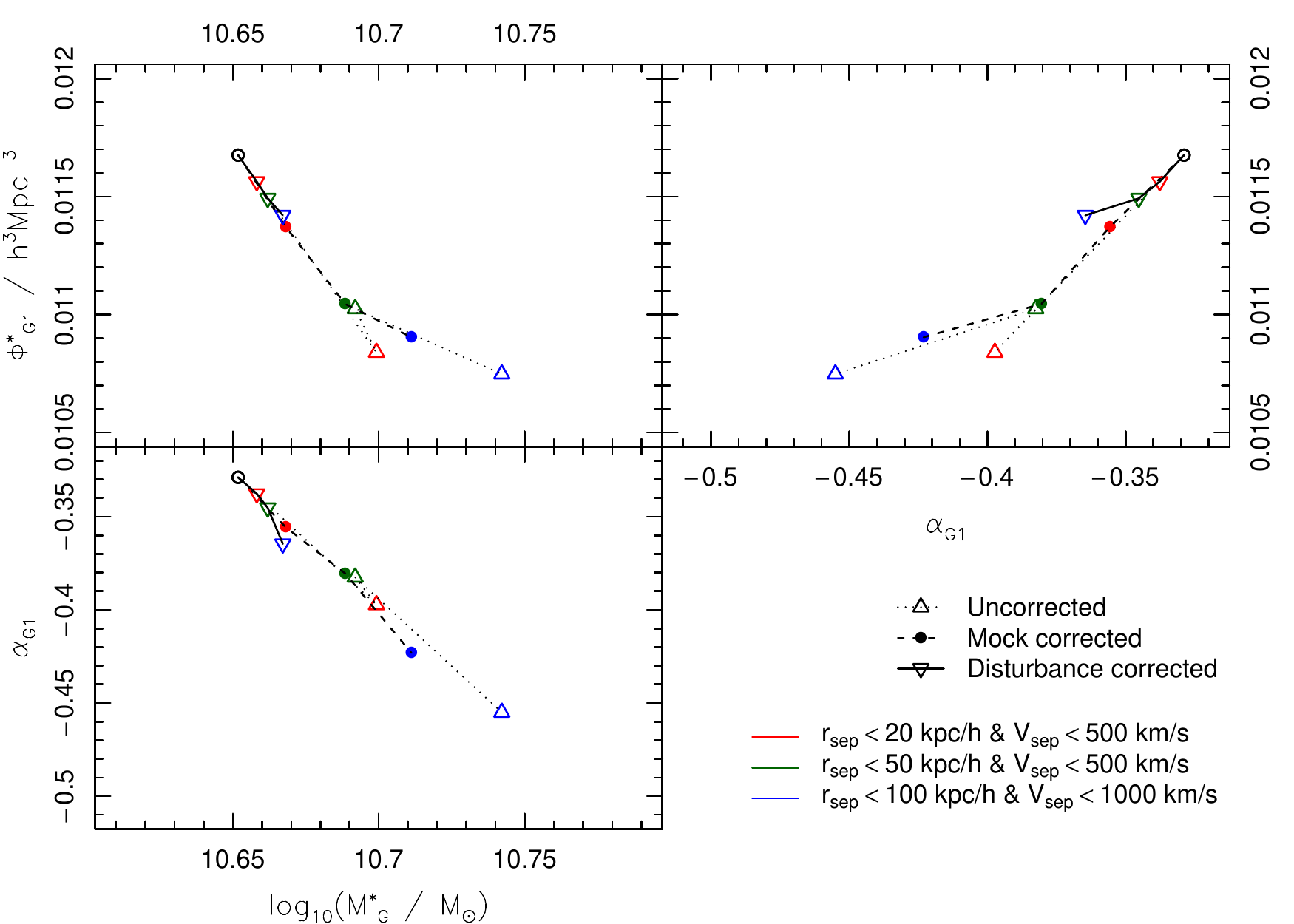}}}
\centerline{\mbox{\includegraphics[width=3.7in]{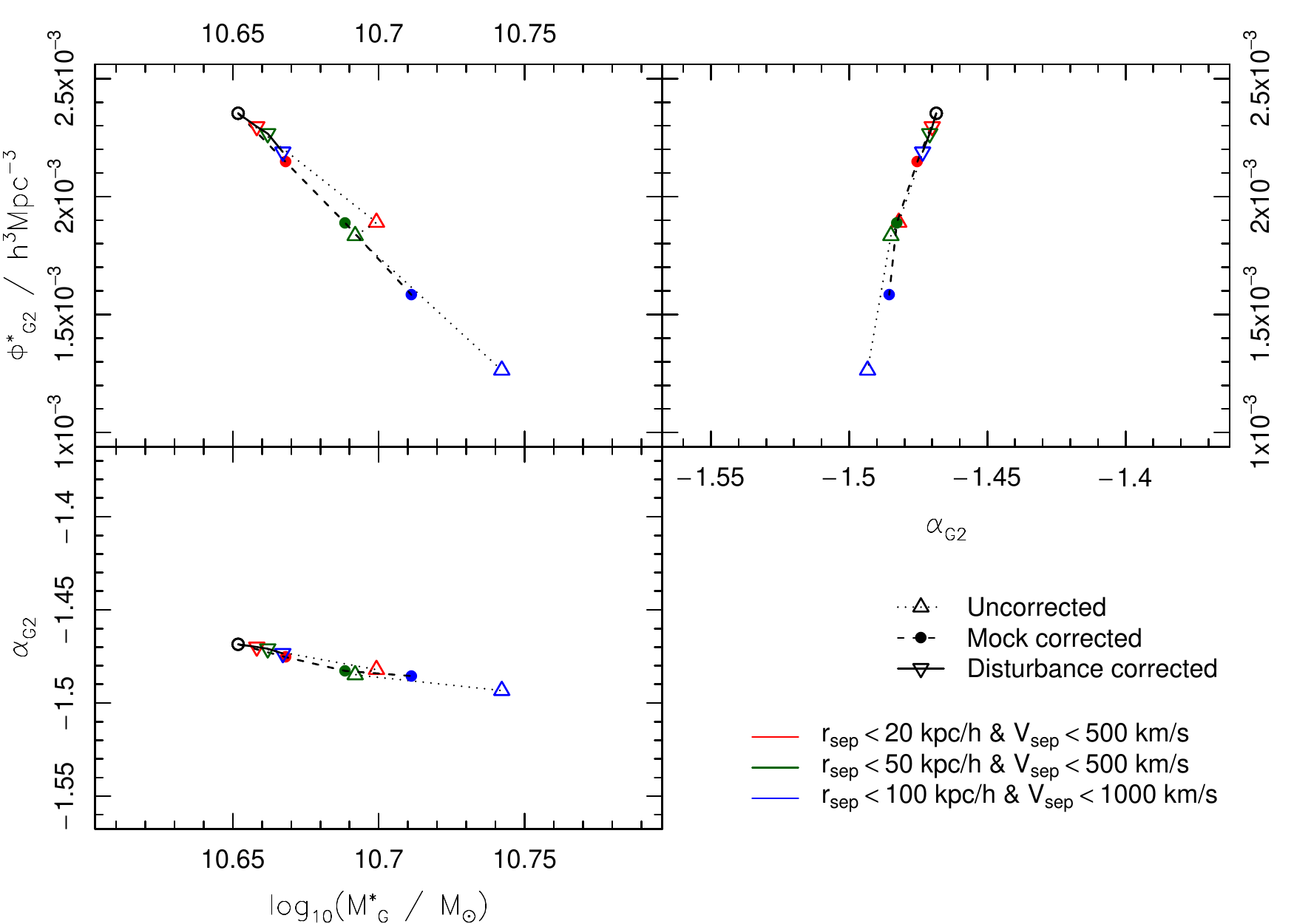}}}
\caption{\small The two sets of panels show the various two-paramter shifts for the Bayesian expectation of the GSMF posterior parameterisation, given different future merger history scenarios. In all panels the original parameter fit is shown by the open black circle.}
\label{fig:paramevo}
\end{figure}

If we assume the most conservative requirement for future mergers (that the galaxies in the close-pair are already visually disturbed) then the bottom panel shows that the net effect of mergers on the GSMF is quite moderate for most of the stellar mass range for all dynamical selection windows (i.e.\ it is hard to identify any change in the GSMF). However, we can deduce from inspection of Figure \ref{fig:numdiffmerge} in combination with Figure \ref{fig:GSMFfuture} that the most massive galaxies stand to have their number densities enhanced by 10s of percent due to future mergers. These massive galaxies are predominantly central galaxies in their own group halos already.

To quantify these effects we have refitted the GSMF using the parameterisation of \citet{bald12} (see Eqn. \ref{eqn:GSMF}) using the data shown in Figure \ref{fig:GSMFfuture} with the same Metropolis MCMC sampling process we used previously. The basic trends are consistent across all close-pair selections and correction schemes: the normalisations moves downwards (mergers produce fewer galaxies overall) and $\mathcal{M}^*$ becomes more massive (mergers produce enhanced mass beyond our current $z=0$ $\mathcal{M}^*$ mass). The two $\alpha$ slopes are slightly more complex. $\alpha_1$ dominates the massive end of the GSMF, and this becomes steeper as mass migrates along the GSMF, i.e.\ more mass is contained in the integral of this Schechter component. $\alpha_2$ dominates the low mass end of the GSMF, and this barely changes--- only becoming slightly steeper as mass migrates efficiently from moderate sightly sub-$\mathcal{M}^*$ masses to beyond $\mathcal{M}^*$.

The near future evolution of $\mathcal{M}^*$ is of particular interest since it is a relatively simple quantity to compare to future simulation work. Considering the $PS_{r20v500}$ sample, the most generous case for future mergers would imply a $\sim0.05$~dex $\mathcal{M}^*$ shift upwards, whilst the most conservative would suggest it might be as little as $\sim0.01$~dex.

\section{Conclusions}
\label{sec:conclusions}

In this work we have used the highly complete close-pair data of GAMA to fully describe the close-pair distribution as a function of the two stellar mass components. We have used these distributions to make a number of significant conclusions:
\newline
\begin{enumerate}

\item This 2D distribution is well described by the multiplication of two Schechter functions with only a single power law component to describe the low mass end. This analytic form can subsequently be used to calculate the pair fraction in any range of interest, and was used to calculate the mean comoving density of stellar mass currently accreting onto galaxies in the present day Universe (see Figures \ref{fig:MergeFitA20V500}, \ref{fig:MergeFitA50V500} and \ref{fig:MergeFitA100V1000}).
\newline
\item The close-pair fractions for major mergers around $\mathcal{M}^*$ galaxies are seen to be broadly consistent with published values in the literature, but they push the mean close-pair fraction towards being systematically higher. We find a small amount of evidence for possible evolution for close-pair fractions between $z=0.05$--$0.2$, although the results are consistent with this quantity remaining flat over this regime (see Figure \ref{fig:majmergeevo}).
\newline
\item The full close-pair distribution was further `true pair' corrected using mock catalogues ($W_{TP-mock}$) and also for signs of visual disturbance ($W_{TP-vc}$). These corrected forms of the 2D distribution were then used to assess the comoving number density of stellar mass entering and produced by mergers. Depending on how conservatively we select our robust merging systems, the fraction of mass accreting on these timescales is between $2.0$\% and $5.6$\% (see Table \ref{tab:massmerge}).
\newline
\item In the two smallest close-pair selection windows explored in this work we see strong evidence that the net effect of mergers below $\mathcal{M}^*$ is to remove galaxies from the GSMF, with these reappearing above $\mathcal{M}^*$ as a measurable excess in number density (see Figure \ref{fig:numdiffmerge}).
\newline
\item Comparing the effect of in-situ star-formation versus mass accretion through mergers, we find that galaxies below $\mathcal{M}^*$ are likely to obtain most of their mass through star-formation, whilst galaxies above $\mathcal{M}^*$ are likely to obtain most of their mass buildup through the accretion of smaller galaxies (see Figure \ref{fig:mergevSF}).
\newline
\item The point of maximal merger activity, and also where the net difference between mergers inputs and outputs is close to zero, is very close to the $z=0$ measurement for $\mathcal{M}^*$ taken from \citet{bald12}. The final results are that we see the strength of the dip in the stellar mass function is likely to become net enhanced by the merger of galaxies currently in close-pairs in the low redshift Universe (see Figure \ref{fig:GSMFfuture}) and that $\mathcal{M}^*$ will be shifted up to more massive galaxies (see Figure \ref{fig:paramevo}).

\end{enumerate}

\section{Future work}
\label{sec:futurework}

The timescale on which dynamically close-pairs will merge is poorly understood \citep{cons06,kitz08,cons09}. Simulation efforts are now underway to better map a given dynamical selection window onto a probability distribution of likely merger timescales. The results of this new work will allow us to better translate the observational results presented in this paper to a typical merger timescale. Until this work in complete we have resisted attempting to categorically assume a timescale on which our different close-pair selections will merge. In a relative sense it is obvious that the $PS_{r20v500}$  selection will merge faster than the $PS_{r100v1000}$ (on average), but it is not clear whether a visually disturbed close-pair in the $PS_{r100v1000}$ selection will (on average) merge faster than a visually undisturbed close-pair in the $PS_{r20v500}$ selection.

There is also uncertainty in the roles of secular evolution in different environments, and the competing effects of mergers both triggering and shutting down star formation \citep{robo13}. An ongoing aim is to build a coherent picture of how mass is assembled throughout the GSMF, and how it naturally segregates into various bimodal (but not necessarily directly correlated) populations of colour, morphology and star formation. Taylor et~al (submitted) is the first in a series of paper that will investigate these interlinked properties.

This close-pair catalogue will be made publicly available at www.gama-survey.org along with other GAMA data products. It has already been used as the source catalogue for various follow on projects (e.g.\ HST GO-13695, PI. Holwerda). Should researchers wish to make use of the catalogue before general release they should directly contact ASGR.

\section*{Acknowledgments}

ASGR acknowledges STFC and SUPA funding that was used to do this work.
GAMA is a joint European-Australasian project based around a spectroscopic campaign using the Anglo-Australian Telescope. The GAMA input catalogue is based on data taken from the Sloan Digital Sky Survey and the UKIRT Infrared Deep Sky Survey. Complementary imaging of the GAMA regions is being obtained by a number of independent survey programs including GALEX MIS, VST KiDS, VISTA VIKING, WISE, Herschel-ATLAS, GMRT and ASKAP providing UV to radio coverage. GAMA is funded by the STFC (UK), the ARC (Australia), the AAO, and the participating institutions. The GAMA website is {\tt http://www.gama-survey.org/}.

Thank you to the anonymous referee. Their helpful comments resulted in a number of improvements over the original manuscript.

\bibliographystyle{mn2e}
\setlength{\bibhang}{2.0em}
\setlength\labelwidth{0.0em}
\bibliography{lgav1}

\begin{appendix}
\section{Details of the Visual Classification Process}
\label{sec:visclassdetails}

\begin{figure*}
\centerline{
\mbox{\includegraphics[width=2.4in]{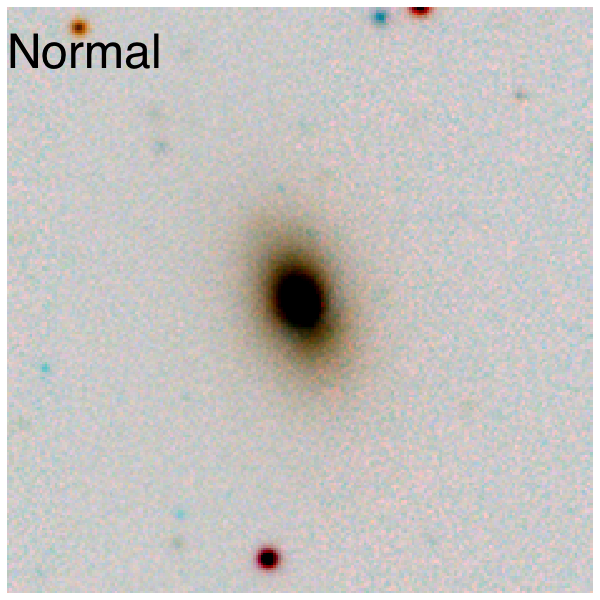}}
\mbox{\includegraphics[width=2.4in]{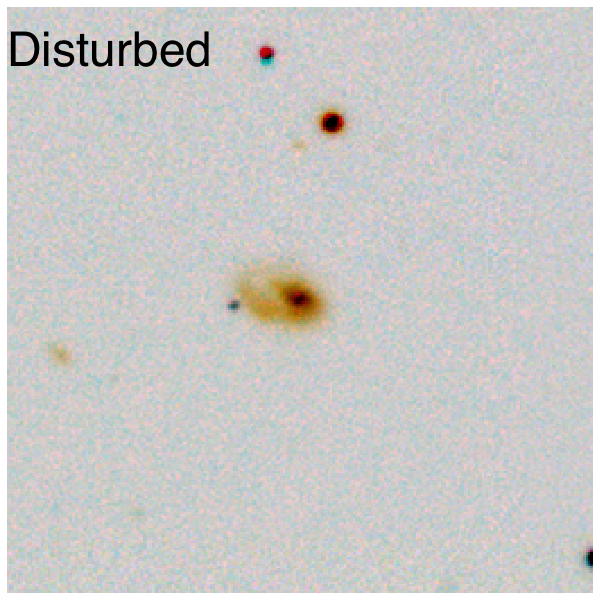}}
\mbox{\includegraphics[width=2.4in]{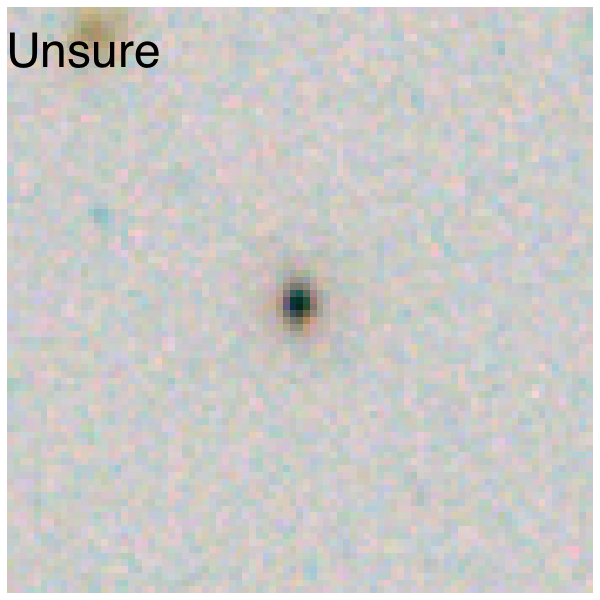}}
}
\caption{\small The three classifications of galaxy disturbance used in this work. All images used are $60\kpch\times60\kpch$ in physical distance and linearly map $g$/$r$/$K$ SDSS and VIKING flux data onto blue, green and red colours which are then inverted. This means visually redder colours correspond to bluer $g-r$ colours in the original images. The left panel is an example of a `normal' galaxy that does not show signs of morphological disturbance, the middle panel is an example of a `disturbed' galaxy and the right panel is an example of a galaxy where our classification would be `unsure' because there are so few pixels of information. Images were randomly selected from the close-pairs subset (i.e.\ all of them have known close-pair companion galaxies, and none are taken from the isolated control sample).}
\label{fig:examplepics}
\end{figure*}

24 GAMA team members volunteered to classify the galaxies, and each observed a batch of $\sim5k$ galaxy images. Each image measured $60\kpch\times60\kpch$ and was centred on the galaxy to be classified (i.e.\ two images were created for each close-pair). In $\sim$80\% of cases the other close-pair galaxy did not appear on the stamp being assessed, and users were asked only to assess the morphological state of the image-centred galaxy. All galaxies were visually classified by at least 4 different random classifiers based on inverted $gr$K colour images, where the simple classes identified were `disturbed' (clear distortion of the light), `normal; (the galaxy is subjectively normal in appearance) and `unsure' (too few pixels to make any classification, or data issue with the image). Figure \ref{fig:examplepics} shows examples of the different galaxy classifications we used for this work. Each observer ($O_{i}$) was then assessed for how consistently they classified galaxies compared to their colleagues.

Due to the subjective nature of the classes and how they were interpreted by observers, there was a substantial amount of overlap between the `unsure' and the `normal' classes: 44\% of galaxies where classifiers all agree the galaxy is not disturbed have a mixture of `unsure' and `normal' classifications. Because of this, and to simplify calculations, the `unsure' class was combined with the `normal' class (both indicating `undisturbed' galaxies). This is a reasonable approach since the redshift classification bias is factored out at a later stage, which is the main cause of `unsure' classifications. i.e.\ the `unsure' class is not a threshold case where the galaxy appears slightly disturbed, but rather there are simply too few pixels to be sure of anything regarding the galaxy morphology.

For each classifier we determined the cumulative number of undisturbed-undisturbed classifications ($D_{UU}$), where they classify a galaxy as undisturbed where the majority of all assessors classify it as undisturbed. Also we calculate the cumulative number of disturbed-disturbed classifications ($D_{DD}$, where they classify a galaxy as disturbed where the majority of all assessors classify it as disturbed). We then calculate classification weights for each classifier, where:

\begin{eqnarray*}
W_{D}(i)&=&D_{DD}(i)/D_{D}(i) \\
W_{U}(i)&=&D_{UU}(i)/D_{U}(i)
\end{eqnarray*}

\noindent where $D_{U}(i)$ is the fraction of all `undisturbed' classifications by classifier i, and $D_{D}(i)$ is the fraction of all `disturbed' classifications by classifier i. Each of their classifications is then replaced by the corresponding classification weight value. This means a classifier who generally agrees with other classifiers regarding `disturbed' classifications will have a high `disturbed' classification weight (near to 1), but if they tend to disagree with `undisturbed' classifications they will have a low `undisturbed' classification weight (near to 0). The classification of each galaxy then becomes the weighted sum of all `disturbed' classifications divided by the weighted sum of all classifications.

As an example, a galaxy might originally have 2 `disturbed' classifications and 2 `undisturbed' classifications. If the two assessors who classified it as `disturbed' have weightings of 0.6 and 0.7 for `disturbed' classifications, and the two assessors who classified it as `undisturbed' have weightings of 0.95 and 0.85 for `undisturbed' classifications  we would end up with 1.3 `disturbed classifications and 1.8 `undisturbed' classifications post weighting. This gives a final `disturbed' score of $1.3/(1.3+1.8)=0.42$, i.e.\ it has a 42\% chance that classifiers, on average, believed it to be disturbed.

\label{sec:vc}
\begin{figure}
\centerline{\mbox{\includegraphics[width=3.7in]{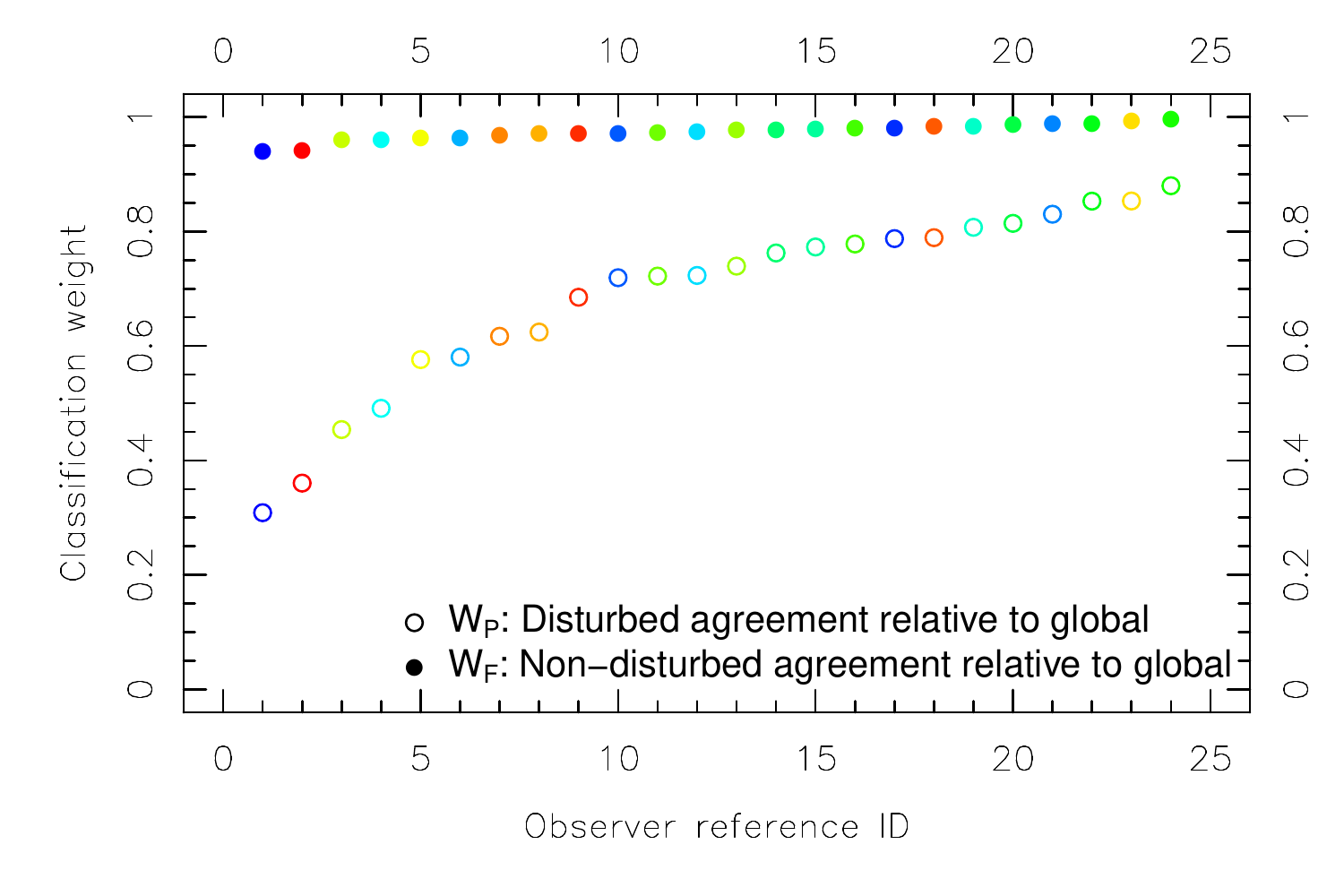}}}
\caption{\small Comparison of classification weights for individual classifiers (24 in total). The classifiers are ordered by their $W_{F}$ weight, i.e.\ how consistently their `undisturbed' classification agrees with that of their fellow classifiers. The y-axis indicates the weight used to re-define classification assessments. If a classifier tends to generously flag galaxies as disturbed compared to the global typical classification (left hand side) their weight for such an assessment is decreased. This is not to say the global choice is always {\it better}, but for this analysis we want all classifications to be biased in the same way.}
\label{fig:disturbresults}
\end{figure}

Figure \ref{fig:disturbresults} shows the `disturbed' and `undisturbed' classification weights for all 24 classifiers. The `undisturbed' weight is uniformly very high. The `disturbed' weight varies quite a lot, from as low as 0.3 for the least representative classifier, up to 0.9 for the classifier who classifies in a manner similar to the global average. This effect is largely an artefact of the data being dominated by `undisturbed' galaxies (even the pairs), so an assessor could be rated as an accurate classifier of `undisturbed' galaxies merely by classifying almost everything as being `undisturbed', but this would of course give them a very poor `disturbed' weight. Looking carefully, the lowest rated classifiers do have a down-turn for both classification types. This is actually important because in ambiguous classification situations the less representative classifiers will not have the casting vote. Since galaxies are assigned randomly to classifiers, an ensemble of galaxies will have a representative mean disturbed fraction as given by the weighted classifications.

\end{appendix}

\end{document}